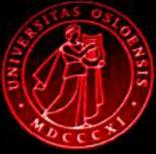

# UiO : University of Oslo

**Pål Sundsøy**

# Measuring patterns of human behaviour through large-scale mobile phone data

Big Data for social sciences

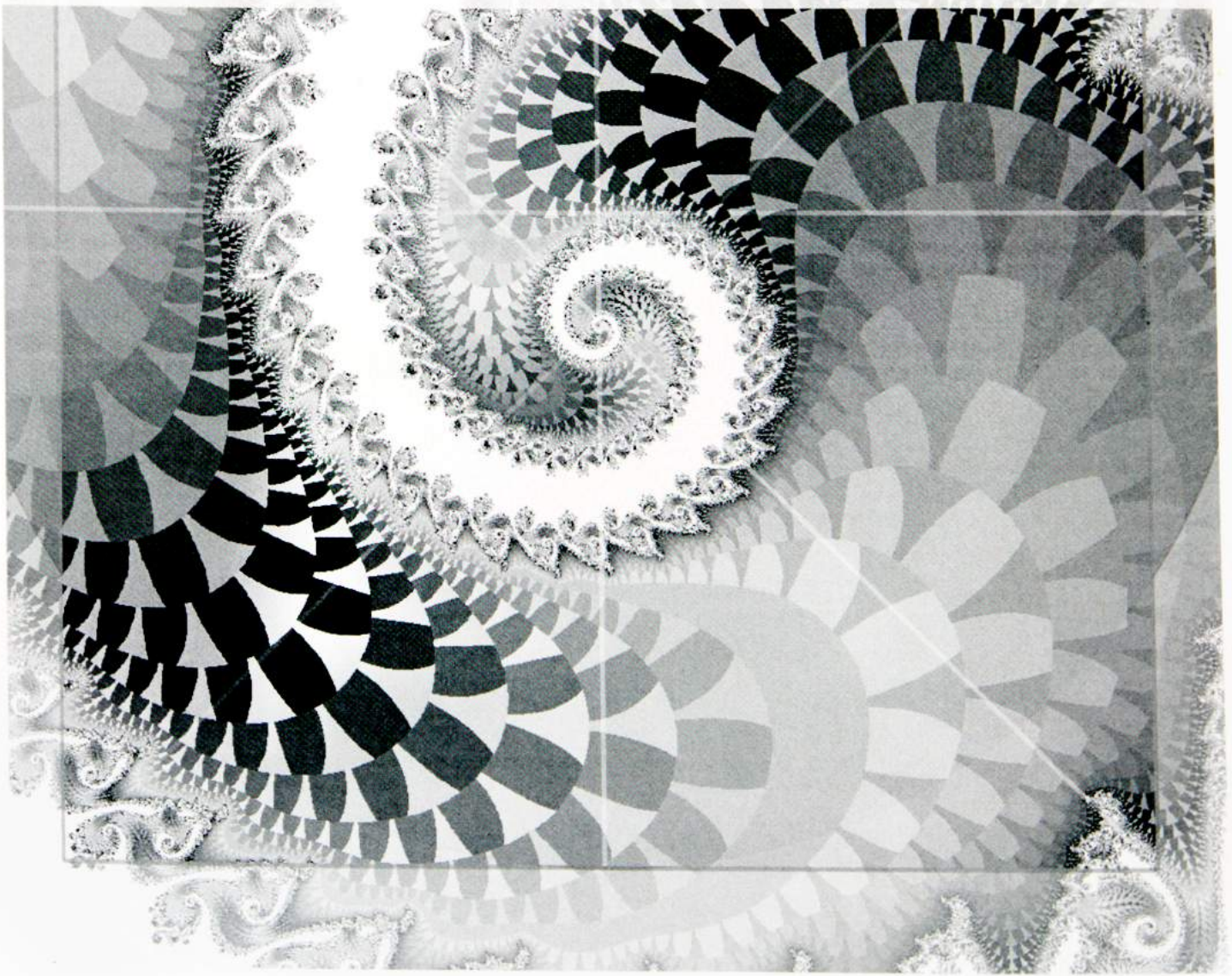

**Faculty of Mathematics and Natural Sciences**　　2017

# Measuring patterns of human behaviour through large-scale mobile phone data

Big Data for social sciences

Pål Sundsøy

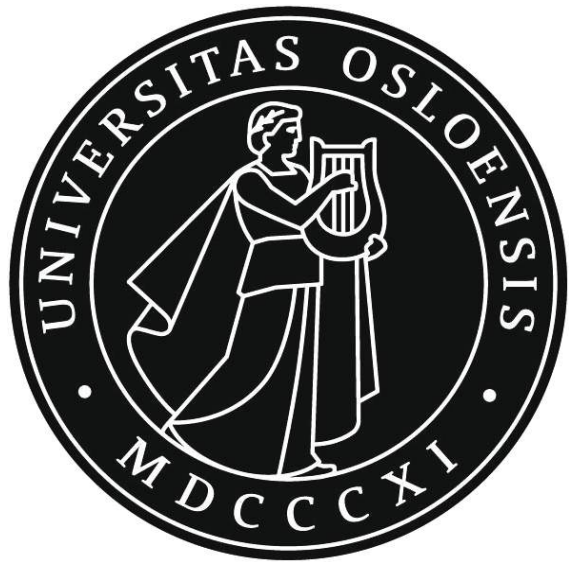

Doctor Philosophiae

Faculty of Mathematics and Natural Sciences

Department of Informatics

UNIVERSITY OF OSLO

February 2017

*'The hope is that as you take the economic pulse in real time, you will be able to respond to anomalies more quickly'*
- Hal Varian, Google Chief Economist (Prof. Emeritus, University of California, Berkley)

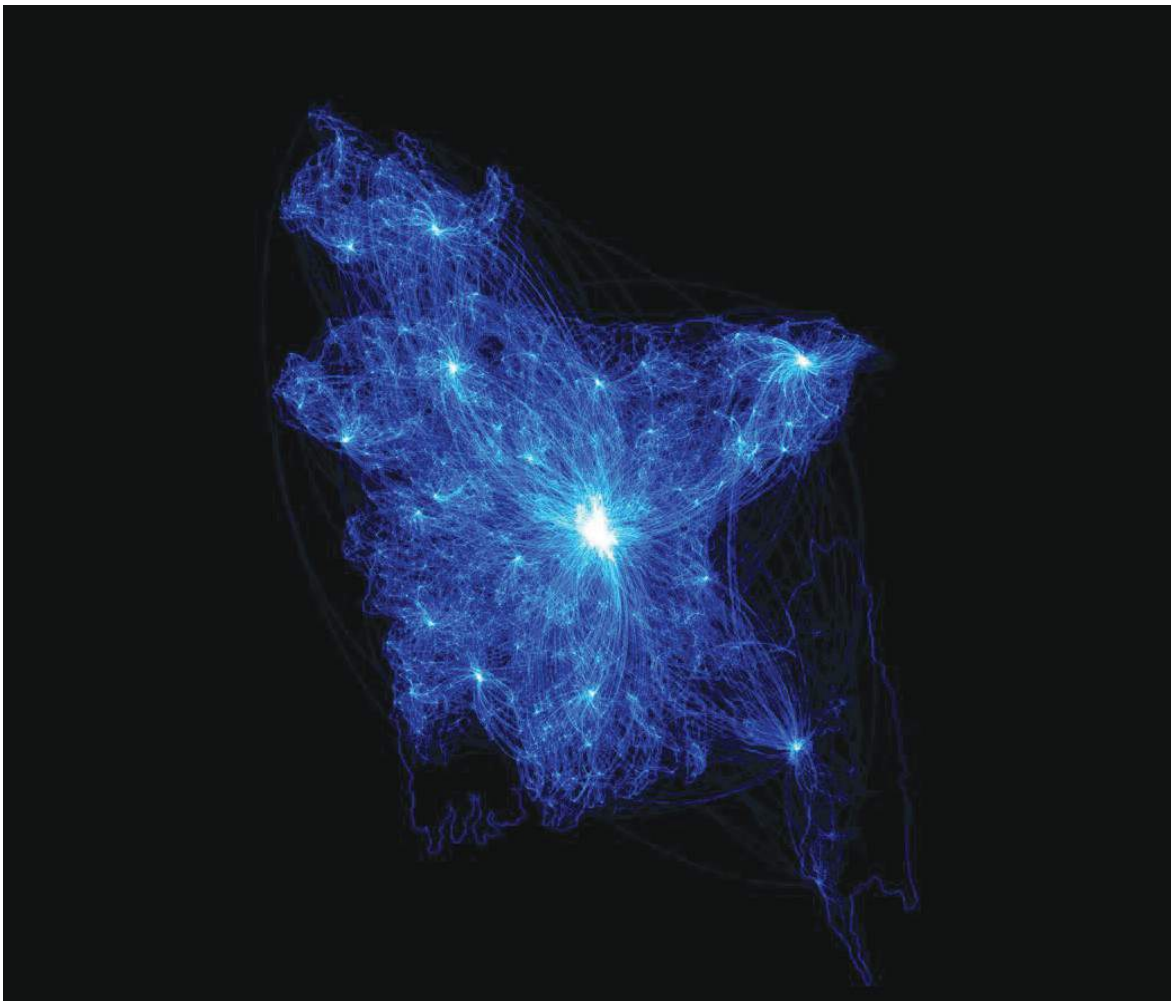

Human migration patterns in Bangladesh, derived from mobility patterns in mobile phone datasets.

Visualisation by Pål Sundsøy.

# List of publications

**1.** **Can mobile usage predict illiteracy in a developing country?**

Preprint available at *arXiv:1607.01337* [cs.AI]. 2016.

**2.** **Deep learning applied to mobile phone data for Individual income classification**

Joint work with Bjelland, J., Reme B.A., Iqbal A. and Jahani, E.

Published in *International conference on Artificial Intelligence: Technologies and Applications* (ICAITA). Atlantic Press. 2016.

**3.** **Mapping Poverty using mobile phone and satellite data**

Joint work with Steele, J.E., Pezzulo, C., Alegana, V., Bird, T., Blumenstock, J., Bjelland J., Engø-Monsen, K., de Montjoye, Y.A., Iqbal, A., Hadiuzzaman, K., Lu, X., Wetter, E., Tatem, A. and Bengtsson, L.

Published in *Journal of The Royal Society Interface 14:20160690*. 2017

**4.** **The activation of core social networks in the wake of the 22 July Oslo bombing**

Joint work with Ling, R., Engø-Monsen, K., Bjelland, J. and Canright, G.

Published in *Social Networks Analysis and Mining ASONAM* (pp. 586-590). 2012.

**5.** **Detecting climate adaptation with mobile network data: Anomalies in communication, mobillity and consumption patterns during Cyclone Mahasen**

Joint work with Lu, X., Wrathall, D., Nadiruzzaman, M., Wetter, E., Iqbal, A., Qureshi, T., Tatem, A., Canright, G., Engø-Monsen, K. and Bengtsson, L.

Published in *Climatic Change*, 138(3-4), pp.505-519. 2016.

**6.** **Comparing and visualizing the social spreading of products on a large-scale social network**

Joint work with Bjelland, J., Engø-Monsen, K., Canright, G. and Ling, R.

Published in *Influence on Technology on Social Network Analysis and Mining, Tanzel Ozyer et. al.* Springer International Publishing. 2012.

**7.** **Big Data-Driven Marketing: How Machine Learning outperforms marketers' gut-feeling**

Joint work with Bjelland, J., Iqbal, A., Pentland, A. and de Montjoye, Y.A.

Published in *International Conference on Social Computing, Behavioral-Cultural Modeling, and Prediction* (pp. 367-374). Springer International Publishing. 2014.

# List of publications not included in the thesis

**8.** Impact of human mobility on the emergence of dengue epidemics in Pakistan

Joint work with Wesolowski, A., Qureshi, T., Boni, M.F., Johansson, M.A., Rasheed, S.B., Engø-Monsen, K. and Buckee, C.O.

Published in *Proceedings of the National Academy of Sciences*, 112(38):11887-92. 2015.

**9.** Improving official statistics in emerging markets using machine learning and mobile phone data

Joint work with Jahani, E., Bengtsson, L., Bjelland, J., Pentland, A. and de Montjoye, Y.A.

In review, *EPJ Data Science*. 2017.

**10.** Unveiling Hidden Migration and Mobility Patterns in Climate Stressed Regions: A Longitudinal Study of Six Million Anonymous Mobile Phone Users in Bangladesh

Joint work with Lu, X., Wrathall, D.J., Nadiruzzaman, M., Wetter, E., Iqbal, A., Qureshi, T., Tatem, A., Canright, G., Engø-Monsen, K. and Bengtsson, L.

Published in *Global Environmental Change* 38, pp.1-7. 2016.

**11.** Small and Even Smaller Circles: The Size of Mobile Phone-Based Core Social Networks in Scandinavia and South Asia

Joint work with Ling, R., Canright, G., Bjelland, J. and Engø-Monsen, K.

Published in *Journal of Intercultural Communication Research* 41(3), pp.320-339. 2012.

**12.** Joy of Giving: Increasing Product Uptake by allowing customers to Forward

Joint work with Bjelland, J., Canright, G., Iqbal, A., Grønnetvet, G., Norton, M. and Reme, B.A.

Article in preparation.

**13.** Handset-centric view of smartphone application use

Joint work with Rana, J., Bjelland, J., Couronne, T., Wagner, D. and Rice, A.

Published in *Procedia Computer Science*, 34, pp.368-375. 2014.

**14.** Small circles: Mobile Telephony and the cultivation of the private

Joint work with Ling, R., Bjelland, J. and Campbell, S.

Published in *The Information Society*, 30(4), pp.282-291. 2014.

**15.** The socio-demographics of texting: An analysis of traffic data

Joint work with Ling, R. and Bertel, T.

Published in *New Media & Society*, 14(2), pp.281-298. 2012.

**16.** **Product adoption networks and their growth in a large mobile phone network**

Joint work with Bjelland, J., Canright, G., Engø-Monsen, K. and Ling, R.

Published in *Advances in Social Networks Analysis and Mining* (pp. 208-216). 2010

**17.** **A Social Network Study of Android VS Apple Smartphone battle**

Joint work with Bjelland, J., Ling, R., Engø-Monsen, K. and Canright, G.

Published in *Advances in Social Networks Analysis and Mining* (pp. 983-987). 2012

**18.** **Using Deep Learning to predict demographics from mobile phone metadata**

Joint work with Felbo, B., Lehmann, S., de Montjoye, Y.A. and Pentland, A.

Article in preparation.

**19.** **Diffusion of Information Through On-Demand Information Seeking Behavior**

Joint work with Riedl, C., Bjelland, J., Canright, G., Iqbal, A., Engø-Monsen, K., Qureshi, T. and Lazer, D.

In review, *PNAS*, 2017.

**20.** **Quantifying socio-economic segregation across continents using mobile phone metadata**

Joint work with Lind, J.T., Kotsadam, A., Reme, B.A. and Bjelland, J.

Article in preparation.

**21.** **Networks and income: Evidence from Individually Matched Income and Mobile Phone Metadata**

Joint work with Jahani, E., Saint-Jacques, G., Bjelland, J., Aral, S. and Pentland, A.

Article in preparation.

# Acknowledgement

This thesis is based on 7 (out of 21) selected papers related to the analysis of human behaviour using large mobile phone datasets. All publications were written after I started at Telenor Research 9 years ago. I am grateful for the opportunity given by my employer to do research on mobile phone datasets.

Doing research on 'Big Data' applied to social sciences has been my main focus during these years, and the interest in this subject has increased gradually throughout my time at Telenor.

Most of the publications are joint work with some very bright individuals. It has been a privilege for me to cooperate with all of you.

I would like to highlight several people. Yves-Alexandre de Montjoye, Eaman Jahani and Alex Pentland at MIT Media Lab; thanks for great research collaboration through several years. You have been a great source of inspiration. Linus Bengtsson and colleagues at Flowminder Foundation; I have really appreciated collaborating with you. Your passion and research dedication towards social good is admirable. Amy Wesolowski/Caroline Buckee at Harvard School of Public Health; thanks for involving me in the work on epidemic spreading. David Lazer at Northeastern; thanks for introducing us to all the great researchers in Boston. Guillaume Saint-Jacques and Sinan Aral at MIT Sloan; thanks for a good and interesting collaboration, bridging the gap between economics and Big Data science. Chris Riedl at Harvard/Northeastern; for a good collaboration on viral spreading. Of course, I also want to give a big thanks to my research colleagues at Telenor Research; Geoffrey Canright and Kenth Engø-Monsen, who introduced me to social network analysis. Rich Ling; you have softened up my (hard physics) view of the world with sociological insight. Johannes Bjelland; thanks for being such a great colleague for 9 years, including all the memorable trips around the world. Bjørn-Atle Reme and Gorm Grønnevet; thanks for the precious advice on economics and good discussions. Asif Iqbal; for helping us with data lobbying when everything seemed hopeless, and for your positivity. Thomas Couronne; thanks for all the nice talks around data visualisation. Juwel Rana and Jo Thori Lind; for valuable advice around thesis structure. I would also like to thank Xin Lu, Linnet Taylor, Sune Lehmann, Bjarke Felbo, Andreas Kotsadam, Taimur Qureshi, Gro Nilsen, Jessica Steele, Andy Tatem, Carla Pezzulo, Erik Wetter, David Wrathall, Weiqing Zhang, Mike Norton, Bjørn Hansen, Astrid Undheim, Christian Tronstad, Hanne-Stine Hallingby, Wenche Nag, Andrew Rice and Daniel Wagner for good discussions.

Oslo, Jan 2017

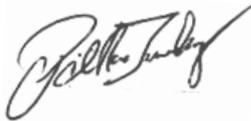

Pål Sundsøy

# Table of contents

# 1 Introduction

Every time a mobile phone customer makes a phone call, sends an SMS or generates Internet traffic there are traces left at the mobile operator. These digital traces, which have traditionally been used mainly for billing purposes, also have other benefits. This thesis is based on 7 publications where mobile phone logs are used to understand large-scale human behaviour. The applications range from improved economic and social wellbeing to marketing, and the level of granularity ranges from the individual to the level of society. Contributions include individual prediction of socio-economic indicators, poverty prediction and understanding human behavioural signals during disasters. Furthermore, how products spread over large social networks is investigated, and how this information can be exploited to perform large-scale marketing experiments. The size of the datasets analysed ranges from 500 million to 300 billion phone records.

Studies 1 and 2 develop scalable predictive models based on mobile phone logs to reliably infer the illiteracy status and income level of individuals. Such insight can be further aggregated to the geographical level to help vulnerable groups in society. Study 3 represents the first attempt to build predictive maps of poverty using a combination of mobile phone and satellite data, with Bangladesh as an example. Knowing where poor people live is a crucial component of poverty eradication, and this study complements expensive approaches that are entirely based on data from traditional surveys with low temporal frequencies. Studies 4 and 5 quantify people's behaviour during larger shocks in society: study 4 reveals human behavioural patterns through the eyes of mobile phone data during the 22$^{nd}$ July 2011 terror attack in Norway, while study 5 shows how people adapt to climate extremes by analysing financial, social and mobility behaviour from 5 million people during a cyclone event in Bangladesh. The aim is to gain understanding and to detect early-warning signals that can help prevent future disasters.

We know that social networks matter when purchase decisions are made. Study 6 is motivated by the question of how people adopt new products and services, and what role the underlying social network structure plays in the process.

Study 7 addresses how social network effects, together with discretionary income and timing, can be modelled and exploited in large-scale marketing experiments in Asia, targeting people with more personalised offers.

## 1.1 Authorship contributions

Authorship contributions for the included articles (I–VII) are specified below.

I. **Can mobile usage predict illiteracy in a developing country?**
   PRS is the sole author.

II. **Deep learning applied to mobile phone data for individual income classification**
   PRS took the initiative, performed the analysis and wrote the paper. JØB, BAR, AI and EJ commented on the draft and suggested improvements for the paper. PRS is the first author.

III. **Mapping poverty using mobile phone and satellite data**
   JS held the main responsibility for the paper and the overall analysis. CP was responsible for survey data management, cleaning and processing, and interpretation and drafting of the final manuscript. PRS were responsible for management of the project from Telenor side, the CDR data management, cleaning, and production of CDR data, and interpretation and drafting of the final manuscript. JB, J.Bj, KE was responsible for interpretation, drafting, and production of the final manuscript. V.A., T.B., Y.M., X.L. and E.W. were responsible for interpretation and production of the final manuscript. A.I. and K.N.H. for handling of income survey data. AJT and LB were responsible for overall scientific management, interpretation and production of the final manuscript. All authors gave final approval for publication.

IV. **The activation of core social networks in the wake of the 22nd July Oslo bombing**
   PRS took the initiative, collected the data, performed the analysis, interpreted the data and wrote the article. RL, JB, KEM, GC interpreted the data, revised the article and approved it. PRS is the first author.

V. **Detecting climate adaptation with mobile network data: Anomalies in communication, mobility and consumption patterns during Cyclone Mahasen**
   XL held the main responsibility for the paper and overall analysis. DW positioned the paper from a development perspective. PRS collected and prepared the data, analysed the financial top-up data and revised the article. MN EW, AI, TQ, AT, GC, KEM, LB interpreted the data, suggested changes and gave the final approval of the article. PRS is the third author.

VI. **Comparing and visualising the social spreading of products on a large-scale social network**
   PRS took the initiative, collected and visualised the data, performed the analysis and drafted the article. RL suggested changes and placed the paper in a sociological context. JB, KEM and GC interpreted, revised and approved the article. PRS is the first author.

VII. **Big Data-driven marketing: How machine learning outperforms marketers' gut feeling**
   PRS and JB coded the experiment and made the model. Several trips to South-East Asia were required to prepare and run the experiment. AI arranged access to data sources. PRS performed the post-analysis and wrote the article. YA and AP suggested improvements, revised the article and positioned the paper. PRS is the first author.

## 1.2 Research objective

The key research question in this thesis can be formulated as follows:

**Question:** Apart from providing basic communication services, what kinds of positive impacts can we create for society and/or individuals using large-scale mobile phone datasets?

## 1.3 Outline of the study

The high-level dissertation roadmap is shown in Figure 1.1. This thesis argues that mobile phone data can be used to:

1. Inform socially beneficial policies.
2. Provide additional insights into human behaviour, with the aim of gaining:
   I. A better understanding of human behaviour and interactions.
   II. Better insights into human behaviour to improve marketing.

The publications in the first category include empirical studies that address challenges in society and how they can be tackled in a different way using mobile phone logs, while complementing existing approaches. The second category addresses how mobile phone logs can be used to obtain new behavioural insights and how such information can be used experimentally for marketing.

**Figure 1.1: Thesis theme overview. Publication number is given for each topic**

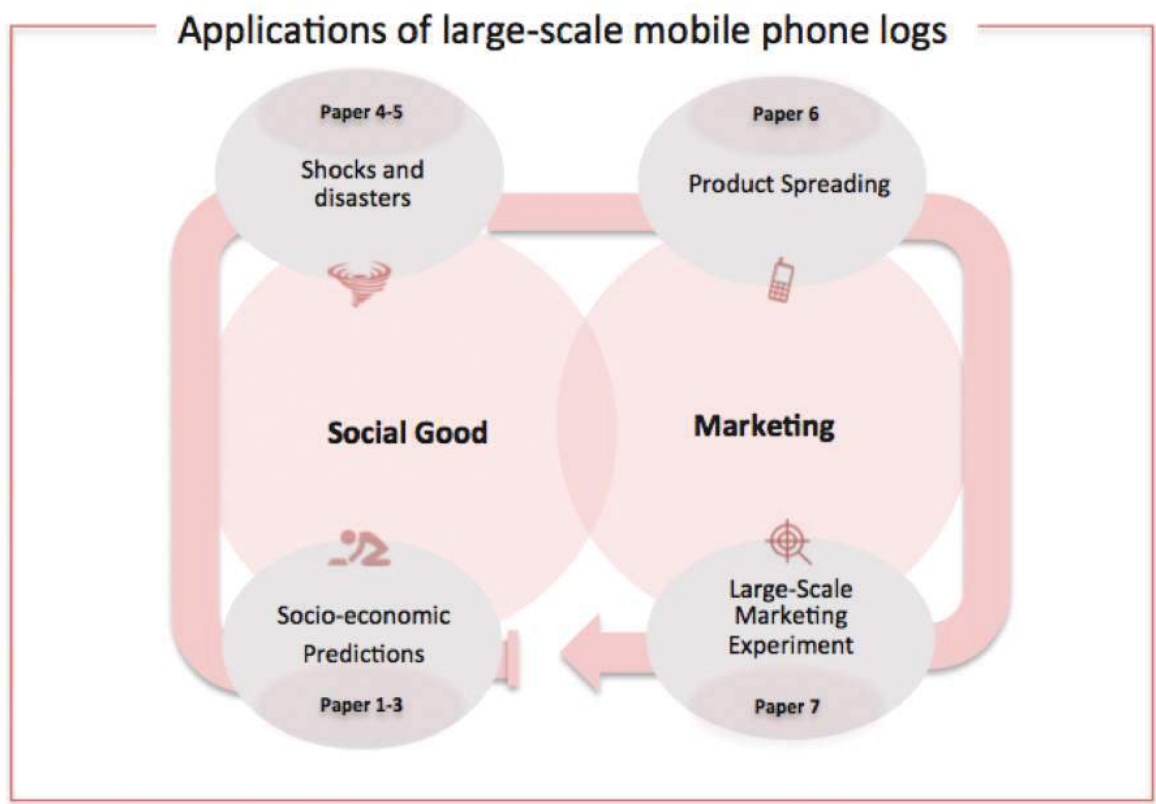

Chapter 2 reviews the literature, with a special emphasis on how mobile phone metadata has been used in the social sciences to date.

In Chapter 3 the research design, tools used and analytical framework are outlined.

Chapter 4 reviews the 7 papers ordered by topic, as shown in Figure 1.1, and introduces the problem, research findings and evaluation for each one.

Chapter 5 discusses the main research question, the limitations and challenges from a holistic point of view, followed by concluding remarks in Chapter 6. All publications are attached after the Bibliography.

# 2 Review of literature

The availability of large datasets, often referred to as 'Big Data', has opened the possibility of improving our understanding of society and human behaviour. The generation and use of large volumes of data is reshaping our social and economic landscapes, creating new industries, products and processes, and producing significant competitive advantages [1].

## 2.1 Big Data for social sciences

Recent research has found that countries could make much more use of data analytics in terms of economic and social benefits if governments did more to encourage investment in Big Data and to promote data sharing and reuse [2]. A consensual definition of Big Data is presented by De Mauro et al. : '*Big Data represents the information assets characterized by such a high volume, velocity and variety to require specific technology and analytical methods for its transformation into value*' [3]. Naturally for Big Data, 'size' is a constantly moving target.

Studies have shown that Big Data has the potential to improve health policies [4], understand large-scale social networks [5], improve the efficiency of poverty prediction [6,7], run large-scale experiments and improve the understanding of urban development by, for instance, considering people's mobility patterns [8]. Data-driven methods have also outperformed traditional marketing approaches [9]. By combining behavioural patterns in Big Data with traditional data it has also been shown to be useful in epidemic spreading predictions [10], and to compliment UN official statistics [11].

In summary, large-scale datasets of human behaviour have the potential to fundamentally transform performing social science research [12].

Traditionally, the broad range of social sciences have focused on explanatory models, whereas researchers from computational sciences have targeted predictive models or observational/'found' data [13]. Conventional statistical and econometric techniques such as regression often work well, but there are issues distinct to big datasets that may depend upon different approaches [14]. While large amounts of data will not overcome the selection problems that make causal inference difficult, this thesis shows that it can provide opportunities to gain understanding and run experiments on a scale that was previously impossible in the social sciences. Recent literature also supports that there are many policy applications where causal inference is not central, or even necessary, and where machine learning is useful for solving prediction problems and generating high social impact [15]. One of the most promising and rich Big Data sources is mobile phone datasets. The mobile ecosystem is a major driver of economic progress and welfare globally. In 2014, the mobile industry generated 3.8% of global GDP [16]. Half of the world's population now has a mobile subscription, and it has been reported that the global penetration rate in 2020 is expected to be 60%. De-identified mobile phone data is a promising source that has the potential to deliver near real-time information of human behaviour on both an individual and societal scale. The next chapter introduces the details of this promising source.

## 2.2 Conceptual framework

### 2.2.1 Mobile phone metadata – what are CDRs?

Whenever a mobile phone call or other transactions are made, a call detail record (CDR) is generated by the mobile operator [17]. A CDR contains, for instance, the start time and duration of a call, but does not provide any information about the content, and is therefore defined as mobile phone *metadata*. Other information recorded in CDRs is which cell tower the caller (and often recipient's) phones were connected to at the time of the call. It is therefore possible to use the CDRs to approximate the location of a user. A sample of CDR fields is presented in Table 2.1. The location is usually found by coupling the cell ID to an external mapping table containing the actual positions (lon,lat) of the tower.

**Table 2.1: Example of set of CDR fields**

| Calling party | Called party | Caller cell ID | Call time | Type | Call duration | IMSI | IMEI |
|---|---|---|---|---|---|---|---|
| 91845206 | 92234065 | 6D45X | 15.05.2016 14:24:50 | Voice | 200 | 4798X | 5840X |
| 91845206 | | A56DE | 15.05.2016 20:10:13 | Internet | | 4798X | 5850X |
| 92234065 | 91845206 | A56DE | 16.05.2016 15:40:25 | SMS | | 4777X | 6382X |

Traditionally, CDRs have mostly been used for billing purposes and the maintenance of business. In recent years researchers from many cultures, data scientists, economists, social scientists and public sector organisations have begun to explore additional applications of de-identified CDRs, where all personal information has been removed.

With regards to Telenor-owned mobile operators, where the data is taken from for use in this thesis, the number of *daily* stored CDR observations varies from 100 million to over 1 billion per country.

One additional promising source of metadata for social science research is airtime credit purchases, or so-called 'top-up transactions', which are used for recharging mobile accounts in pre-paid markets like Asia and Africa [18]. Each purchase contains the user ID, top-up amount, date/time of top-up and often the location information of the retailer used. See the example in Table 2.2. Retailers, used by customers as a touch-point to refill their account, can range from basic kiosks on the street level to stores in large shopping malls.

**Table 2.2: Example of a set of airtime purchase transaction fields**

| Buyer | Retailer | Retailer location | Time of purchase | Recharge amount |
|---|---|---|---|---|
| 8849039482 | 880348403 | 6D45X | 25.05.2016 20:10:13 | 300 |

Even if airtime purchases are stored separately they are often placed under the CDR umbrella, due to their transactional nature.

### 2.2.2 Human behavioural traces

Mobile phone metadata contains longitudinal digital traces of human behaviour which have proved valuable to guiding development policies and humanitarian action [19,20]. At least three dimensions can be measured: financial activity, mobility and social interactions.

**Figure 2.1: Dimensions measured by mobile phone metadata (CDR)**

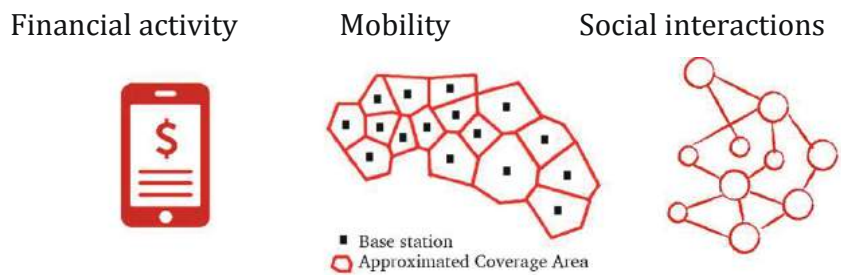

Financial activity      Mobility      Social interactions

**Financial activity:** When people in developing countries have more money to spend, they tend to spend a large portion of it on a luxury good such as cell phone communication, specifically by topping up their mobile airtime credit [21]. As we will see, several studies have already been run on socio-economic wellbeing using mobile phone data.

**Social interactions**: Social interactions include the number of 'friends' a person has, and how central the person is in the entire social network. In the context of mobile phone metadata the social network will be limited by the 'call graph' derived from phone logs, which is known to be a good proxy for the real social network [22]. The social interactions might also be used to understand how services and products diffuse on a large scale (or spread virally) throughout society [23].

**Mobility:** Since mobile phone users send and receive calls and messages through various cell towers, it is possible to reconstruct the movement patterns. This information may be used to understand daily rhythms of commuting to and from home, work, school and markets, but also have applications in modelling anything from a disease spreading to the movements of a disaster-affected population [24-26]. Typical derived personal-level features include most-used cell site, the radius of gyration (mobility radius) and total distance travelled within a given time period.

In addition to the three dimensions above, personal interests or content information, such as app usage, can be derived from on-board sniffer apps or deep packet inspection data [27]. Such information is however still very limited in use, and extends beyond the definition of metadata. There are also serious privacy implications [28].

The next section highlights examples of how mobile phone metadata has been applied to social sciences.

*Socioeconomics*

Socioeconomics is the social science that studies how economic activity affects and is shaped by social processes. Eagle et al. quantified the correlation between network diversity and a population's economic wellbeing [29]. The findings revealed that the diversity of individuals' relationships is strongly correlated with the economic development of communities. The assumption that more diverse ties correlate with better access to social and economic opportunities was untested at the population level. It concludes that frequently making and receiving calls with contacts outside one's immediate community is correlated with higher socio-economic class. Similar results were later verified by Jahani et al., which investigates the differences between social networks of the rich and the poor based on individually matched income data [30].

Additionally, CDRs have also been shown to provide proxy indicators for assessing regional poverty levels, as proven by studies in Cote D'Ivoire [31] and Rwanda [6]. They can also complement national surveys in estimating the changes associated with a growing economy, by exploiting the relationships between socio-economic factors and cell-phone usage [32]. It has also been hypothesised that airtime purchases are correlated with socio-economic status, but it has been difficult to validate this with external reliable data [33]. Monitoring airtime purchases for trends can be useful for detecting early impacts of an economic crisis, as well as for measuring the impact of programmes designed to improve livelihoods and food security [34]. As shown later, someone's handset type might also be a good indicator of their economic wellbeing.

*Population density*

Measuring population density in different regions can be explored by using the number of people who are calling each tower. By using CDRs, population density has been mapped out in France and Portugal [35] and Cote d'Ivoire [36]. In underdeveloped countries census data is costly and difficult to obtain, and existing data is often outdated, so CDRs can therefore provide updated information on the actual density of population in such regions. Population density has implications for economic growth and policies; the effects of population density and other socio-economic factors on poverty rates were studied in [37].

*Unemployment*

The findings by Toole et al. highlight the potential of mobile phone metadata to improve forecasts of critical economic indicators, such as unemployment [38]. The researchers found that CDRs, specifically changes in mobility and social interactions, can be used to predict unemployment rates up to four months before the release of official reports and more accurately than using historical data alone. This research can potentially identify macroeconomic statistics faster and with much finer spatial granularity than traditional methods of tracking the economy.

*Disasters and societal shocks*

Disasters initiate a complex chain of events that can disrupt the local economy, and in some cases the national economy [39]. It has been shown that CDRs can be used to monitor extreme situations and predict the movements of people after natural disasters. Bagrow et al. studied the reactions of people to different emergency situations, such as a plane crash, bombing and earthquake [40]. During the Haiti earthquake, Bengtsson et al. identified mobility patterns by analysing CDRs, providing more post-analysis of population migration. They found that the destinations of people who left the capital during the first three weeks after the earthquake were correlated with the locations where they had strong social bonds [41,42]. This research indicated that relief efforts could be planned more precisely, as population movement patterns may be significantly more predictable than has been previously understood.

*Infectious diseases*

Infectious diseases have serious consequences for people and the economies they affect. Diseases, such as dengue fever and cholera, often receive little attention because their real costs are poorly understood. Costs can be measured in loss of productivity, in dollars spent on healthcare interventions and in people's health and quality of life [43]. As CDR can help follow people's movements, these movements can also provide information about how a disease could spread across a country. The dynamics depend on the disease and how it can be transmitted, and therefore different models based on mobility have been suggested. Studies have validated the use of mobile phone data as a proxy for modelling epidemics [26]. Mobility patterns have been identified in Kenya, where regional travel patterns of millions of subscribers were mapped and related to areas in which malaria had a higher probability of spreading [10]. One shortcoming in this area of research comes from the current difficulty of gaining access to high-quality ground truth data to compare the results with. Gaining access to mobile phone data from more than one country has also proven to be very difficult, as exemplified by the Ebola outbreak [44]. The largest CDR analytics project on epidemics to date was performed on a dengue outbreak in Pakistan in 2013, where mobility patterns of 40 million subscribers were combined with climatic suitability indices and epidemiological data [45]. High-resolution fine-scale risk maps predicting epidemics were produced, providing a platform to allow local public health departments to prepare for epidemics.

*Transportation and economic development*

Transportation has substantial effects on economic growth, but the relationship between transportation and the economy are poorly understood [46]. Several studies on CDRs using mobility patterns are based on the subject of transport planning, and avoiding traffic jams and road accidents. Berlingerio et al. mapped new routes to decongest Abidjan's crowded roads, which could potentially reduce travel time by 10% [47]. Another study estimated the flow of residents between each pair of intersections in a city [48]. A recent study used CDRs to track commuters during peak morning rush hour in five cities [49]. People try to travel from home to work as quickly as possible, but simulations showed that as much as 30% of the total time lost to congestion is caused solely by what the authors refer to as selfish routing. They suggest that social route planning could make driving less problematic. For more information on the research in this field the reader is referred to [50].

*Marketing*

Marketing is a wide term that touches upon many disciplines. From a societal point of view, marketing provides the link between the material requirements in society and its economic response patterns. One factor that influences marketing strategies is social forces and peer-influence [51]. As social networks are becoming more explicit through other technologies, understanding how peer-influence creates and sustains behavioural congestion is also becoming more feasible [52]. For instance, Trusov et al. has demonstrated that, on average, approximately one fifth of a user's friends actually influence his or her activity level on a specific website [53].

In particular, mobile phone metadata provides the opportunity to quantify the effects of social interactions on marketing (and vice versa) on a scale that has never been done before. It is known that the neighbourhood of an individual influences their decisions [54]. A study by Hill et al. quantified the neighbours' probability of adopting an undisclosed technological service [55], and Ahorony et al. showed that common app installations were overrepresented for pairs of users who often have physical meetings [56]. Furthermore, Risselada et al. showed that the neighbours' influence on product adoption evolves over time, depending on the time since introduction of the product in the market [57].

Product adoption has also been recently studied in other contexts, such as the adoption of Mobile Money [58]. Billions of people around the world live without access to banks or other formal financial institutions, and mobile money platforms, which deliver basic financial services over the mobile phone network, are believed to improve the lives of the poor. Unfortunately in many countries the adoption rates are still very low. A recent research study addresses how machine learning can be used to predict passive and active mobile money usage, using behavioural information from CDRs [59]. The results highlight key correlations of mobile money use in three development countries, as well as the potential for such methods to drive adoption. However, the models developed in one country do not perform very well in other countries, which may indicate, in the context of mobile money, that each population has a unique signature in terms of which metrics are good predictors of adoption.

# 3 Methodology

This dissertation combines methods from multiple disciplines, including social network analysis, statistical analysis, machine learning and visualisation. The approach is data-centric, where inductive reasoning is used to make generalisations from specific observations. Extensive travelling, especially in South-East Asia, has been required to collect and analyse the datasets. This chapter elaborates on the data collection, tools and analytical framework used.

## 3.1 Research design

### 3.1.1 Data collection

The data was collected from subsidiaries of the Telenor Group [60]. Telenor Group has operations in 13 countries in South-East Asia, Eastern Europe and Scandinavia (Fig 3.1), covering over 200 million mobile subscribers. Local data warehouses store the raw data, which is maintained by business intelligence teams. Research datasets are de-identified and either analysed locally, or transferred to a research data warehouse in Norway. The collection process is shown in Figure 3.2. In the case of external research collaboration, de-identified data are shared under special non-disclosure agreements.

**Figure 3.1: Global presence of Telenor in 13 countries (blue)**

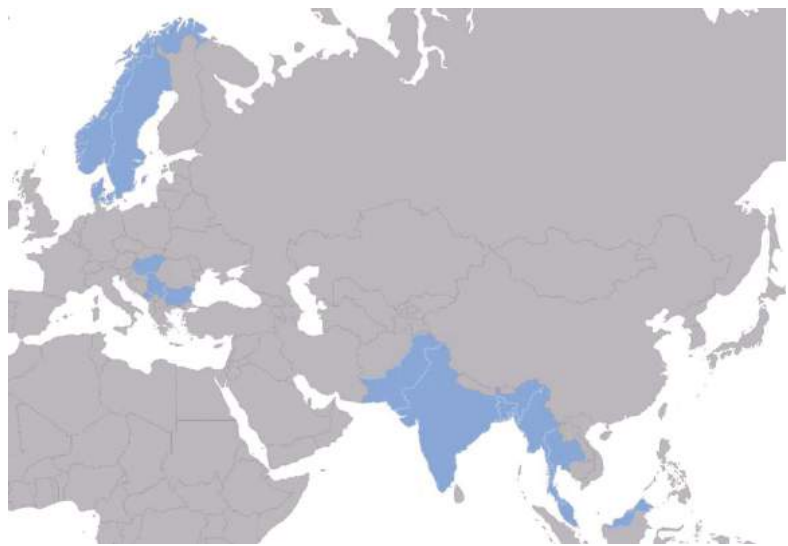

**Figure 3.2: Data collection. Network information is collected from mobile users (A) and stored in local data warehouses (B). De-identified information (CDR, purchases) is transferred to the research data warehouse via a secure channel**

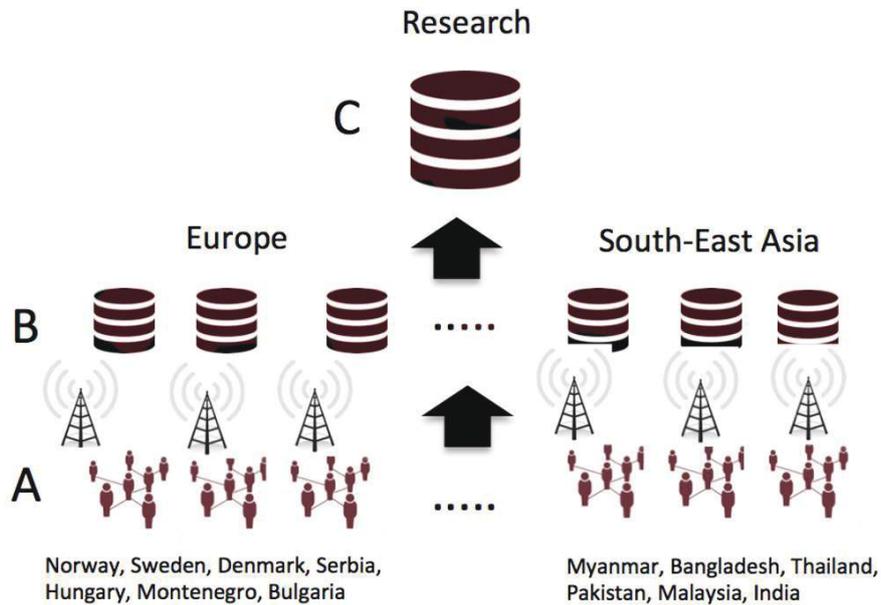

### 3.1.2 Tools

A combination of several tools is used in this thesis. All analysis originates from second-resolution raw data, where processing requires a significant amount of computational power. Most datasets are pre-processed in-database using SQL, either on Oracle, Teradata or PostGres platforms, on high-performance Linux servers [61]. Considering the size of the datasets, SQL is also used for efficiency purposes for specific analytical and algorithmic tasks. Other tools include Python, R and specific tools for data analysis and machine learning [62,63]. For visualisation purposes, tools include Autodesk Maya for 3D modelling [64], and Gephi/Cytoscape for network visualisations [65]. QGIS is mainly used for spatial analysis and geographic visualisations [66].

### 3.1.3 Analytical framework

The analytical framework is considered as multi-disciplinary. Fig 3.3 provides an overview of the type of data sources and the *main* methodology used in each of the included publications.

**Figure 3.3: Per-paper overview of data sources and methodology**

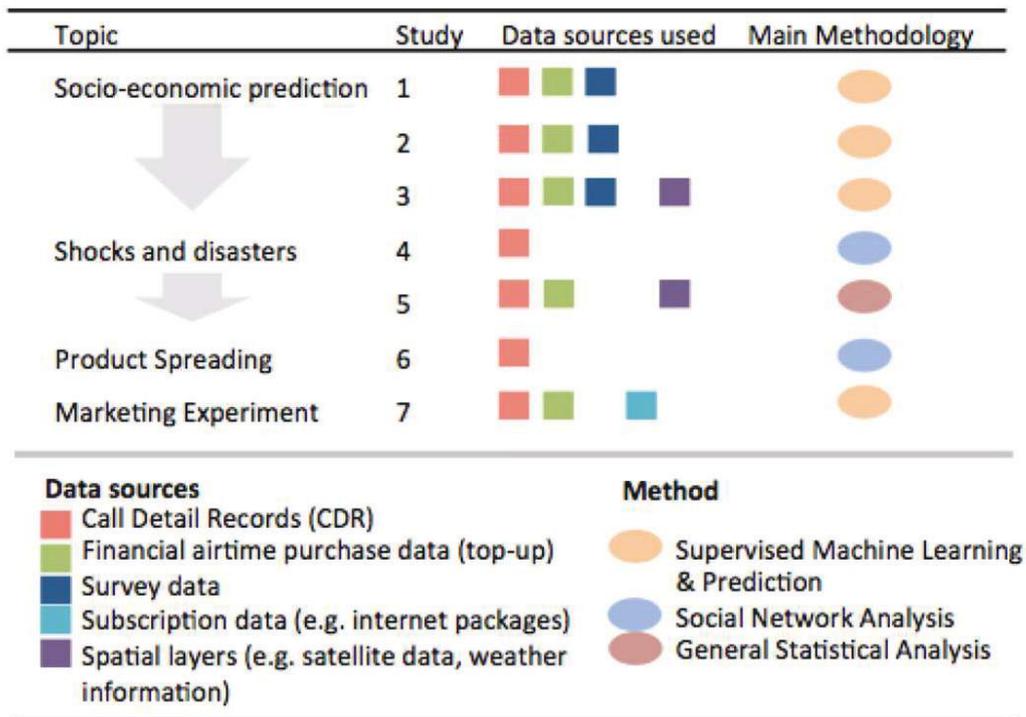

Papers 1, 2, 3 and 7 use supervised machine learning and/or prediction methods to infer a function from the training data that is further used for prediction. These studies make use of surveys or other subscriber information as ground truth information. An important note is that the feature generation processes (which generate the variables) are often based on other types of analysis, including mobility and social network analysis. Paper 3 uses techniques from machine learning, such as cross-validation and test-set, but within a Bayesian statistical modelling framework [67]. The methodology in papers 4 and 6 are mainly social network analysis. Paper 5 applies general statistical analysis and anomaly detection. All of the studies use CDR data as the main data source, while some of the studies make use of additional data sources such as financial airtime purchase information, subscription information and spatial layers, including satellite information. Data visualisation has also been vital, but is not mentioned in the methodology as shown in Figure 3.3.

### 3.1.4 Social network analysis

Mobile operators often have access to a huge portion of the social network in a given country, and these sources are therefore good candidates for social science research, e.g. for studying social influence and purchase decisions. Understanding the nature of relationships and connections between entities is important for understanding a range of phenomena throughout multiple disciplines. Social network analysis (SNA) has broad and successful applications in economics epidemiology, sociology, biology and criminology [68]. The building block in the field of SNA is graphs, residing in graph theory, employed to represent the structure of interactions among people or any type of entities [69]. From a network perspective, it is the structure of the network and how the structural properties affect behaviour that is informative, and not simply the characteristics of the actors in the network. Analysis of large social networks is a non-trivial task that introduces challenges due to long processing time and large computational resources.

*Basic elements in network analysis*

A social network is defined as graphs representing social relationships between people or organisations. Each node, also called a vertex or an actor, in a graph represents an individual person or a group of people. The connection between two individual nodes is referred to as an edge or tie. Two important concepts are components and centrality.

**Figure 3.4: Example social network**

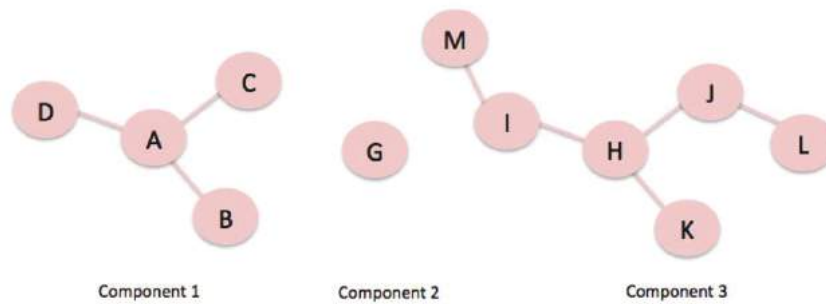

*Social component:* A component is a portion of the network where all actors are connected, directly or indirectly, by at least one tie. By definition, each isolate is a separate component. Figure 3.4 consists of a social network of three social components. Node G is situated in an isolated component, while component 3 is the largest connected component (LCC) in the network.

**Centrality:** Centrality measures identify the most prominent actors, that is those who are extensively involved in relationships with other actors. The 'importance' of actors in networks is indicated by centrality. The most used centrality measures include degree centrality, closeness centrality, betweenness centrality and eigenvector centrality [70]. By using Figure 3.4 as an example, we see that Nodes H and A both have three close friends each, indicating a degree centrality of three. If we use another centrality metric, such as eigenvector centrality (EVC), we will see that Node H suddenly has a higher EVC than A. The reason is that EVC takes into account the global network, where the friends of friends also matters.

From mobile phone datasets we can study a weighted social network, where the weight (or intensity) $w_{ij}$ of an edge connecting person $i$ and and person $j$ is defined as the aggregated time that the two users spend talking to one another. Most often, in communication networks the edge weight (i,j) is taken as the total number of calls, or as the aggregated duration of calls between $i$ and $j$ during the period under investigation. Previous studies have shown that they give an equivalent quantification of edge weight [71].

### 3.1.5 Machine learning

Machine learning picks its algorithms from different academic disciplines [72]. It is closely related to and often overlaps with computational statistics, which is a discipline that also focuses on prediction-making through the use of computers, and generalising from examples [73]. It is also considered as a sub-field of artificial intelligence. Practically, machine learning can be described as the algorithmic part of a data mining process, where the data preparation step is often the most tedious one. This type of data mining process model, which describes commonly used approaches to tackle problems, can be best visualised with the CRISP-DM framework [74]. A more detailed scheme of the isolated machine learning process is visualised in Figure 3.5. Here the process has been divided into 3 phases. The first phase is the pre-processing phase, where the features are generated from raw data, and the data are split into train and test.

**Figure 3.5: Process of machine learning**

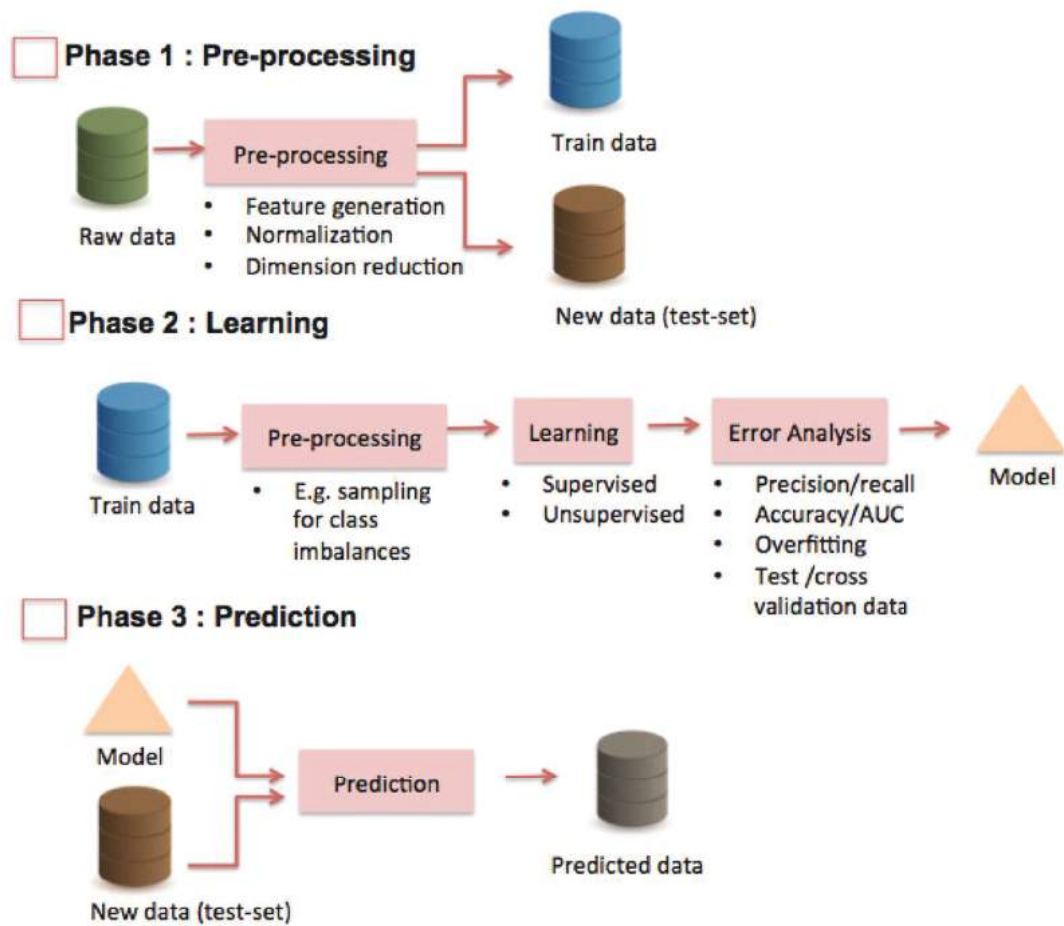

The second phase is the computational learning phase, where techniques such as cross-validation are used to generate the final model [75]. Cross-validation is a technique for assessing how the results will generalise to an independent dataset, prevents the model from overfitting and provides an insight on how the model will generalise to an independent dataset, which is used in phase 3. There are also several ways to evaluate the performance of a machine learning model: popular choices include accuracy, recall, precision and 'area under the curve' (AUC) [76]. For instance, accuracy measures the proportion of true positives and negatives in the whole dataset. It is calculated for a given threshold; for example, logistic regressions return positive or negative depending on whether the logistic function is greater or smaller than a threshold, usually set to 0.5 by default. AUC measures how true positive rate (recall) and false positive rate trade off, and is an evaluation of the classifier as threshold varies over all possible values. The interpretation of AUC is the probability that a randomly chosen positive example is ranked above a randomly chosen negative example, according to the classifier's internal value for the examples. The third phase is where the model is applied to unseen data (test-set), predictions are made and performance metrics are evaluated. If the performance is significantly lower than the train set, an overfitted model is used, and a re-calibration of the model is needed.

By the learning style (phase 2), the machine learning algorithms can be mainly divided into the following types: supervised learning and unsupervised learning.

*Supervised learning*

In supervised learning, each sample in the dataset is a pair of an input vector and an external output vector (or value) that we are trying to predict [77]. Supervised learning is closely related to regression or classification in econometrics. By analysing a training set under a supervised learning algorithm an inferred function can be generated. The inferred function, e.g. the training model, can be then used to map or predict new samples. Both classification and regression are supervised learning where there is an input vector X and an external output Y, and the task T is to learn the experience E from the input X to the output Y. Typical supervised learning algorithm types are shown in Table 3.1.

**Table 3.1: Examples of supervised learning algorithms [78-81].**

| Linear regression | Non-linear regression and classification | Regression and classification trees |
|---|---|---|
| • Ordinary linear regression<br>• Partial least squares regression<br>• Penalised regression | • Multivariate adaptive regression splines<br>• Support vector machines (SVM)<br>• Artificial neural networks (incl. deep learning)<br>• K-nearest neighbours | • Bagging tree<br>• Random forest (RF)<br>• Boosted trees/gradient boosted trees |

*Unsupervised learning*

In unsupervised learning there is no external output (or label), and we only possess the input vector. The aim is to find similarities among samples in the unlabelled dataset. Typical algorithms include clustering (e.g. k-means), latent variable models and blind signal separation techniques [82]. Principal component analysis (PCA) is an example of a blind signal separation technique that can be used to explain key features of the data, and reduce the data from a high-dimensional space. Often the feature selection itself is embedded into learning algorithms. Unsupervised learning is closely related to density estimation and clustering in econometrics. In this thesis supervised machine learning is mainly considered, as access to labelled data is available, and it is therefore possible to evaluate predictions.

### *A comment on econometrics vs. supervised machine learning*

The main objective in supervised machine learning is to provide accurate *predictions* of the variables of interest. Even though these techniques are extremely powerful for forecasting, it can be very difficult to interpret the underlying structure implied by them.

On the contrary, the main objective in econometrics is to provide an *explanation* of various observed outcomes. The goal is often to produce reliable estimates of parameters that describe economic systems, to provide an understanding of the underlying process that determines equilibrium outcomes. The estimation process is based on conditions implied by economic theory. Such a structural approach is beneficial when we want to know what happens when 'the world' changes (used for e.g. auctions, pricing). Machine learning can not easily predict the effect of intervention (how y changes as some x change).

Hal Varian argues that there are several things that econometricians can learn from machine learners and vice versa [14]. Machine learning introduces the train-test-validate (including cross-validation) concept to avoid model overfitting, and there are several non-linear estimation techniques, as mentioned above. Variable selection methods are also developed to deal with large amounts of data.

However, casual inference including confounding and instrumental variables is seldom considered in machine learning, even though some efforts have been made to combine the two approaches [83,84]. In addition, time series are often decomposed into trend and seasonal components to look at deviations from expected behaviour. The concept of cross-validation does not work directly on time series. Kleinberg et al. show how machine learning adds value over traditional regression approaches in solving prediction problems, and argue that causality is not always important when dealing with policy applications [15].

# 4 Key contributions

This chapter summarises the main findings in each of the 7 papers included in this thesis. Section 4.1 focuses on using mobile phone logs in a context that can benefit society as a whole. Contributions include individual prediction of socio-economic indicators, poverty prediction and understanding human behavioural signals during disasters. Section 4.2 investigates how products spread over large social networks, and further how to use these key findings in a large-scale marketing experiment in Asia, giving people access to more personalised offers.

## 4.1   Contributions to social good

### 4.1.1   Studies 1–3: Predicting socio-economic indicators

Section 2.2.2 discussed how mobile phone metadata has been used to study socio-economic behaviour. Here we build on these findings; in study 1 and 2 prediction models are developed that can be used to infer individual characteristics of users, such as illiteracy status and income, by looking only at the users' mobile phone behaviour. The fact that most phones in the developing world are prepaid means that the data lacks very basic information about the individual. This prevents numerous uses of this data in development economics research and social sciences. More importantly, it prevents the development of humanitarian applications such as the use of mobile phone data to target aid towards the most vulnerable groups during a crisis. For development purposes insight can be learned on the individual scale, before it is aggregated on a spatial scale of mobile towers.

**Figure 4.1: Studies 1–3: topic, data sources and country**

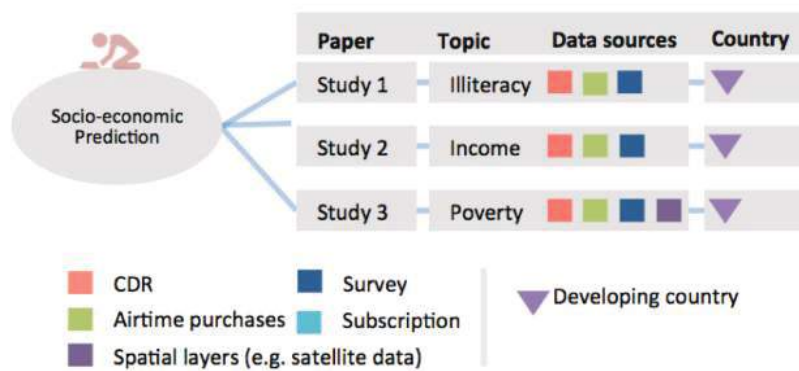

Study 3 addresses how mobile phone data can be used to predict multiple dimensions of poverty in Bangladesh using both mobile phone metadata and additional spatial information, such as satellite information, as model input. As indicated in Figure 4.1, all the studies use CDR and survey information as their main data sources.

*Study 1: Can mobile usage predict illiteracy in a developing country?*

The aim with study 1 is to investigate whether large-scale mobile phone metadata, in the form of CDRs and airtime purchases (top-up) can support quantifying individual and spatial illiteracy in a developing country. Geographical mapping of illiteracy is crucial to know where illiterate people are, and where to distribute resources to improve education. In underdeveloped countries such mapping can be based on out-dated household surveys with low spatial and temporal resolution. One in five people worldwide struggle with illiteracy, and it is estimated that illiteracy costs the global economy more than $1 trillion dollars each year [85].

By deriving a broad set of mobile phone indicators reflecting users' financial, social and mobility patterns, as introduced in section 2.2.2, we show how supervised machine learning can be used to predict individual illiteracy. On average the model performs 10 times better than random guessing with 70% accuracy. Feature investigation indicates that the most frequent cell tower and incoming SMS are the superior predictors, followed by diversity of communication partners and Internet volume (Figure 4.2a). Furthermore, the investigation extends to how individual illiteracy can be aggregated and mapped geographically at cell tower resolution (Figure 4.2b).

**Figure 4.2: a) Most important predictors from mobile phone metadata data for predicting illiteracy; b) geographical mapping of illiteracy, top predictors and the cell tower distribution in one major city**

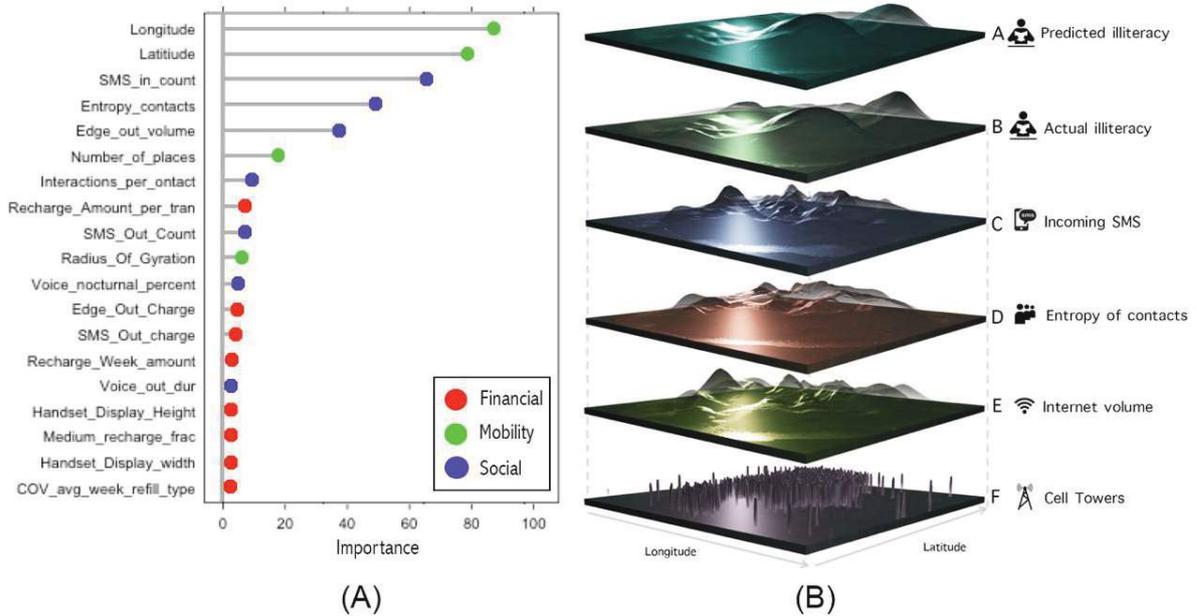

This study shows how illiteracy can be predicted from mobile phone logs, purely by investigating users' metadata. An important policy application of this work is the prediction of regional and individual illiteracy rates in underdeveloped countries where official statistics are limited or non-existing. Further work is required to investigate the findings up against population illiteracy and also verify the robustness of indicators in other countries.

*Study 2: Deep learning applied to mobile phone data for individual income classification*

One promising supervised classification method for social sciences is deep learning, sometimes referred to as 'deep neural networks', as introduced in section 3.1.5. For social sciences the main advantage of deep learning is to avoid cumbersome feature engineering, and let the algorithm itself decide the features (based on near raw data as input).
Deep learning has had breakthroughs in computer vision [86] and speech recognition [87]. The aim of study 2 is to understand whether deep learning can be beneficial for useful prediction tasks on mobile phone data, where classic machine learning algorithms are often under-utilised due to time-consuming country and domain-specific feature engineering (as shown in paper 1). Specifically the aim is to see how well the socio-economic status of an individual can be predicted, with a comparison to traditional data mining models as a benchmark.

A simple deep learning architecture is implemented and is compared with traditional data mining models as benchmarks. On average this model achieves 77% AUC on test data using location traces as the sole input. In contrast, the benchmarked state-of-the-art data mining models include various feature categories such as basic phone usage, top-up pattern, handset type, social network structure and individual mobility. The traditional machine learning models achieve 72% AUC in the best-case scenario. Figure 4.3a shows the predictive performance, as measured by the AUC on test set, where the true positive vs. false positive rate is plotted for deep learning (DL), gradient boosting machines (GBM) and random forest (RF). The top predictive indicators in the random forest model and given in 4.3b and coloured by their respective variable family.

**Figure 4.3: a) Income prediction; predictive deep learning (DL), gradient boosting machines (GBM) and random forest (RF); b) top predictors for random forest model**

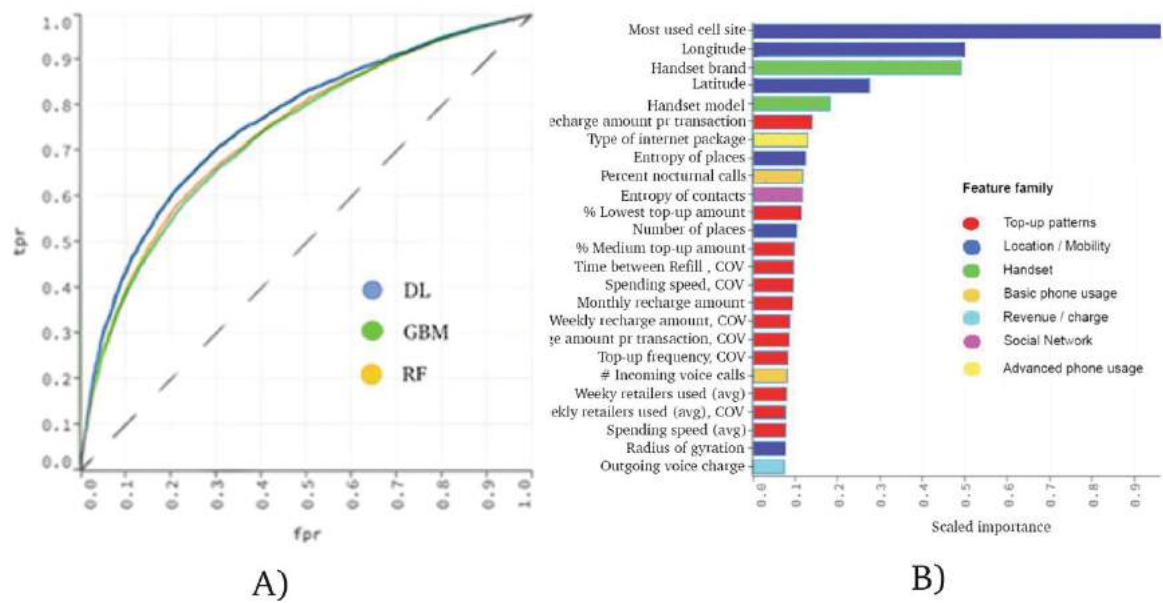

Even though the traditional model is not meant to be interpretable, Figure 4.4b gives some insight into the most predictive features. We especially note the importance of location dynamics, handset brand and airtime credit purchase patterns for predicting income.

*Location dynamics:* Where the user spends most of his time is a good signal of his income. This indicates that our models have detected regions of low economic development status. This is also in line with the deep learning model.

*Handset brand:* In the country of the study, minimal and more affordable handset brands are very popular among the lower income quantiles, while expensive smartphones are considered as a huge status symbol.

*Top-up pattern:* Interestingly, the recharge amount per transaction is more predictive than the total recharge amount. It can be observed that individuals from the lower income quantiles usually top-up with lower amounts when they first fill up their account.

The approach suggests that deep learning approaches could be an effective tool for predicting economic indicators based on mobile communication patterns. The disadvantage is that this is a more 'black box' approach and harder to interpret than traditional models.

### Study 3: Predicting multiple dimensions of poverty using mobile phone and satellite data

In 2015, approximately 700 million people lived in extreme poverty, defined as living on less than $1.90 a day [88]. To end poverty in all its forms everywhere is one of the selected targets of the UN Sustainable Development Goals [89]. As mentioned in section 2.2.2, CDRs have been shown to provide proxy indicators for assessing regional poverty levels in Cote D'Ivoire and Rwanda. Paper 3 builds on these approaches, and extends the approach by also including spatial layers such as satellite information, in addition to predicting multiple dimensions of poverty, with Bangladesh as an example.

Eradication of poverty requires national and subnational quantification and monitoring over the next 15 years, and the challenge is to establish appropriate, effective and timely measurements. Existing approaches to estimate multi-dimensional poverty rely on census data collected with limited temporal frequency. Alternate measures are needed to update estimates in the time between censuses. Here the aim is to investigate whether mobile phone data, combined with satellite data, can complement existing approaches for predicting multiple dimensions of poverty. An objective is to know which sources are most promising in rural vs. urban areas.

In this work, a Bayesian geostatistical modelling framework is used, combining data from GPS-located household surveys, satellite and other spatial layers, and mobile phone metadata to predict asset-, consumption- and income-based metrics of poverty at high resolution. This reveals that models employing a combination of mobile operator data and satellite variables provide the highest predictive power and lowest uncertainty (highest for wealth index: $r^2=0.78$).

**Figure 4.4: National level prediction maps for mean wealth index. Maps were generated using mobile phone features, remote sensing data and Bayesian geostatistical models. The lower map restricts the focus on the poorest cells.**

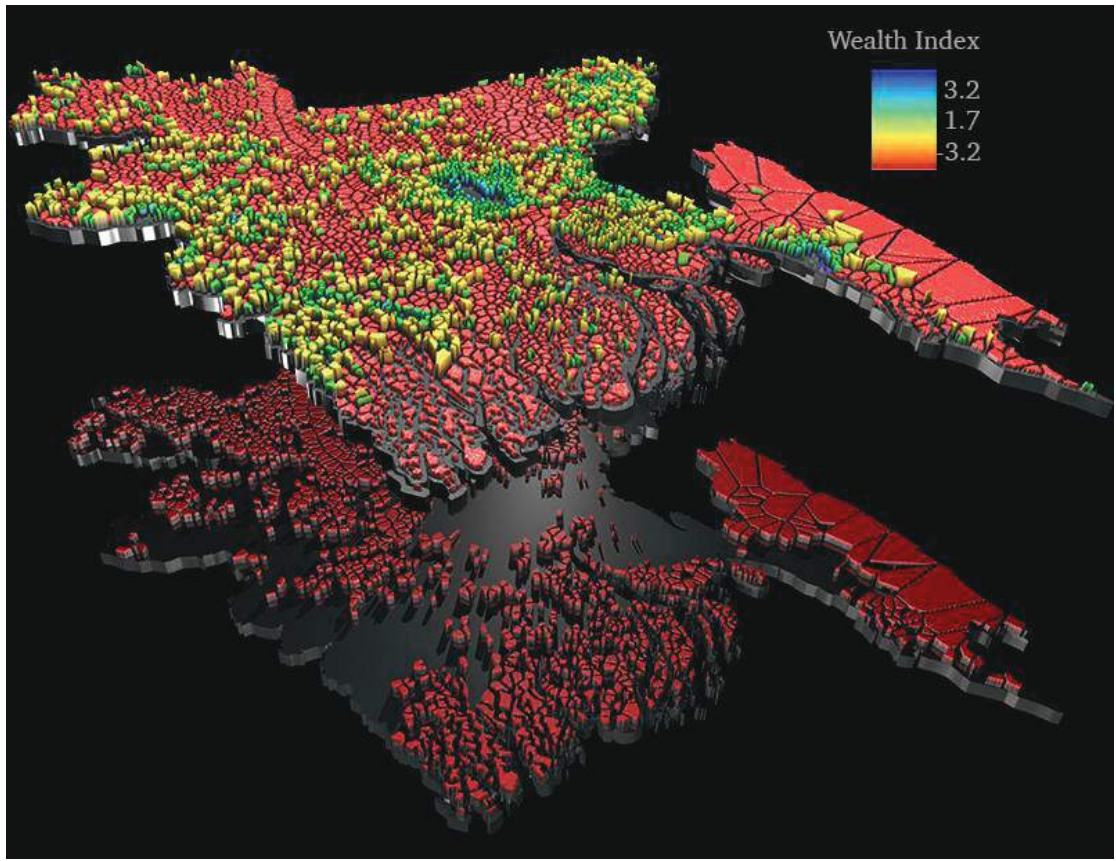

National, urban and rural models were built to predict poverty rates from three separate geo-referenced household survey datasets. Indicators such as night-time lights, transport time to the closest urban settlement and elevation were important, both nationally and in rural models; climate variables were also important in rural areas. Distance to roads and waterways were significant within urban and rural strata. In general, the addition of CDR data did not change the selection of satellite features at any level. Top-up features derived from recharge amounts and tower averages were significant in every model. Percentage of nocturnal calls, and count and duration of SMS traffic were significant nationally. Mobility and social network features were important at all strata, but only in rural models when combined with satellite data. In urban areas, SMS traffic was important, whereas multimedia messaging and video attributes were key in rural areas. The outputs correlate strongly with previous poverty estimates for Bangladesh, highlighting the value of such cell phone-satellite-driven models in producing high-resolution poverty maps that can be rapidly updated. The findings based on this research can be utilised for real-time monitoring and decision-making to more effectively reduce poverty.

### 4.1.2 Papers 4–5: Understanding systemic shocks and disasters in society

A shock is an unexpected or unpredictable event that affects the economy, either positively or negatively. Shocks are typically produced when accidents or disasters appear. Systemic shocks will prompt hundreds or thousands of individuals or households to react in roughly similar ways. Section 2.2.2 discussed how mobile phone data have been used to study shocks and disasters, such as in the case of Haiti [41].

**Figure 4.5: Studies 4–5: topic, data sources and country**

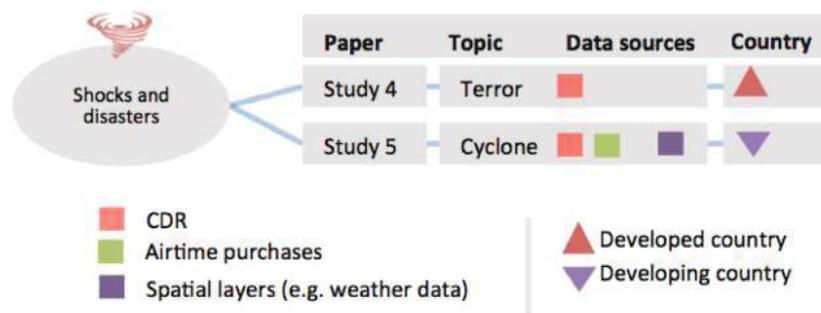

The studies introduced in this section are related to two extreme events, as shown in Figure 4.5. The first study uses CDR information to investigate the terror incident that hit Oslo, Norway, in 2011. The next study investigates an extreme weather event – cyclone Mahasen, which hit Bangladesh in 2013.

*Study 4: The activation of core social networks in the wake of the 22nd July Oslo bombing*

Study 4 examines human behavioural patterns on Friday 22nd July 2011 when a powerful bomb exploded in Regjeringskvartalet (the centre of national administration) in central Oslo, Norway. It killed 8 people and seriously injured almost 100 others.

Earlier work has presented qualitative results on the 9/11 catastrophe, emphasising the need to reach out to the closest tie [90]. Another study examined the geographical distribution of traffic after a terrorist bombing in Israel [91]. This work differs from the others in the sense that large-scale mobile phone logs are used to understand communication patterns around the disaster. In such events we feel the need to check the wellbeing of family and friends, to organise assistance, and to make sense of the situation. Mobile phone logs can illuminate the ways these needs are met in the face of disaster.

Empirical mobile traffic data illuminate exceptional behaviour immediately after the bombing in Oslo; in the minutes after the bombing people called ties that were close socially and perceived to be in danger, that is, people who were close to the bombing point. The main findings: (1) individuals first focus on their single closest contact ('best friend'), but soon after switch to spending more mobile communication resources than average on contacts ranked 2–5; (2) a large increase (over typical) in traffic is clear to and from, and not least within, the affected area (Oslo). In some cases this was more than a 300% increase immediately after the bombing. Interestingly, a marked increase in traffic also occurred for relationships where both persons were outside Oslo. All of these results illustrate the importance of social contact in this highly unusual situation.

**Figure 4.6: Fraction of active subscribers contacting their 1–5 closest relations vs. time for all active subscribers. All curves are normalised by average communication for the same contact number (1–5)**

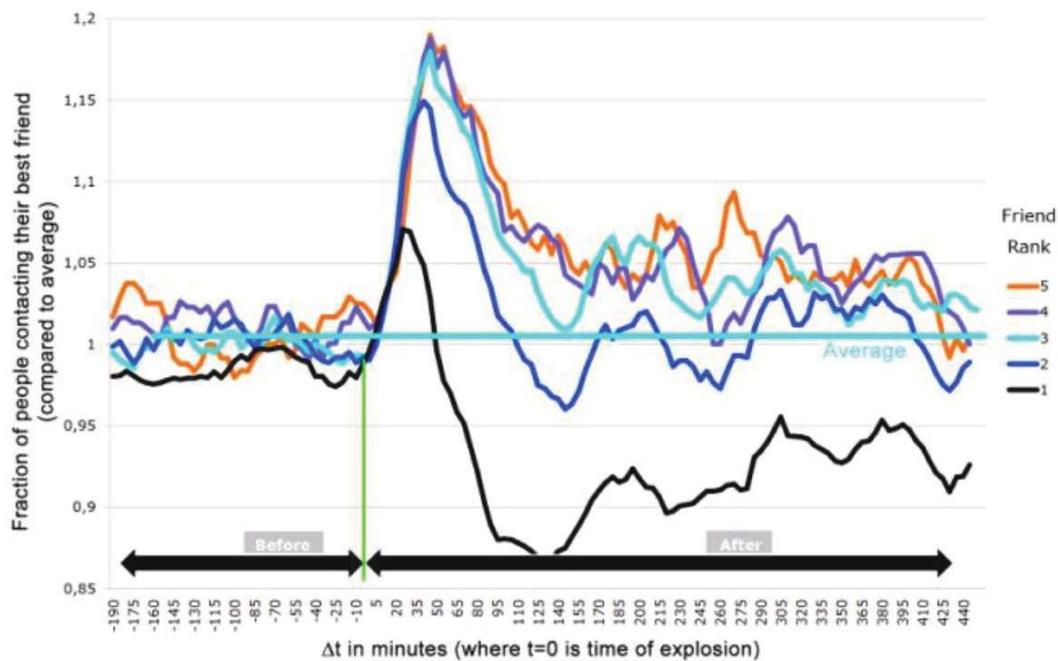

This paper underlines how the mobile phone is an instrument of the intimate sphere. The situation on the 22nd July in Oslo is a prime example of an unexpected situation where individuals use the mobile to get critical information on their loved ones in their core network. The study has several limitations. These include the fact that the data are taken only from one operator in Norway, and the data is therefore not generalisable to all of Norway. The location-based data use also the postal code of the subscriber, and not the actual location of the phone at the time of the bombing. This is a result of privacy issues.

### *Study 5: Detecting climate adaptation with mobile network data: anomalies in communication, mobility and consumption patterns during Cyclone Mahasen*

Other shocks in society include natural disasters. Extreme weather events have always had and will continue to have significant consequences for society and the economy. Climate projections indicate that changing extreme weather patterns are very likely to increase exposure to those events. Researchers have also found that a temperature change will leave the average income around the world 23% lower in the year 2100 than without climate change [92]. The following paper quantifies the impact of an extreme weather event in Bangladesh using CDR data. Weather events in Bangladesh already have a major impact on the economic performance and livelihoods of millions of poor people, and climate change is likely to drive migration from environmentally stressed areas.

In this study mobility, economic and social patterns are analysed during cyclone Mahasen, which hit Bangladesh in 2013. The aim is to investigate whether data from mobile phone metadata may be a useful tool to prioritise locations in which rapid needs assessment is performed after a cyclone. The aim is also to investigate whether anomaly detection can help to understand 'signals' of response in the population exposed to the cyclone.

The results show that anomalous patterns of calling frequency correlate with rainfall intensity at the local scale, likely providing a spatiotemporal indicator of users' physical exposure to the storm.

**Figure 4.7: The temporal and spatial distribution of anomalies in airtime purchase anomalies. The threshold detection was set at three standard deviations from the mean of baseline**

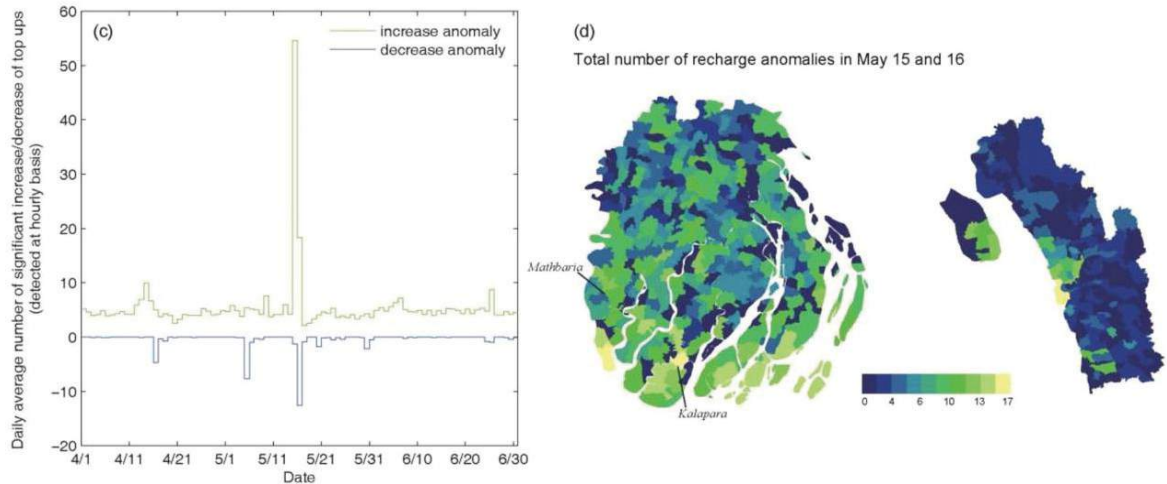

In addition, the results show that mobile recharge purchases increase in impact zones before landfall, representing preparations for potential environmental threats. The temporal and spatial recharge purchase anomalies are shown in Figure 4.7. Anomalous patterns of mobility are also identified during evacuation and storm landfall, indicating how people respond to storm forecasts and early warnings.

Detecting anomalous usage patterns from mobile network data is a promising avenue for researching human behavioural responses to environmental impacts across large spatiotemporal scales. Based on comparisons with rainfall measurements at landfall, and considering the considerable weakening of cyclones as they pass over land, calling frequency and population movement anomalies seem to be the best indicators of physical interaction and impact of the cyclone. The anomaly detection technique applied to CDRs, as presented here, overcomes some of these challenges, and demonstrates the potential value of CDR as a complement to current cyclone impact assessment tools to improve the accuracy, timeliness, and cost-effectiveness of cyclone impact assessments. Data from CDRs may be very useful as a tool to prioritise locations in which rapid needs assessment are performed after cyclone landfall, with the potential to drastically reduce the time to reach those most in need. Primary limitations of the study involve representativeness of the data for the general population. The indicators found should however reflect natural human response to shocks, but it is important to stress that CDR data should not be used alone to indicate where post-disaster assistance is needed.

## 4.2 Contributions to behavioural insight and marketing

Building on the marketing literature from section 2.2.2 and the methodology introduced in section 3.1, the last two studies investigate how the social network matters when purchase decisions are made.

**Figure 4.8: Studies 6–7: topic, data sources and country**

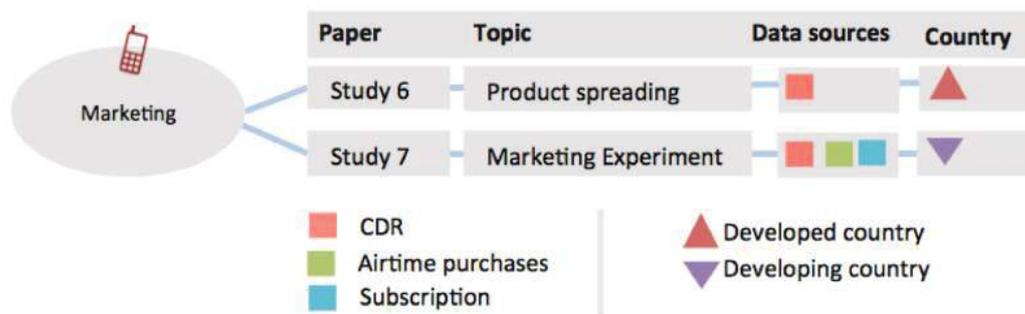

**Study 6: Product spreading over large-scale social networks**

Most of the existing research on product diffusion in networks has been focused on a single product, with a static snapshot of the network. Here the interest is to know how the social network among product adopters develops over time. In addition the aim is to understand how the product diffusion depends on the underlying social network. In short this study looks for evidence of social spreading effects. The starting point of the analysis covers more than 2.5 million subscribers and 50 billion CDR records collected over 9 quarters.

By combining the social interactions, derived from CDRs, with the adoption history of a product of interest, how different products are spread over the social network can be observed. Several products have been studied, including the iPhone, iPad, video telephony, Doro and Android phones. For each product the evolution of fraction of adopters is shown by social component size, and centrality distributions by product is presented, as measured by eigenvector centrality of the adopters. All products show various degrees of social spreading effects. This work reveals in particular strong social effects in the adoption of iPhones and iPads. The Doro handsets have very weak social spreading, and will probably never 'take off'. Video telephony also has strong spreading effects, but early takeoff was prevented by an external factor, here a new price model.

**Figure 4.9: Time evolution of the iPhone adoption network (largest-connected component). One node represents one subscriber. Node colour represents iPhone model: red = 2G, green = 3G, yellow = 3GS**

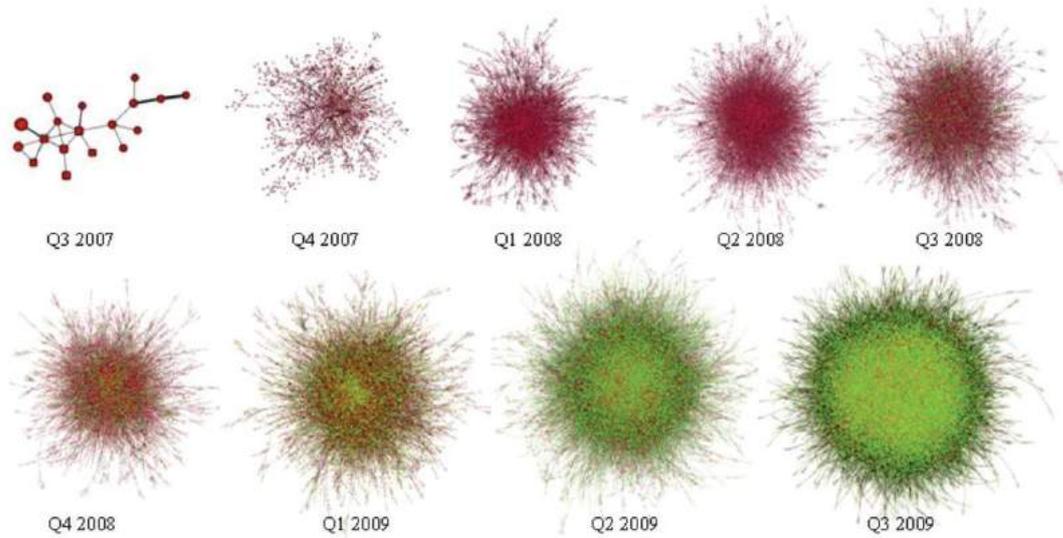

To verify social spreading effects the probability $p_k$ that a subscriber has adopted a product is measured, given that k of the subscriber's friends have adopted the product. Figure 4.10 shows the following: if we know that a subscriber has one friend using a Doro phone, that subscriber's probability of using one themself is roughly twice the adoption probability for a subscriber with no adopting friends.

**Figure 4.10: Adoption probability $p_k$ vs. the number k of adopting friends for three products. In each case a monotonic growth of $p_k$ with k is visible, indicating that some kind of social spreading is occurring**

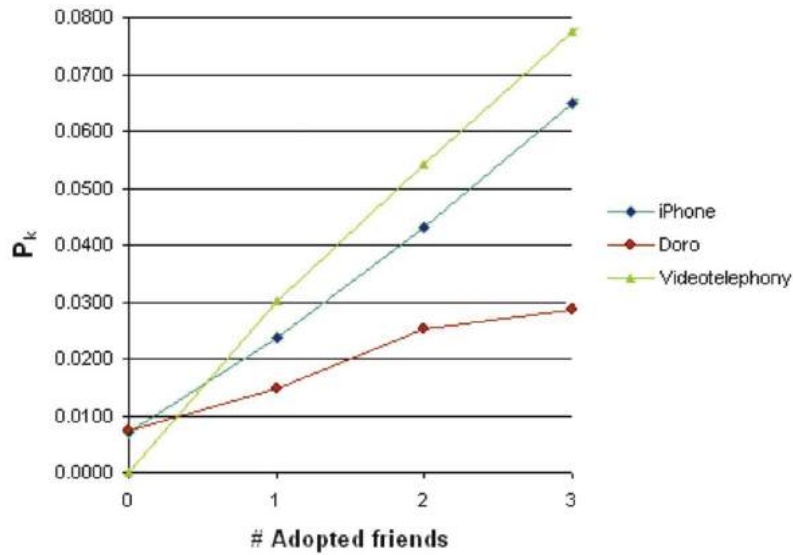

Through visualisations, supported by analysis, a glimpse is provided of how products spread over large-scale social networks. These measurement methods give new and useful insight into how and why the different services have performed so differently. Initially social spreading effects were unexpected for Doro, which is a handset marketed towards the elderly. The findings quantify and suggest that such effects exist for most products. The strength of the effects also produce the possibility that such knowledge could be investigated further in marketing experiments.

### Study 7: Large-scale marketing using machine learning

Paper 6 quantified the adoption probability as a function of friends that have adopted a product. In paper 7 this insight is experimentally exploited in a campaign in Asia, using Internet packages as the product of choice. Economic growth in Asia is steadily increasing, but the overall Internet penetration is very small, which contributes to an economic and educational gap in comparison to more developed countries [93]. Mobile operators in this geographical region typically run thousands of text-based campaigns a year, resulting in customers receiving several promotional texts per month. This is a major problem for operators, since they are considered as spammers rather than providing people with useful information. For this particular operator, the policy is to not send more than one text per customer every 14 days. It is therefore crucial to personalise the offers towards the customers. This paper evaluates a data-driven approach to marketing against the operator's current best practice in a large-scale 'Internet data' experiment. The experiment will compare the conversion rates of the treatment and control groups after one promotional text.

**Figure 4.11 a) Conversion rate in the control (best practice) and treatment (data-driven approach) groups; b) the percentage of converted people who renewed their data plan after using the volume included in the campaign offer**

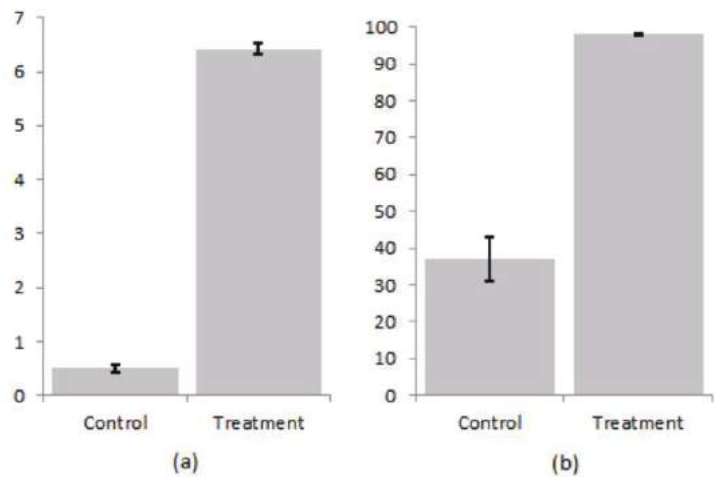

The findings show that the data-driven approach using machine learning and SNA leads to higher conversation rates than the best-practice marketing approach.

The study also shows that historical natural adoption data can be used to train models when campaign response data is unavailable. New features to identify which customers are the most likely to convert into mobile Internet users were created. The features fall into the following categories: discretionary income, timing and social learning. In total 50 million customers were scored, and the treatment group was composed of the top 250,000 customers with the highest score. The conversion rate of an Internet data campaign was increased by 13 times compared to current best practice (the control group). The model also shows very good properties in the longer term, as 98% of the converted customers in the treatment group renewed their mobile internet packages after the campaign, compared to 37% in the control group.

**Table 4.1: Top 10 most predictive indicators to classify natural converters, ranked by importance in the model**

| Rank | Type | Description |
| --- | --- | --- |
| 1 | Social learning | Total spending on data among close social graph neighbours |
| 2 | Discretionary income | Average monthly spending on texts (binned) |
| 3 | Discretionary income | Average monthly number of texts sent (binned) |
| 4 | Discretionary income | Average monthly spending on value added services over text (binned) |
| 5 | Social learning | Average monthly spending on data among social graph neighbours |
| 6 | | Data-enabled handset according to IMEI (Yes/No) |
| 7 | Social learning | Data volume among social graph neighbours |
| 8 | Social learning | Data volume among close social graph neighbours |
| 9 | Timing | Most used handset has changed since last month |
| 10 | | Amount of 'accidental' data usage |

The cross-validated model only relies on a few key variables. Only 20 features out of the initial 350 were selected for the final model. Table 4.1 shows the top 10 most useful features to classify natural converters.

The social learning features computed for this study turn out to be very helpful to help classify the natural converters. The total spending on data among the closest social graph neighbours is the most predictive feature. Using social features in selecting customers for the offer might have improved the retention rates. The results lead to speculation that the value customers derive from mobile data increases when their neighbours are also mobile data users. In other words, it is expected that a network externality effect exists in mobile Internet data. This means that selecting customers whose closest neighbours are already using mobile data might have locally used this network effect to create the lasting effect observed with a very high retention rate in the second month. It is also worth noting that the marketing team believed that average revenue per user (ARPU) would be the most predictive feature for Internet uptake. It can be seen however that the low-ARPU segment was slightly overrepresented, and the text- and data-focused discretionary spending variables are selected as being more important in the model. These findings open up exciting avenues of research within data-driven marketing and customer understanding. Such an approach has now been implemented in the respective business unit, and is greatly reducing spamming by providing the customer with more relevant offers. Such an approach can also be used to micro-target outreach or drive uptake of e.g. agricultural technologies, health-seeking behaviour or mobile money.

# 5 Discussion

Here the overall research question and challenges are addressed from a holistic perspective. The purpose is to analyse the contribution of this thesis, both in detail and from a societal perspective.

## 5.1   On the research question

**Q:** Apart from providing basic communication services, what kind of positive impact can we create for society and/or individuals using large-scale mobile phone datasets?

**A:** As shown in this thesis, new techniques could be adopted to achieve near-real-time insights into people's wellbeing and target aid interventions to vulnerable groups. The results show that mobile phone data can be used for making reliable socio-economic predictions on the individual and geographical level. The way you use a phone can also reveal something about 'who' you are, for example, whether you are illiterate or your income level. Such insight can further be aggregated to the geographical level to help vulnerable groups in society. Additional results show that mobile phone data can be combined with satellite information to estimate multi-dimensional poverty at high spatial resolution, using Bangladesh as an example. Eradication of poverty requires national and subnational quantification and monitoring, and existing poverty estimates rely on census data collected with limited temporal frequency. How traditional survey information may be complimented by adoption of this methodology has also been addressed.

Further results show that mobile phone data can be used to understand human behaviour during large systemic shocks and disasters in society. These results were exemplified by investigating the terror attack that hit Norway in 2011, and by analysing financial, social and mobility patterns from millions of people during a large cyclone event that hit Bangladesh in 2013. Such knowledge can be used to both gain understanding and detect early-warning signals that can help to prevent future disasters. Understanding how the diffusion of information occurs may also help us to understand the complex network in a modern society. In summary, these findings can be used to inform socially beneficial policies that include improved understanding of crisis behaviour, accurate mapping of service needs and faster tracking and response. As obtaining survey information is costly and slow, there is a strong need for better approaches.

The following results are motivated by the questions of how people adopt new products and services, and what role the underlying social network structure plays in this process. Results indicate that social networks strongly matter when purchase decisions are made, and that these effects can be exploited to boost uptake of products and services. By including social patterns and machine learning techniques in a large-scale marketing experiment in Asia, the adoption rate was increased by 13 times compared to current best practice for a given product. This approach, which has now been adopted by the company, greatly reduces spam and provides customers with more relevant offers. At the same time the method provides an efficient way of increasing Internet penetration in a highly underdeveloped country. The same methodology may also be used for other services, as mentioned in section 4.2.2. In summary, these results provide additional insights into human behaviour with the aim to understand social interactions and improve marketing.

## 5.2  Challenges and limitations

Using de-identified mobile phone logs represents many opportunities, but also throws up big challenges around public trust, privacy and potential abuses of data.

### 5.2.1  Privacy challenges

Privacy challenges can be easy to overlook when confronted with the challenges of poverty, disease and basic economic growth. They are however critical to realise the great potential of innovations to help address these critical problems.

Privacy is defined by ITU as the 'right of individuals to control or influence what information related to them may be disclosed' [94]. Personal information that needs to be protected is central to the privacy framework. The OECD defines personal data as 'any information relating to an identified or identifiable individual (data subject)' [95]. The result of this approach has been the policy of 'informed consent' practiced by most companies to inform users of what data are being collected and how they are used. In the world of 'Big Data' this method has been argued to be impractical [96]. The main reason is that the greatest potential often lies in secondary applications long after the data has been collected. It is not possible for companies to know a priori all the potential uses and to continuously seek permission from all users.

Other concerns relate to data anonymisation. Recent research on 'computational privacy' has shown that de-identification of users in CDR is possible, with high accuracy, if only four user data points are available [97]. In practice this is difficult as the real-world identities of users are unknown, but the authors argue that cross-referencing to other data sources might help to de-identify the individual. Such risks are currently lower in underdeveloped countries, as the amount of data stored on each user is still low. As addressed in papers 1 and 2, the great majority of mobile phone users in the developing world are prepaid, with minimal registration associated. In some countries, such as Pakistan, this might change, as mobile phone users must now go through a biometric verification before buying a SIM card. This creates a solid link between the owner of the SIM card and their actual identity [98].

One additional concern is when data go beyond metadata and move into the space of content information [99]. Linking non-personal data to an actual individual can then be made easier [100]. In particular this can be a problem when several sources of Big Data are coupled with other data sources to generate new insight.

### 5.2.2 Analysis and interpreting data

Establishing a generalisation of findings from large-scale analysis might often be difficult, as several data sources are often involved and the chance of data quality issues increases. This section addresses some of the challenges related to analysis and data interpretation.

*Data cleaning process*

Given multiple raw sources of data, it is important to understand how the data has mutated along the pathway into the final datasets. Some mobile phone operators choose to include the complete route of traffic that has been forwarded. This implies that there are multiple records in the CDRs for the same traffic sessions. The result is, for instance, errors when conducting SNA, where edge strengths are overvalued. Another concern is redundant locations, generated by the so-called ping-pong phenomenon where a user connects to several towers frequently. This might introduce errors when inferring the mobility patterns of users [101].

*Selection bias and ground truth data*

A known concern is sampling selection bias. A large dataset may make the sampling rate irrelevant, but it does not make it representative [102]. The fact is that those who use a mobile phone are not necessarily a representative sample of the larger population being considered. This issue is particularly relevant when considering how mobile phone data may be used for monitoring, economic forecasting and development. Research studies are often based purely on data from one mobile operator, and depending on the type of data one can expect wealthier or poorer, and educated or uneducated individuals. Even if data from all operators in a country were available, nearing the total population, this is still not the whole population.

The main problem with results based on non-representative samples is that they lack external validity, which is the degree to which an internally valid conclusion can be generalised beyond the sample.

*Change of behaviour and data sources*

As people adopt new technology, one challenge is the change of structure and content of mobile phone datasets. The signals of human behaviour in mobile phone datasets might change over time and place, analogously to stock markets where signals will change over time and new signals will appear or disappear. As communication signals might be transferred to other platforms, the CDR might not capture the social network as well as before [103]. The choice of data source(s) and when to re-train the models will be vital when considering the research questions. Another challenge is the change in human behaviour itself, which can be subject to self-censorship (turning the phone off in certain areas) and creation of multiple personas (SIM-switching).

*Causation and correlation*

Big Data analysis draws much of its methodology from artificial intelligence and machine learning, which is primarily about prediction and correlations. Most often the datasets are observational and therefore are not able to measure causality. Hal Varian stated that 'there are often more police in precincts with high crime, but that does not imply that increasing the number of police in a precinct would increase crime' [14]. While some researchers have predicted the end of theory and hypotheses-testing [104], others are more sceptical [105]. Sendihl Mullainathan believes that inductive science (algorithmically mining data sources) will not threaten deductive science (hypothesis testing). He believes that greater volume makes Big Data induction techniques more effective, while more variety makes them less effective. There will therefore be a need for explaining behaviour (deductive science) rather than just predicting behaviour. At the same time it is worth mentioning that causal modelling is possible in the Big Data regime by conducting large-scale experiments. Mobile phone operators utilise such techniques to learn about product usage and pricing.

### *The role of 'small' data*

Many of the publications in this thesis rely on a combination of survey data and large-scale datasets. To be able to create understanding and models that reflect ground realities, 'small' survey data will still be important in the future. For many studies it is crucial to verify the underlying assumptions in Big Data using survey data.

### 5.2.3 The future of artificial intelligence in social sciences

Throughout this thesis machine learning and prediction methodology have been applied in a positive context, without much of a focus on potential negative pitfalls. Of course there are both short and long term implications of this emerging field that need to be considered from all angles. Artificial intelligence and machine learning are now being implemented in driverless cars [106], commercial drones [107], financial trading [108], autonomous weapons [109] and monitoring patients in care [110]. As such applications emerge it is important to be transparent around the decision-making process, especially since intelligent machines sometimes make errors too. In the field of social sciences this includes always validating the methodology to actual ground truth data, and using it as a complementary source of insight. It also includes updating analytical models, as researchers pointed out in relation to Google Flu trends, which made inaccurate forecasts for 100 out of 108 weeks [111].

Certainly, the development of artificial intelligence will increasingly make it a part of our daily lives. Currently we do not know if human-level intelligence will be incorporated into artificial intelligence, but it is important to consider it as a possible outcome of today's research.

# 6 Conclusion

This thesis has highlighted the economic and social benefits of using large-scale mobile phone metadata. This data paints a picture of an individual's patterns of behaviour and interactions. When applying this capability across a nation covering millions of people, it is possible to collapse a very detailed picture of the entire population. There are still many challenges to overcome, especially when addressing the privacy implications of Big Data, but mobile phone records can be mutually beneficial to both the private sector and the government.

This thesis has addressed how this information can be used to inform socially beneficial policies on socioeconomics, poverty and disasters on the society level. More targeted and timely policy actions can be taken by governments to address the underlying problems, which would not be possible with the limited and lagged insights revealed by traditional official statistics.

Furthermore, the possibility of using new behavioural insights has been explored, by exploiting large-scale social networks to build models and target the right individuals in marketing experiments. By incorporating social information, product adoption is significantly increased and spam is reduced.

Hopefully, this dissertation will contribute to generating debates and interest among a wide range of audiences, to advancing the understanding of analysis of large-scale datasets and their applications, and most importantly to making such analyses possible and beneficial for society.

# Publications

**Paper 1**
Can mobile usage predict illiteracy in a developing country?..

**Paper 2**
Deep learning applied to mobile phone data for Individual income classification

**Paper 3**
Mapping Poverty using mobile phone and satellite data

**Paper 4**
The activation of core social networks in the wake of the 22 July Oslo bombing

**Paper 5**
Detecting climate adaptation with mobile network data: Anomalies in communication, mobillity and consumption patterns during Cyclone Mahasen

**Paper 6**
Comparing and visualizing the social spreading of products on a large-scale social network

**Paper 7**
Big Data-Driven Marketing: How Machine Learning outperforms marketers' gut-feeling

# Paper 1

# Can mobile usage predict illiteracy in a developing country?

Pål Sundsøy

Telenor Group Research, Big Data Analytics
Snarøyveien 30,1331 Fornebu, Norway

**Abstract.** The present study provides the first evidence that illiteracy can be reliably predicted from standard mobile phone logs. By deriving a broad set of mobile phone indicators reflecting users' financial, social and mobility patterns we show how supervised machine learning can be used to predict individual illiteracy in an Asian developing country, externally validated against a large-scale survey. On average the model performs 10 times better than random guessing with a 70% accuracy. Further we show how individual illiteracy can be aggregated and mapped geographically at cell tower resolution. Geographical mapping of illiteracy is crucial to know where the illiterate people are, and where to put in resources. In underdeveloped countries such mappings are often based on out-dated household surveys with low spatial and temporal resolution. One in five people worldwide struggle with illiteracy, and it is estimated that illiteracy costs the global economy more than $1 trillion dollars each year [*1*]. These results potentially enable cost-effective, questionnaire-free investigation of illiteracy-related questions on an unprecedented scale.

## 1 Introduction

Functional illiteracy means a person may be able to write simple words, but cannot apply these skills to tasks such as filling out a job application, reading a medicine label or balancing a chequebook [*2*]. Illiterates are often trapped in a cycle of poverty with limited opportunities for employment or income generation [*3*]. They also have higher chances of poor health, turning to crime and dependence on social welfare or charity when available [*4*]. Mapping of literacy statistics is currently based on tedious household surveys with a low spatial and temporal frequency [*5*]. The increasing availability and reliability of new data sources, and the growing demand of comprehensive, up-to-date international literacy data are therefore of high priority [*6*]. One of the most promising rich Big Data sources are mobile phone logs [*7*], which have the potential to deliver near real-time information of human behaviour on individual and societal scale [*8*]. Several research studies have used large-scale mobile phone metadata, in the form of call detail records (CDR) and airtime purchases (top-up) to quantify various socio-economic dimensions. Eagle et al. [*9*] quantified the correlation between network diversity and a population's economic well-being. The findings revealed that the diversity of individuals' relationships is strongly correlated with the economic development of communities. The assumption that more diverse ties correlate with better access to social and economic opportunities was untested at the population level. It concludes that frequently making and receiving calls with contacts outside of one's immediate community is correlated with higher socio-economic class. CDRs have also shown to provide proxy indicators for assessing regional poverty levels, as shown in studies done in several countries [*10,11,12*]. It has also been hypothesized that airtime purchases is correlated with socio-economic status [*13*], and this has later been extended and verified also using external reliable data [*14*] . Monitoring airtime transactions for trends and sudden changes can also be useful for detecting early impact of economic crisis [*15*], as well as for measuring impact of programmes designed to improve livelihoods and food security [*16*].

The rest of this paper is organized as follows: In section 2 we describe the methodological approach, including the features and modelling approach. In section 3 we describe the results. In chapter 4 we discuss the limitations from a holistic perspective, while we finally draw our conclusions in section 5.

## 2 Approach

We collect educational status, including illiteracy, from 76 000 individuals in a low HDI Asian country based on two large-scale surveys run by a professional agency on behalf of a large mobile operator. The sample includes 6.8% illiterates, 40% primary degree, 26% SSC, 17.6% HSC, 5.6% bachelor, 3.5% master and 0.13% other degrees (incl. Ph.D.). The survey is representative for all mobile users in the country, where more than 85% have a mobile phone. In general, half of the world's population now have a mobile phone [*17*]. The individual records are de-identified and coupled with 6-month raw mobile phone metadata, including CDR and airtime purchases. Content information, such as SMS content, is never accessed by the operator or any researcher. The educational status is re-labelled to a binary classifier, namely illiterates and non-illiterates. This section describes the features and the machine learning algorithm used for our prediction.

### 2.1 Features

A structured dataset consisting of 160 mobile phone features are built, and categorized into three dimensions: (1) financial (2) mobility and (3) social features, as shown in Table 1. The features are custom made and include various parameters of the corresponding distributions such as weekly or monthly median, mean and variance.

**Table 1.** Sample of features from mobile phone metadata used in model

| Dimension | Features |
|---|---|
| Financial 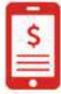 | **Airtime purchases:** Recharge amount per transaction, Spending speed, fraction of lowest/highest recharge amount, coefficient of variation recharge amount etc |
| | **Revenue:** Charge of outgoing/incoming SMS, MMS, voice, video, value added sevices, roaming, internet etc. |
| | **Handset:** Manufacturer,brand, camera enabled, smart/feature/basic phone etc |
| Mobility 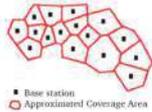 | Home district/tower, radius of gyration, entropy of places, number of places visited etc. |
| Social 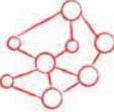 | **Social Network:** Interaction per contact, degree, entropy of contacts etc. |
| | **General phone usage:** Out/In voice duration, SMS count, Internet volume/count, MMS count, video count/duration, value added services duration/count etc. |

### 2.2 Model algorithm

Several model algorithms were tested such as neural networks, random forest and support vector machines. Based on performance, a gradient boosted machines model (GBM) ended up as the final model [*18*]. Here the base classifiers are built sequentially, and the algorithm combines the new classifier with ones from previous iterations in an attempt to reduce the overall error rate. The main motivation is to combine several weak models to produce a powerful ensemble. To compensate class imbalance, the minority class in the *training set*, that includes the illiterates being 6.8%, is up-sampled. The minority class is then randomly sampled, with replacement, to be the same size as the majority class. A 10-fold cross-validation is used as re-sampling technique. Feature importance scores are calculated via a backwards elimination feature selection routine that looks at reductions in the generalized cross-validation (GCV) estimate of error. It tracks the changes in GCV, for each predictor and accumulates the reduction in the statistic when each predictor's feature is added to the model. In our set-up, each model is trained and tested using a 75/25 split. All results are reported for the test-set.

## 3   Results

### 3.1   Individual illiteracy

Figure 1 shows the final features and their contribution in predicting illiteracy. Concretely, 19 of our features were related to illiteracy and were all included in the final GBM classifier. The model predicted whether phone users were illiterate with an accuracy of 70.1% (95% CI: 69.6-70.8). The deviation of accuracy from the training set was only 3.8%, which disregard model overfitting. The true positive rate (sensitivity/recall) was 71.6% and true negative rate (specificity) 70%. Given the original baseline of 6.8% in the test set, we predict on average 10 times better than random.

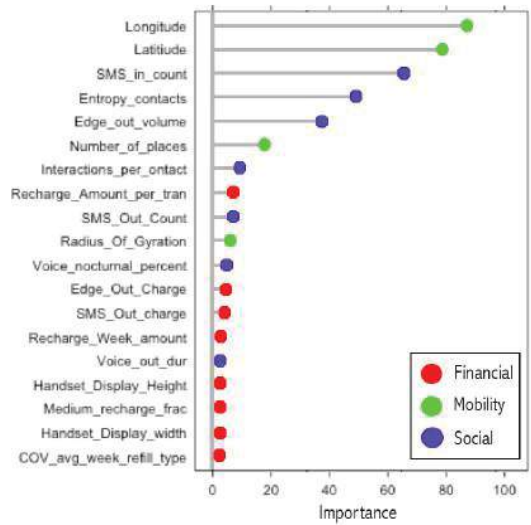

**Fig. 1.** Top features in the GBM Model colored by their respective feature family

An investigation of the most important predictors, as seen in Fig 1, reveals some interesting associations. We especially notice that most frequently used longitude and latitude stand out as good predictors: where the people spend most of their time is a good signal of their education level. One explanation can be that this signal indicates that the model catch regions of low economic development status, e.g. slum areas where illiteracy is high. Another important feature is the number of incoming SMS, which outperforms outgoing SMS: one hypothesis is that people do not send SMS' to persons that they know are illiterate. Moreover, we see that entropy of contacts is important – illiterates tend to concentrate their communication on few people. This is also in line with Eagle's work on geographical level [9], which shows that economic well-being is correlated with social diversity. Further we see that illiterates have limited use of internet (predictor 5), and their mobility pattern is limited to a few base stations (predictor 6).

The corresponding density distributions of the top 6 predictors are shown in Fig 2. We notice sharp peaks in the location distributions (Fig 2 A and B), which indicate that larger urban areas where (on average) people are more literate are picked up. This has been verified by checking the coordinate range. The dip in the Internet volume distribution (Fig 2 E) can be explained by fixed volume Internet packages that the mobile operator provides.

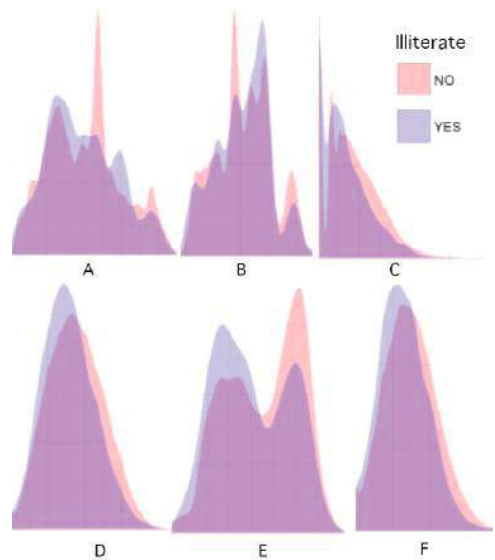

**Fig. 2.** Density distributions of top 6 mobile phone predictors of individual illiteracy A) Longitude B) Latitude C) Incoming SMS (log) D) Entropy of contacts E) Internet volume (log) F) Number of places visited

### 3.2 Geographical illiteracy mapping

A natural next step is to move from individual illiteracy to geographical illiteracy. In big Asian cities there are often thousands of mobile towers that can be used as "sensors" to estimate illiteracy rates in the areas covered by the towers. In the rural areas where towers are less dense, interpolation techniques can be utilized to include information from the neighbour towers. In figure 3 A) we have mapped out the predicted illiteracy rate per tower, in one of the larger cities.

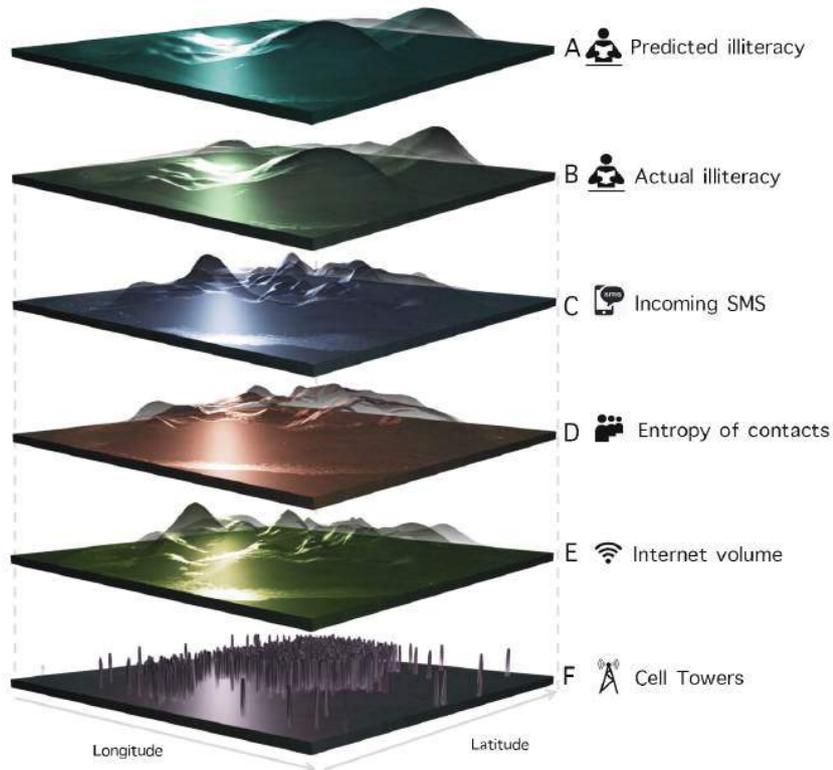

**Fig. 3.** Geographical mapping of illiteracy, top predictors and the cell tower distribution in one major citiy. Height (z-axis) is proportional to the tower averages for each given metric.

The individual illiteracy rates are here calculated by using the test set, aggregated and averaged to tower level, and then further spatially interpolated, using an IDW algorithm [19], to average out the noise of local variations between towers. The *actual* illiterate rates in Fig 3b is calculated by using the training set as ground truth. We notice three large pockets of larger illiteracy rates in the city. By also including distributions of the top predictors (Fig 3 c-e) it is possible to visually observe spatial correlations. E.g. we can observe a large area of high SMS activity (Fig 3c,left) that can be associated with low illiteracy rates. Such a visual approach might help the interpretation of findings: In most areas we will observe anti-correlated behaviour of illiteracy and the top predictors, but given the non-linearity of the GBM model, where several terms interact, this is not always the case. E.g. it has been reported from the mobile operator that some groups of illiterate people have learned to answer SMS when special mobile offers are marketed. Such learnings can quickly go viral, and spread to nearest social and spatial neighbourhood. A traditional linear model will not easily pick up these complex non-linear interactions.

## 4  Discussion and challenges

One general concern in such studies is always the sampling selection bias. A large data set may make the sampling rate irrelevant, but it doesn't necessarily make it representative [20]. The fact that the people who use mobile phone are not necessarily a representative sample of the larger population considered. This issue is especially of high relevance when considering how mobile phone data may be used for monitoring, economic forecasting and development. Research studies are often based purely on data from one mobile operator, and depending on the type of data one can expect wealthier or poorer, more males than females and uneducated or educated individuals. Even if data from all operators in a country were available, nearing the total population, it is still not the whole population. In our study we consider data from *one* large operator, where people has to own a mobile phone to be counted. We argue, however, that there should be a good correlation between *population literacy* and *sample literacy*, since the sample is large, and most people in the country own a mobile phone. Additional sources for external validation, except the two large-scale surveys, were not easily obtainable.

Another challenge is that behavioural "signals" might change over time and place. As traditional mobile communication might be transferred to other platforms (e.g. WhatsApp, Facebook messenger) the mobile phone operators might not be able to capture the social network as well as before. The model input features will therefore change accordingly. The choice of data sources and the frequency of model re-training will be vital when considering the research questions. Also, especially in development countries, people might switch between various SIM-cards to save money, and several persons might use the same phone. In countries where this is a problem, the prediction algorithm might get confused, and model performance lead to higher number of false positives. By dealing with such issues a priori, e.g. including questions around it in validation survey, we should expect an even higher accuracy.

The last challenge addresses correlation and causation. Most often the datasets are (by nature) observational and can therefore not measure causality [*21*]. In our study we don't explain behaviour (deductive science), rather just predicting behaviour. Illiteracy might have many causes, such as living conditions, including poverty, parents with little schooling, lack of books at home, learning disabilities, doing badly or dropping out of school. Where you live, or what cell tower you use most, do not alone decide your educational status. If an illiterate move from one area to another area, with few illiterates, this doesn't cause the person more literate. Oppositely, moving from a well-educated area into a poor neighbourhood, with overrepresented illiteracy, doesn't make a person more illiterate. As the model consists of a non-linear combination of many predictors and weightings, there is a reduced risk of such errors, and less chance of false positives or negatives: it is not one or two predictors alone that decide whether a person is illiterate. The same goes for the geographical level. If illiteracy in one area increases, we might see a drop in the incoming SMS, social entropy and Internet volumes – and the complex relationships between the 19 features in the model make the final call whether an early warning signal should be sent out. Of course there is a trade-off on accuracy between such a "black box" approach and an interpretable linear model. Consequently we do not claim to have found causal patterns between mobile phone usage and illiteracy – but potentially useful predictive signals for NGO's who need complementary estimates to detect illiteracy on a higher spatial and temporal resolution.

## 5  Conclusion

This study shows how illiteracy can be predicted from mobile phone logs, purely by investigating users' metadata. By deriving economic, social and mobility features for each mobile user we predict individual illiteracy status with 70% accuracy. Further we show how individual illiteracy can be aggregated and mapped geographically with high spatial resolution on cell tower level. Feature investigation indicates that home cell tower and incoming SMS are the superior predictors, followed by diversity of communication partners and Internet volume.

An important policy application of this work is the prediction of regional and individual illiteracy rates in underdeveloped countries where official statistics is limited or non-existing. For future work we would like to address the issues mentioned in chapter 4, and see how our method reflect the overall population illiteracy, as well as verifying the robustness of predictors in other countries. We would also like to combine mobile phone data with other spatial sources to address how poverty mapping can be done more efficiently to support the UN sustainable development goal of eradication of extreme poverty within 2030.

# Paper 2

# Deep learning applied to mobile phone data for Individual income classification

Pål Sundsøy[1][*], Johannes Bjelland[1], Bjørn-Atle Reme[1], Asif M.Iqbal[1] and Eaman Jahani[2]

[1]Telenor Group Research, Snarøyveien 30,1331 Fornebu, Norway
[2]Massachusetts Institute of Technology, The Media Laboratory, USA
[*]pal-roe.sundsoy@telenor.com

*Abstract*— Deep learning has in recent years brought breakthroughs in several domains, most notably voice and image recognition. In this work we extend deep learning into a new application domain - namely classification on mobile phone datasets. Classic machine learning methods have produced good results in telecom prediction tasks, but are underutilized due to resource-intensive and domain-specific feature engineering. Moreover, traditional machine learning algorithms require separate feature engineering in different countries. In this work, we show how socio-economic status in large de-identified mobile phone datasets can be accurately classified using deep learning, thus avoiding the cumbersome and manual feature engineering process. We implement a simple deep learning architecture and compare it with traditional data mining models as our benchmarks. **On average our model achieves 77% AUC on test data using location traces as the sole input. In contrast, the benchmarked state-of-the-art data mining models include various feature categories such as basic phone usage, top-up pattern, handset type, social network structure and individual mobility. The traditional machine learning models achieve 72% AUC in the best-case scenario.** We believe these results are encouraging since average regional household income is an important input to a wide range of economic policies. In underdeveloped countries reliable statistics of income is often lacking, not frequently updated, and is rarely fine-grained to sub-regions of the country. Making income prediction simpler and more efficient can be of great help to policy makers and charity organizations – which will ultimately benefit the poor.

*Deep learning; Mobile phone data; poverty;household income ; Big Data Analytics; Machine learning; Asia, Mobile network operator; metadata; algorithms*

## I. INTRODUCTION

Recent advances in Deep Learning [1][2] have made it possible to extract high-level features from raw sensory data, leading to breakthroughs in computer vision [9][10][11] and speech recognition [12][13]. It seems natural to ask whether similar techniques could also be beneficial for useful prediction tasks on mobile phone data, where classic machine learning algorithms are often under-utilized due to time-consuming country and domain- specific feature engineering [6].

Our work investigates how we can separate individuals with high and low socio-economic status using mobile phone call detail records (CDR). Finding good proxies for income in mobile phone data could lead to better poverty prediction – which could ultimately lead to more efficient policies for addressing extreme poverty in hardest hit regions.

With this in mind, we perform a large-scale country-representative survey in a low HDI Asian country, where household income is collected from approximately 80 000 individuals. The individual records are de-identified and coupled with raw mobile phone data that span over 3 months. This dataset allow us to build a deep learning model, as well as a benchmarking model using custom feature engineering and traditional data mining algorithms.

From the household income we derive two binary classifiers (1) below or above median household income and (2) below or above upper poverty level. The income threshold for poverty level is derived from official statistics and based on average national household income and household size. Our survey classifies participants into 13 household income bins – where bin 1 and 2 correspond to below upper poverty level.

The rest of this paper is organized as follows: In section 2 we describe the features and models used for benchmarking our deep learning approach. Section 3 describes the deep learning approach itself. In section 4 we compare the results of the two approaches. Finally, we draw our conclusions in section 5.

## II. BEST PRACTICE – TRADITIONAL DATA MINING

This section describes the features and the standard machine learning algorithms used as our benchmark. The data preparation phase in the data mining process is a tedious task, that requires specific knowledge about both the local market and the various data channels as potential sources for input features. Typically the data warehouse architecture and the data format vary between the operators and third-party software packages, making it hard to create generalizable feature-sets.

### A. Features

We build a structured dataset consisting of 150 features from 7 different feature families, see Table 1. The features are custom-made and range from basic phone usage, top-up pattern, social network and mobility to handset usage. The features include various parameters of the corresponding distributions such as weekly or monthly median, mean and variance.

**TABLE 1** SAMPLE FEATURES USED IN TRADITIONAL MACHINE LEARNING BENCHMARKS.

| Feature family | Feature examples |
|---|---|
| Basic phone usage | Outgoing/incoming voice duration, sms count etc. |
| Top-up transactions | Spending speed, recharge amount per transaction, fraction of lowest/highest recharge amount, coefficient of variation recharge amount etc |
| Location/mobility | Home district/tower, radius of gyration, entropy of places, number of places etc. |
| Social Network | Interaction per contact, degree, entropy of contacts etc. |
| Handset type | Brand, manufacturer, camera enabled, smart/feature/basic phone etc |
| Revenue | Charge of outgoing/incoming SMS, MMS, voice, video, value added sevices, roaming, internet etc. |
| Advanced phone usage | Internet volume/count, MMS count, video count/duration, value added services duration/count etc. |

### A. Models

Ensemble methods have been proven as powerful algorithms when applied to large-scale telecom datasets [14][6]. They combine predictions of several base classifiers built with a given learning algorithm in order to improve robustness over a single classifier. We investigate two different state-of-the-art ensemble methods:

*1) Random forest*, where we build several independent classifiers and average their predictions, thus reducing the variance. We choose grid search to optimize the tree size.

*2) Gradient boosting machines* (GBM) where the base classifiers are built sequentially. The algorithm combines the new classifier with ones from previous iterations in an attempt to reduce the overall error rate. The main motivation is to combine several weak models to produce a powerful ensemble.

In our set-up, each model is trained and tested using a 75/25 split. For the response variable related to poverty level we introduce misclassfication cost. Since few people are below the poverty level (minority class), a naive model will predict everyone above poverty line. We therefore apply a misclassification cost to the minority class to achieve fewer false positives and adjust for the ratio between the classes.

### III. DEEP LEARNING

In this section we describe the input data and the structure of our deep learning model.

### B. Features

Classification results of the traditional learning algorithms are inherently limited in performance by the quality of the extracted features [7]. Deep learning can instead reproduce complicated functions that represent higher level extractions, and replace manual domain-specific feature engineering.

Earlier studies have shown a good correlation between location/human mobility and socio-economic levels [15-17]. Using this as a motivation we build a simple vector whose length corresponds to the number of mobile towers (8100 dimensions), and the vector elements correspond to the mobile phone activity at the given tower – shown in Table 2.

**TABLE 2** INPUT VECTORS TO DL ALGORITHM BEFORE NORMALIZATION

| Hashed Phone number | Tower 1 | Tower2 | …. | Tower8100 | Below/above poverty level |
|---|---|---|---|---|---|
| 1 | 0 | 67 | …. | 16 | 0 |
| 2 | 7 | 0 | …. | 0 | 1 |
| . | . | . | …. | . | . |
| . | . | . | …. | . | . |
| 80K | 0 | 9 | …. | 6 | 0 |

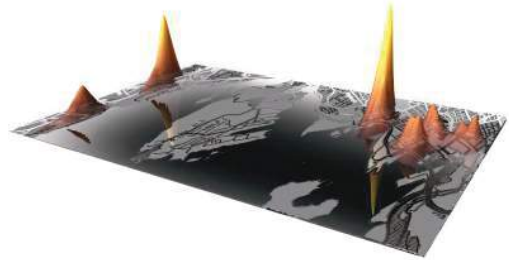

**FIGURE 1** LOCATION ACTIVITY MAP. THE MAP PROVIDES INFORMATION ABOUT EACH USERS' TOWER ACTIVITY.

A visual representation of Table 2 can be seen in Fig 1, where each subscriber is represented by a location density map.

### A. Models

We use a standard multi-layer feedforward architecture where the weighted combination of the n input signals is aggregated, and an output signal $f(\alpha)$ is transmitted by the connected neuron. The function $f$ used for the nonlinear activation is rectifier $f(\alpha) \approx \log(1+e^{\alpha})$. To minimize the loss function we apply a standard stochastic gradient descent with the gradient computed via back-propagation. We use dropout [4][5] as a regularization technique to prevent over-fitting. For the input layer we use the value of 0.1 and 0.2 for the hidden layers. Dropout secures that each training example is used in a different model which all share the same global parameters. This allows the models to be averaged as ensembles, improving generalization and preventing overfitting. The split between train and test set, as well as the introduced misclassification cost for poverty level, are similar to the benchmark models.

## IV. RESULTS

Our deep neural network is trained to separate individuals with high and low socio-economic status. We evaluate our models' performance with AUC [2][3] achieving 77% and 74% AUC on test set when the classifier is predicting above/below median household income and above/below poverty level respectively (Fig 2, Table 3). The corresponding AUC on the train sets are respectively 80% and 77%, showing that our model does not overfit significantly. Our results indicate that the DL model achieves a higher performance than those of multi-source RF and GBM models, which vary between 71-72% and 68-69% for above/below median household income and poverty level respectively.

Next, we feed the RF and GBM models with the same representation as fed into the DL algorithm - achieving RF performance of AUC 64% and GBM performance of AUC 61%. The performance of these models greatly suffers as the number of input features increase without increasing the training size. We conclude that given a fixed training sample, the traditional models are not able to learn a complicated function that represents higher level extractions, but perform better when using manual domain-specific feature engineering.

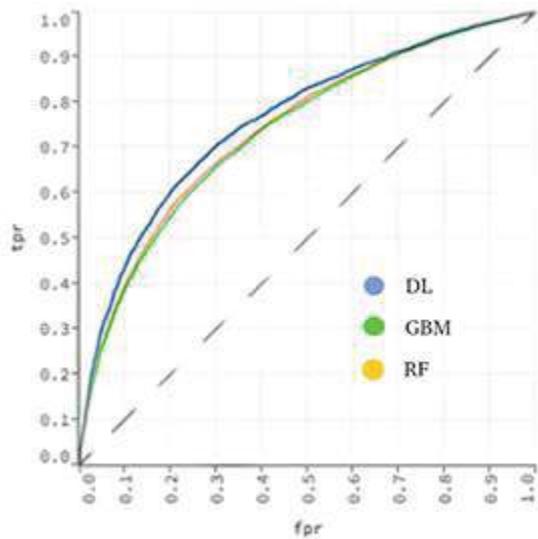

**FIGURE 2** AUC ON TEST SET SHOWING THE TRUE POSITIVE VS FALSE POSITIVE RATE FOR DEEP LEARNING (DL), GRADIENT BOOSTING MACHINES (GBM) AND RANDOM FOREST (RF).

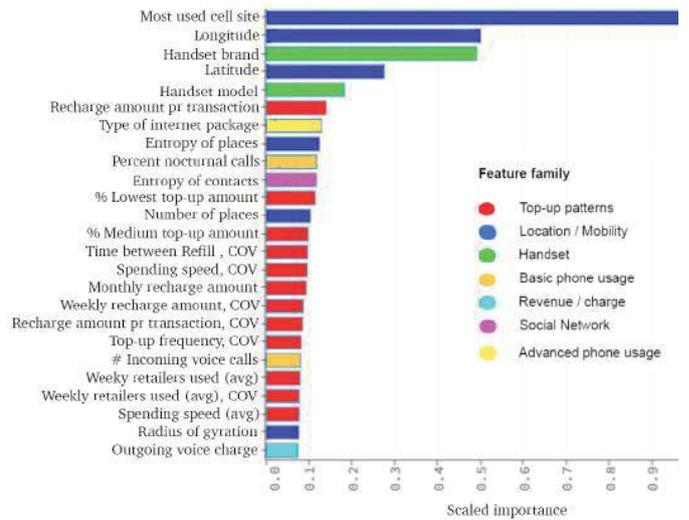

**FIGURE 3** TOP FEATURES IN RF MODEL COLORED BY THEIR RESPECTIVE FEATURE FAMILY.

**TABLE 3** MODEL PERFORMANCE (AUC) ON TEST SET

| Model | AUC test set Below/above poverty level | AUC test set Below/above median income |
|---|---|---|
| DL | **0.74** | **0.77** |
| GBM | 0.68 | 0.71 |
| RF | 0.69 | 0.72 |

A posteriori inspection of the top features selected by our traditional models leads to some interesting qualitative insights. We observe a similar pattern in the top features selected by the RF and GBM model. Figure 3 shows the top features in our RF model. We notice the importance of three feature families:

*1) Location dynamics:* Where the user spends most of his time is a good signal of his income. This indicates that our models have detected regions of low economic development status.

*2) The handset brand:* In the country of our study, minimal and more affordable handset brands are very popular among the lower income quantiles, while expensive smartphones are considered as a huge status symbol.

*3) The top-up pattern:* Interestingly, the recharge amount per transaction is more predictive than the total recharge amount. We observe that individuals from the lower income quantiles usually top-up with lower amounts when they first fill up their account.

## V. CONCLUSION

In order to predict household income based on mobile phone communication and mobility patterns, we implemented a multi-layer feedforward deep learning architecture. Our approach introduces a novel data representation for learning neural networks on real CDR data.

Our approach suggests that multi-layer feedforward models are an effective tool for predicting economic indicators based on mobile communication patterns. While capturing the complex dependencies between different dimensions of the data, deep learning algorithms do not overfit the training data as seen by our test performance.

Furthermore, our deep learning model, using only a single dimension of the data in its raw form, achieves a 7% better performance compared to the best traditional data mining approach based on custom engineered features from multiple data dimensions. Even though such an automated approach is time-saving, many of the classic machine learning approaches have the advantage of being interpretable. However, since a large portion of a data mining process is data preparation, there is a big demand to automate this initial step.

As future work, we would like to investigate the performance implications of including temporal aspects of raw CDRs in our models and the data representation. In addition, we will work on finding a general representation of telecom data that can be used for various prediction tasks.

An important application of this work is the prediction of regional and individual poverty levels in low HDI countries. Since our approach only requires de-identified customer and tower IDs, we find this method more privacy preserving compared to traditional data mining approaches where the input features may reveal sensitive information about the customers.

# ◆ Paper 3

# INTERFACE

rsif.royalsocietypublishing.org

Research 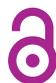

# Mapping poverty using mobile phone and satellite data

Jessica E. Steele[1,2], Pål Roe Sundsøy[3], Carla Pezzulo[1], Victor A. Alegana[1], Tomas J. Bird[1], Joshua Blumenstock[4], Johannes Bjelland[3], Kenth Engø-Monsen[3], Yves-Alexandre de Montjoye[5], Asif M. Iqbal[6], Khandakar N. Hadiuzzaman[6], Xin Lu[2,7,8], Erik Wetter[2,9], Andrew J. Tatem[1,2,10] and Linus Bengtsson[2,7]

[1]Geography and Environment, University of Southampton, University Road, Building 44, Southampton, UK
[2]Flowminder Foundation, Roslagsgatan 17, Stockholm, Sweden
[3]Telenor Group Research, Oslo, Norway
[4]School of Information, University of California, Berkeley, CA, USA
[5]Data Science Institute, Imperial College London, London, UK
[6]Grameenphone Ltd, Dhaka, Bangladesh
[7]Public Health Sciences, Karolinska Institute, Stockholm, Sweden
[8]College of Information System and Management, National University of Defense Technology, Changsha, Hunan, People's Republic of China
[9]Stockholm School of Economics, Saltmätargatan 13-17, Stockholm, Sweden
[10]John E Fogarty International Center, National Institutes of Health, Bethesda, MD, USA

JES, 0000-0001-6741-1195; PRS, 0000-0003-4596-3209; VAA, 0000-0001-5177-9227; TJB, 0000-0003-1345-6480; JB, 0000-0002-1813-7414; KE-M, 0000-0003-1618-7597; Y-AdM, 0000-0002-2559-5616; LB, 0000-0001-9712-2669

Poverty is one of the most important determinants of adverse health outcomes globally, a major cause of societal instability and one of the largest causes of lost human potential. Traditional approaches to measuring and targeting poverty rely heavily on census data, which in most low- and middle-income countries (LMICs) are unavailable or out-of-date. Alternate measures are needed to complement and update estimates between censuses. This study demonstrates how public and private data sources that are commonly available for LMICs can be used to provide novel insight into the spatial distribution of poverty. We evaluate the relative value of modelling three traditional poverty measures using aggregate data from mobile operators and widely available geospatial data. Taken together, models combining these data sources provide the best predictive power (highest $r^2 = 0.78$) and lowest error, but generally models employing mobile data only yield comparable results, offering the potential to measure poverty more frequently and at finer granularity. Stratifying models into urban and rural areas highlights the advantage of using mobile data in urban areas and different data in different contexts. The findings indicate the possibility to estimate and continually monitor poverty rates at high spatial resolution in countries with limited capacity to support traditional methods of data collection.

## 1. Background

In 2015, approximately 700 million people lived in extreme poverty [1]. Poverty is a major determinant of adverse health outcomes including child mortality [2], and contributes to population growth [3], societal instability and conflict [4]. Eradicating poverty in all its forms remains a major challenge and the first target of the Sustainable Development Goals (SDGs) [5]. To eradicate poverty, it is crucial that information is available on where affected people live. Such data improve the understanding of the causes of poverty, enable improved allocation of resources for poverty alleviation programmes, and are a critical component for monitoring poverty rates over time. The latter issue is especially pertinent for efforts aimed at reaching the SDGs, which need to be monitored at national and subnational levels over the coming 15 years [5].

The definition of poverty and the measurement methods used to identify poor persons are part of a longstanding discussion in development economics [6–9]. Different approaches exist to calculate indicators of living standards, including the construction of unidimensional and multidimensional indices, as well as the use of monetary or non-monetary metrics. A further discussion for living standard indices regards the methods used to set appropriate thresholds (poverty lines) under which a person is defined as poor [10–12]. Monetary-based metrics identify poverty as a shortfall in consumption (or income) and measure whether households or individuals fall above or below a defined poverty line [13,14]. By contrast, asset-based indicators define household welfare based on asset ownership (e.g. refrigerator, radio or bicycle), dwelling characteristics, and access to basic services like clean water and electricity [15]. Moreover, poverty indicators can capture the status of a household or individual at a given point in time, or identify chronic versus transient poverty over time [14,16–18].

Every approach used to calculate indicators of living standards for a population has its advantages and disadvantages, and each indicator discerns different characteristics of the population. Consumption data can be highly noisy due to recall error or because expenditures occurred outside the period captured in surveys, but provide a better shorter-term concept of poverty [19,20]. Asset-based measures have been regarded as a better proxy for the long-term status of households as they are thought to be more representative of permanent income or long-term control of resources [20–22]. The same population can be ranked quite differently along a poverty distribution when comparing consumption and asset-based measures and many assumptions are necessarily accepted in order do such comparisons. These include assumptions that the data represent the same populations in the same time period; that the indicators are well matched in their wording and response options; and that the poverty measures have a similar distribution of responses [20,23]. Furthermore, it is difficult to compare asset-based measures to income or consumption as it is not straightforward to link the productive potential of a household to their assets owned; this can be particularly relevant in rural areas where the return on physical assets can be strongly environmentally related and interactions among assets may be important [24]. These factors necessitate a flexible approach to modelling poverty as indicators representing asset-based, consumption-based and income-based measures are not necessarily expected to produce similar results.

While numerous high-resolution indicators of human welfare are routinely collected for populations in high-income countries, the geographical distribution of poverty in low- and middle-income countries (LMICs) is often uncertain [25]. Small area estimation (SAE) forms the standard approach to produce sub-national estimates of the proportion of households in poverty. SAE uses statistical techniques to estimate parameters for sub-populations by combining household survey and census data to use the detail in household surveys and the coverage of the census. Common variables between the two are used to predict a poverty metric across the population [26–28]. These techniques rely on the availability of census data, which are typically collected every 10 years and often released with a delay of one or more years, making the updating of poverty estimates challenging. Recently, there are promising signs that novel sources of high-resolution data can provide an accurate and up-to-date indication of living conditions. In particular, recent work illustrates the potential of features derived from remote sensing and geographic information system data [29–35] (hereafter called RS data) and mobile operator call detail records (CDRs) [36–39]. However, the predictive power in integrating these two data sources, and their ability to estimate different measures of poverty has not been evaluated.

RS and CDR data capture distinct and complimentary correlates of human living conditions and behaviour. For example, RS data of physical properties, such as rainfall, temperature and vegetation capture information related to agricultural productivity, while distance to roads and cities reflects access to markets and information. Similarly, monthly credit consumption on mobile phones and the proportion of people in an area using mobile phones indicate household access to financial resources, while movements of mobile phones and the structure and geographical reach of the calling networks of individuals may be correlated with remittance flows and economic opportunities [39–41].

RS and CDR data are generated at different spatial scales, which further complement each other. The CDR indicators used in this study are derived from data aggregated at the level of the physical cell towers to preserve the privacy of individual subscribers. Thus, the spatial resolution of these data is determined by tower coverage, which is larger in rural areas and fine-scaled in urban areas. By contrast, RS data can be relatively coarse in urban areas and only capture physical properties of the land. As RS and CDR data are continually collected, the ability to produce accurate maps using these data types offers the promise of ongoing subnational monitoring required by the SDGs.

Here, we use overlapping sources of RS, CDR and traditional survey-based data from Bangladesh to provide the first systematic evaluation of the extent to which different sources of input data can accurately estimate three different measures of poverty. To date, the predictive power in integrating these data sources, and their ability to estimate different measures of poverty, has not been evaluated. We use hierarchical Bayesian geostatistical models (BGMs) to construct highly granular maps of poverty for three commonly used indicators of living standards: the Demographic and Health Surveys (DHS) Wealth Index (WI); an indicator of household expenditures (Progress out of Poverty Index, PPI) [42] and reported household monetary income. We additionally compare our results with previous poverty estimates for Bangladesh at coarser and finer resolutions.

## 2. Material and methods

### 2.1. Spatial scale and data processing

All data used in this study were processed to ensure that projections, resolutions and extents matched. The spatial scale of analysis was based on approximating the mobile tower coverage areas using Voronoi tessellation [43] and models were built on the scale of the Voronoi polygons (figure 1). This allowed us to maintain the fine spatial detail in mobile phone data within urban areas, as Voronoi polygon size, and corresponding spatial detail, varies greatly from urban to rural areas (minimum 60 m, maximum 5 km) as shown in the figure. All datasets were then summarized to spatially align with these polygons. In practice, each polygon was assigned RS and CDR values representing the mean, sum or mode of the corresponding data. The survey data are matched to the Voronois based on the GPS located lat/long of PPI data, the lat/long representing the centroid of each DHS cluster, and the home tower of each income survey respondent. Where multiple points from the same output (WI, PPI and income) fell within the same polygon, we used the mean aggregated value.

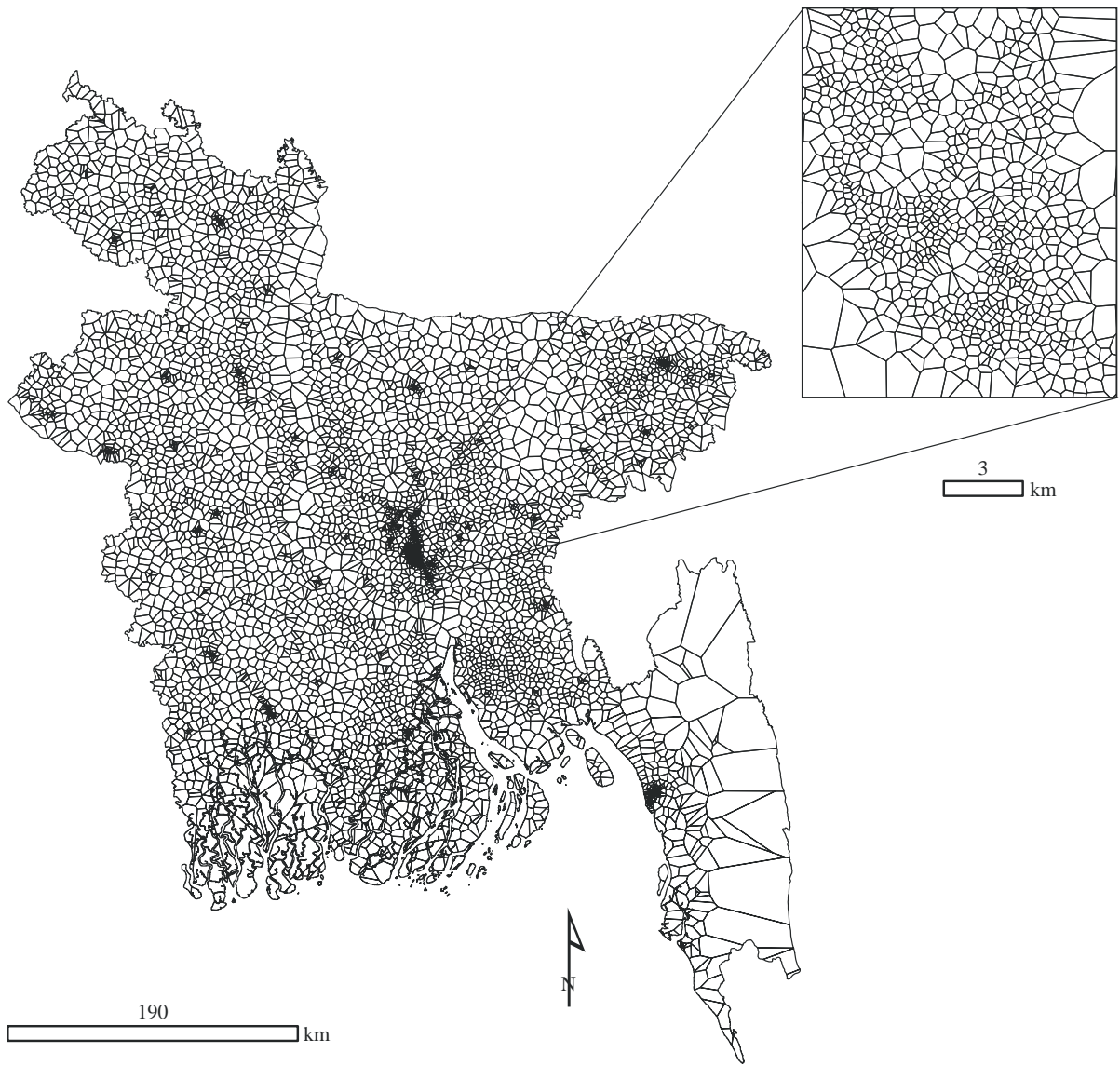

**Figure 1.** Spatial structure of Voronoi polygons based on the configuration of mobile phone towers in Bangladesh. The zoom window shows the spatial detail of Dhaka.

## 2.2. Poverty data

We used three geographically referenced datasets representing asset, consumption and income-based measures of wellbeing in Bangladesh (see the electronic supplementary material, figure S1 and section A.1). These data were obtained from three sources: the 2011 Bangladesh DHS, the 2014 FII survey [44] with data collected on the PPI (www.progressoutofpoverty.org) and national household surveys conducted by Telenor Group subsidiary Grameenphone (GP) between November 2013 and March 2014 collecting household income data.

The DHS WI is constructed by taking the first principal component of a basket of household assets and housing characteristics such as floor type and ceiling material, which explains the largest percentage of the total variance, adjusting for differences in urban and rural strata [45]. A final composite combined score is then used as a WI whereby each household is assigned its correspondent quintile in the distribution and each individual belonging to the same household shares the same WI score. A higher score implies higher socioeconomic status (range = −1.45 to 3.5). Here, we used aggregated average WI scores per primary sampling unit (PSU) for 600 PSUs (207 in urban areas and 393 in rural areas) to estimate the mean WI of sampled populations residing in each Voronoi polygon.

The PPI is a measurement tool built from the answers to 10 questions about a household's characteristics and asset ownership, scored to compute the likelihood the household is living above or below a poverty line. In Bangladesh, these poverty score-card questions were determined using data from the 2010 Household Income and Expenditure Survey (HIES) [42,46], and used in a nationally representative survey of 6000 Bangladeshi adults undertaken in 2014 [44]. Together with basic demographics and access to financial services information, the 10 questions needed to construct the PPI were collected. These data were used to assign a poverty measure to each individual interviewed: the likelihood they have *per capita* expenditure above or below a poverty line. Here, we estimate the mean likelihood (range = 12.3–99.7%) of populations residing in each Voronoi polygon to be below the $2.50 a day poverty line.

Income data were obtained from two independent, sequential household surveys run by GP. For each survey, face-to-face interviews were conducted with 90 000 individuals, and their corresponding household income was collected, together with basic demographic information for each survey participant (e.g. gender, age, profession, education) and phone usage. Respondents were directly asked about income and were requested to place themselves within pre-set income bins. Among GP subscribers, CDRs were successfully linked to phone numbers for 76 000 participants. Here we converted income bins to USD (range = 0–1285$) and modelled the average USD for each Voronoi polygon.

### 2.3. CDR and RS data

CDR features were generated from four months of mobile phone metadata collected between November 2013 and March 2014. GP subscribers consented to the use of their data for the analysis. GP, the largest mobile network operator in Bangladesh, had 48 million customers at the time of the analysis, with a network covering 99% of the population and 90% of the land area [47]. CDR features range from metrics such as basic phone usage, top-up patterns, and social network to metrics of user mobility and handset usage. These features are easily made available in data warehouses and do not rely on complex algorithms. They include various parameters of the corresponding distributions such as weekly or monthly median, mean and variance (see the electronic supplementary material).

We further identified, assembled and processed 25 raster and vector datasets into a set of RS covariates for the whole of Bangladesh at a 1 km spatial resolution. These data were obtained from existing sources and produced *ad hoc* for this study to include environmental and physical metrics likely to be associated with human welfare [31,33,48–50] such as vegetation indices, night-time lights, climatic conditions, and distance to roads or major urban areas. A full summary of assembled covariates is provided in the electronic supplementary material.

### 2.4. Covariate selection

Prior to statistical analyses, all CDR and RS covariate data were log transformed for normality. Bivariate Pearson's correlations were computed for each pair of covariates to assess multicollinearity, and for high correlations ($r > 0.70$), we eliminated covariates that were less generalizable outside Bangladesh. For example, population data are widely available (e.g. www.worldpop.org.uk/) but births data may not be; similarly, volumes of calls could be computed and compared across countries, but charges may be country-specific.

To identify the set of predictors most suitable for modelling the WI, PPI, and income data, we employed a model selection stage as is common in statistical modelling [51]. For this we used non-spatial generalized linear models (glms), implemented via the R *glmulti* package [52,53], to build every possible non-redundant model for every combination of covariates. Models were built on a randomly selected 80% of the data to guard against overfitting. Models were chosen using Akaike's information criterion (AIC), which ranks models based on goodness of fit and complexity, while penalizing deviance [52]. A full IC-based approach such as this allows for multi-model inference. Where multiple glms had near-identical AIC values, we selected the model with the fewest number of covariates. For the CDR data only, we used forward and backward stepwise selection ($p = 0.05$) prior to model selection to reduce the initial CDR inputs from 150 to 30 or less. The covariate selection process was completed for all three poverty measures for national, urban and rural strata, and using RS-only, CDR-only and CDR–RS datasets (27 resulting models). This allowed us to explore differences in factors related to urban and rural poverty, as well as to explicitly compare the ability of RS-only, CDR-only and CDR–RS datasets to predict poverty measures. The resulting models were then used in the hierarchical Bayesian geostatistical approach (see the electronic supplementary material, tables S2a–c).

### 2.5. Prediction mapping

Using the models selected by the previous step, we employed hierarchical Bayesian geostatistical models (BGMs) to predict the three poverty metrics at unsampled locations across the population. We chose BGMs as they offer several advantages for addressing the limitations and constraints associated with modelling geolocated survey data. These include straightforwardly imputing missing data, allowing for the specification of prior distributions in model parameters and spatial covariance, and estimating uncertainty in the predictions as a distribution around each estimate [54,55].

Additionally, we needed to account for spatial autocorrelation in the data as they are aligned to the tower locations, which are clustered across varying spatial scales (described in §2.1 and figure 1). BGMs can achieve this through incorporating a spatially varying random effect. Here, the Voronoi polygons themselves form the neighbourhood structure for this spatial random effect, and neighbours are defined within a scaled precision matrix [56]. The matrix represents the spatially explicit processes that may affect poverty estimates. It is passed through a graph function in the model which assumes the neighbour relations are connected [57], that is all adjacent polygons share a boundary. This function accounts for the spatial covariance in the data by allowing observations to have decreasing effects on predictions that are further away.

Here, all BGMs were implemented using integrated nested Laplace approximations (INLA) [58], which uses an approximation for inference and avoids the computational demands, convergence issues and mixing problems sometimes encountered by MCMC algorithms [59]. The model is fit using R-INLA, with the Besag model for spatial effects specified inside the function [60,61]. In the Besag model, Gaussian Markov random fields (GMRFs) are used as priors to model spatial dependency structures and unobserved effects. GMRFs penalize local deviation from a constant level based on the precision parameter $\tau$, where the hyperpriors are loggamma distributed [56]. The hyperprior distribution governs the smoothness of the field used to estimate spatial autocorrelation [56]. The spatial random vector $\mathbf{x} = (x_1, \ldots, x_n)$ is thus defined as

$$x_i | x_j, i \neq j, \tau \sim \mathcal{N}\left(\frac{1}{n_i}\sum_{i \sim j} x_j, \frac{1}{n_i \tau}\right),$$

where $n_i$ is the number of neighbours of node $i$, $i \sim j$ indicates that the two nodes $i$ and $j$ are neighbours. The precision parameter $\tau$ is represented as

$$\theta_1 = \log \tau,$$

where the prior is defined on $\theta_1$ [60]. The geostatistical models defined for the WI, PPI and income data were applied to produce predictions of the each poverty metric for each Voronoi polygon as a posterior distribution with complete modelled uncertainty around estimates. The posterior mean and standard deviation for each polygon were then used to generate prediction maps with associated uncertainty (figure 2 and electronic supplementary material, figures S2–S6). Model performance was based on out-of-sample validation statistics calculated on a 20% test subset of data. Pearson product-moment correlation coefficient ($r$) (or

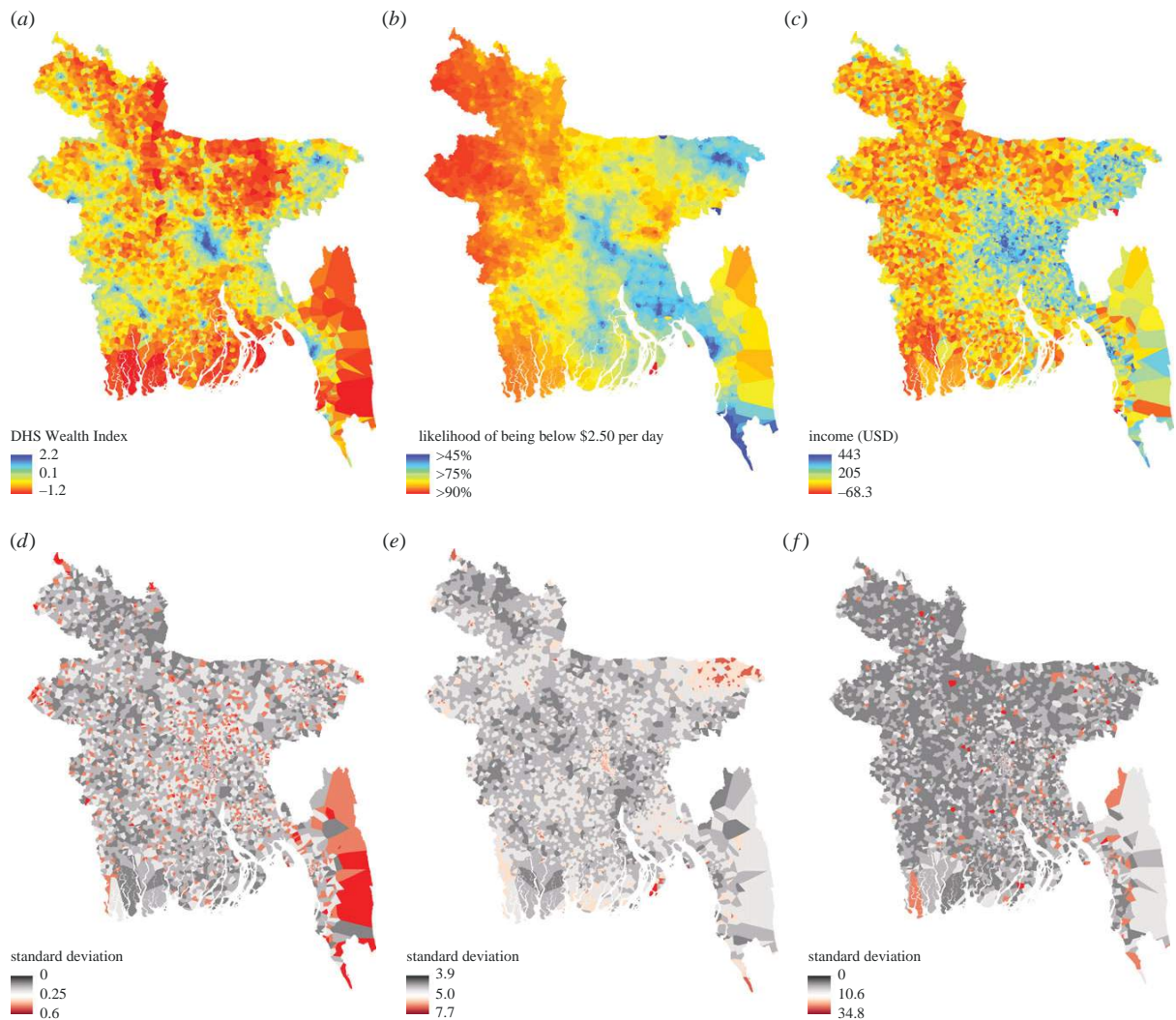

**Figure 2.** National level prediction maps for mean WI (*a*) with uncertainty (*d*); mean probability of households being below $2.50/day (*b*) with uncertainty (*e*); and mean USD income (*c*) with uncertainty (*f*). Maps were generated using call detail record features, remote sensing data and Bayesian geostatistical models. The maps show the posterior mean and standard deviation from CDR–RS models for the WI and income data (*a*,*c*), and the RS model for the PPI (*b*). Red indicates poorer areas in prediction maps, and higher error in uncertainty maps.

Spearman's rho ($\rho$) for $n < 100$), root-mean-square-error (RMSE), mean absolute error (MAE) and the coefficient of determination ($r^2$) were calculated for all BGMs. Finally, because glms do not incorporate prior information for model parameters, we ran each model through INLA while excluding the random spatial effect to obtain non-spatial Bayesian estimates and compare model fit and performance due to the explicit spatial process.

## 3. Results

We find models employing a combination of CDR and RS data generally provide an advantage over models based on either data source alone. However, RS-only and some CDR-only models performed nearly as well (table 1). While the combined CDR–RS model performed well in both urban ($r^2 = 0.78$) and rural ($r^2 = 0.66$) areas, and at the national level ($r^2 = 0.76$), the performance of RS-only and CDR-only models was more context-dependent. For example, PPI and income models did not improve predictions in urban areas, but in rural areas the RS-only models performed nearly as well for both indicators. The fine spatial granularity of the resultant poverty estimates can be shown in figure 2, which shows the predicted distribution of poverty for all three measures. Spatially, the models exhibit higher uncertainty where fewer data are available, such as the peninsular areas surrounding Chittagong in the southeast where mobile towers are sparse. We also find that explicitly modelling the spatial covariance in the data was critically important. This resulted in improved predictions, lower error and better measures of fit based on cross-validation and the deviance information criteria (DIC), a hierarchical modelling generalization of the AIC [62] (electronic supplementary material, tables S3 and S4).

Separating estimation by urban and rural regions further highlights the importance of different data in different contexts (electronic supplementary material, tables S2*a*–*c*). Night-time lights, transport time to the closest urban settlement, and elevation were important nationally and in rural models; climate variables were also important in rural areas. Distances to roads and waterways were significant in urban and rural strata. In general, the addition of CDR data did not change the selection of RS covariates at any level. Top-up features derived from recharge amounts and tower averages were significant in every model, affirming their importance in poverty work. People predicted to be poorer top-up their phones more frequently in small amounts. Per cent nocturnal calls, and count and duration of SMS traffic were significant nationally. Mobility and social network features were important in all three strata. In urban areas, SMS traffic was important, whereas multimedia messaging and video attributes were key in rural areas.

**Table 1.** Cross-validation statistics based on a random 20% test subset of data for all Bayesian geostatistical models.

| poverty metric | model | $r^2$ | RMSE |
|---|---|---|---|
| whole country | | | |
| DHS WI | CDR–RS | 0.76 | 0.394 |
| | CDR | 0.64 | 0.483 |
| | RS | 0.74 | 0.413 |
| PPI | CDR–RS | 0.25 | 57.907 |
| | CDR | 0.23 | 58.562 |
| | RS | 0.32 | 57.439 |
| income | CDR–RS | 0.27 | 105.465 |
| | CDR | 0.24 | 107.155 |
| | RS | 0.22 | 108.682 |
| urban | | | |
| DHS WI | CDR–RS | 0.78 | 0.424 |
| | CDR | 0.70 | 0.552 |
| | RS | 0.71 | 0.433 |
| PPI | CDR–RS | 0.00 | 60.128 |
| | CDR | 0.03 | 60.935 |
| | RS | 0.00 | 60.384 |
| income | CDR–RS | 0.15 | 168.452 |
| | CDR | 0.15 | 172.738 |
| | RS | 0.05 | 176.705 |
| rural | | | |
| DHS WI | CDR–RS | 0.66 | 0.402 |
| | CDR | 0.50 | 0.483 |
| | RS | 0.62 | 0.427 |
| PPI | CDR–RS | 0.18 | 57.397 |
| | CDR | 0.17 | 57.991 |
| | RS | 0.21 | 57.162 |
| income | CDR–RS | 0.14 | 81.979 |
| | CDR | 0.13 | 82.773 |
| | RS | 0.23 | 76.527 |

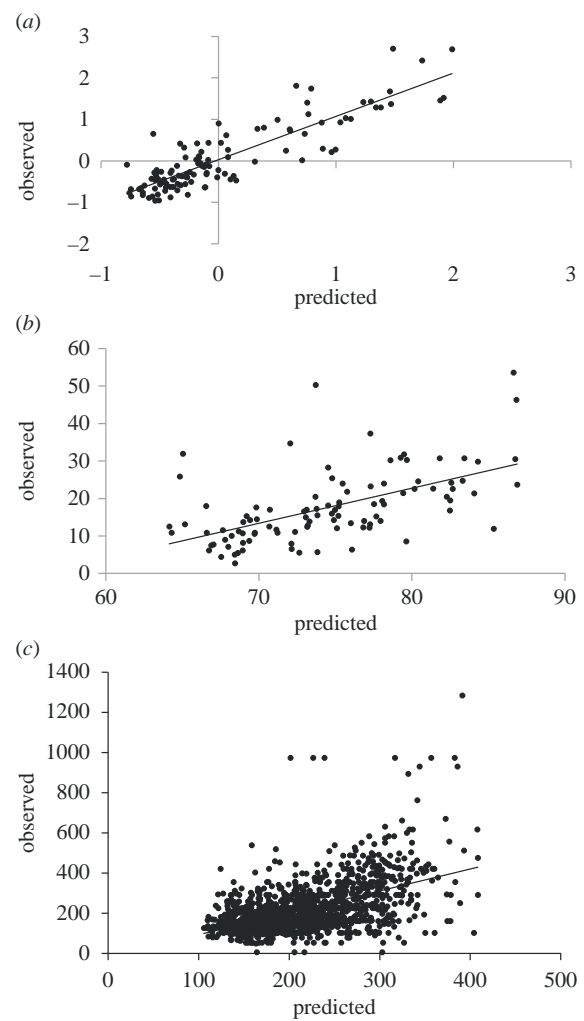

**Figure 3.** Out-of-sample observed versus predicted values for (a) DHS WI using mobile phone and remote sensing data: $r^2 = 0.76$, $n = 117$, $p < 0.001$, RMSE = 0.394; (b) progress out of Poverty Index using remote sensing data: $r^2 = 0.32$, $n = 100$, $p < 0.001$, RMSE = 57.439; and (c) income using mobile phone and remote sensing data: $r^2 = 0.27$, $n = 1384$, $p < 0.001$, RMSE = 105.465.

Models were most successful at reconstructing the WI to model poverty ($r^2 = 0.76$); consumption-based and income-based poverty proved more elusive. WI models have better fit, lower error and higher explained variance based on out-of-sample validation (figure 3). Combined CDR–RS data produced the best WI models and lowest error ($r^{2\,(\text{CDR–RS})} = 0.76$, $r^{2\,(\text{RS})} = 0.74$, $r^{2\,(\text{CDR})} = 0.64$; RMSE$^{(\text{CDR–RS})} = 0.394$, RMSE$^{(\text{RS})} = 0.413$, RMSE$^{(\text{CDR})} = 0.483$). However, for the PPI models, the best model predicting the probability of falling below \$2.50/day was the RS-only model (figure 2b,e, $r^{2\,(\text{RS})} = 0.32$; RMSE$^{(\text{RS})} = 57.439$). The model discerns many urban areas but also predicts areas with very low poverty likelihood and high uncertainty outside urban areas, especially around Sylhet in the northeast. Income predictions (figure 2c,f) show greater variation across the country, and the best national model was for combined CDR–RS data ($r^{2\,(\text{CDR–RS})} = 0.27$, RMSE$^{(\text{CDR–RS})} = 105.465$).

The resulting predictions line up well with existing SAE estimates for Bangladesh, and with high-resolution maps of slum areas in Dhaka. The urban CDR–RS model has the highest explained variance for any model ($r^{2\,(\text{CDR–RS\_urb})} = 0.78$) and the urban CDR-only model outperforms the national CDR-only model ($r^{2\,(\text{CDR\_urb})} = 0.70$). Precision and accuracy are slightly lower, but the improved correlation highlights the advantage of using CDRs within a diverse urban population. To explore this further, we compared our WI predictions against a spatially explicit dataset of slum areas in Dhaka [63] (figure 4). We find the mean predicted WI of slum and non-slum areas to be significantly different, $t_{615} = -17.2$, $p < 0.001$, predicting slum areas to be poorer than non-slum areas.

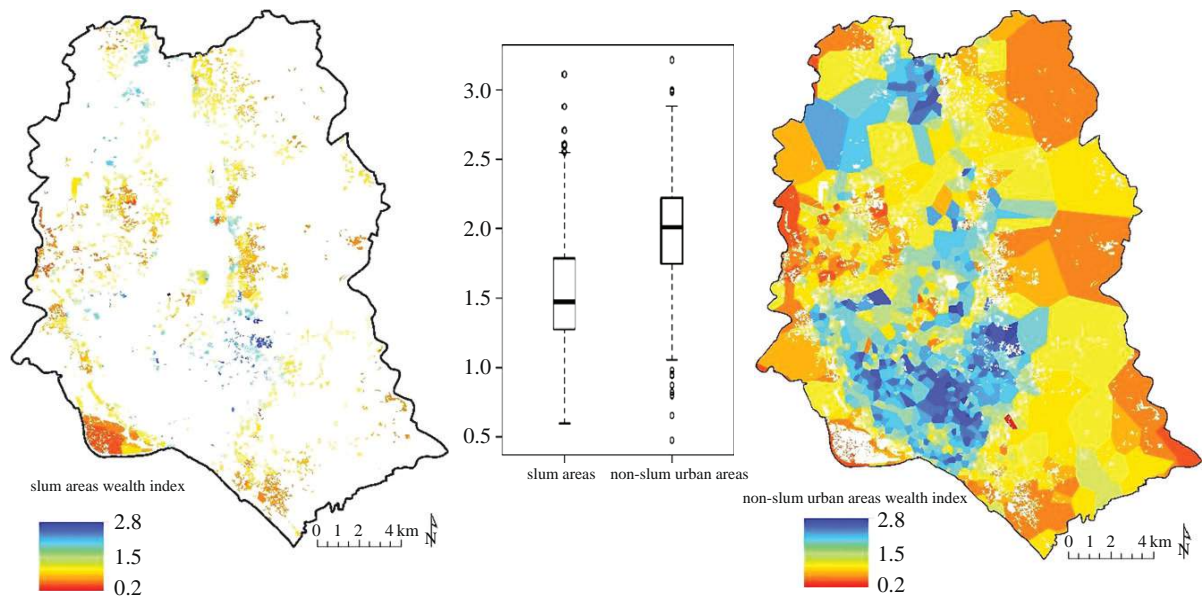

**Figure 4.** Comparison of predicted mean DHS WI values between slum and non-slum areas in Dhaka as delineated by Gruebner et al. [63] $t_{615} = -17.2$, $p < 0.001$. The 95% confidence interval using Student's t-distribution with 615 degrees of freedom is $(-0.48, -0.38)$.

To compare our method to previous poverty estimates at administrative level 3 (upazila), we used the same methodology at the lower spatial resolution, using the upazila boundaries to form the random spatial effect in the model, and covariates from the best national level model for each poverty measure. We find strong correlations ($r = -0.91$ and $-0.86$ for the WI; 0.99 and 0.97 for the PPI; and $-0.96$ and $-0.94$ for income, respectively, $p < 0.001$ for all models) between our upazila predictions and earlier estimates of poverty derived from SAE techniques based on data from the 2010 Household Income and Expenditure (HIES) survey and 2011 census [64] (figure 5). The r-values reported for WI and income are negative at administrative level 3 because as the proportion of people below the poverty line as estimated by Ahmed et al. decreases, the WI value and income in USD of the sampled population increases. That is, people who are wealthier as estimated by the WI and income data are also less likely to live below the poverty line according to earlier estimates. The geostatistical method presented here thus accurately maps heterogeneities at small spatial scales while correlating well with earlier coarser estimates. All remaining WI, PPI and income prediction maps are provided in the electronic supplementary material.

## 4. Discussion

This work represents the first attempt to build predictive maps of poverty using a combination of CDR and RS data. The results demonstrate that CDR-only and RS-only models perform comparably in their ability to map poverty indicators, and that integrating these data sources provides improvement in predictive power and lower error. These results are promising as the CDR data here produce accurate, high-resolution estimates in urban areas not possible using RS data alone. As such, CDRs potentially allow for estimation of wealth at much finer granularity—including the neighbourhood or even the household or individual—than the current generation of RS technologies [36]. While CDRs are proprietary data, they are increasingly used in research, and have formed the basis for hundreds of published articles over the past few years [65]. They also provide significant advantages in temporal granularity: CDRs update in real-time versus RS data, which update far less frequently. Although in this study we have not used dynamic validation data, it is a clear future application for CDRs in real-time to better comprehend the dynamic nature of poverty.

The higher accuracy of predictions for the asset-based WI over other poverty metrics is presumably due to several factors. The predictive power for assets has been shown to be higher than for consumption [35] in addition to the aforementioned issues of survey question wording and response options [20,23]. Further, income and consumption can vary hugely by day, week, and can be related to changes in household size, job loss or gain, piecework or harvest outcomes. Assets and housing characteristics are generally considered more stable [20–22]. For the datasets used in this study, WI data are based on clusters of households, and this sampling strategy provides more robust estimates and less variability than the individually based PPI and income data. Greater success in predicting the WI is also presumably due to the WI measuring a wider range of living standard across the population. That is, the full range of distribution from poorest to wealthiest in the population is represented in these data. Alternatively, by considering a streamlined 10 questions, the PPI is meant to identify the poorest individuals in a population. Similarly, in the income data, there were very few respondents in higher income categories.

The higher error associated with CDR-only models is not surprising considering the noise inherent in these data. CDR features are derived from daily and weekly measurements aggregated over short temporal intervals, while RS covariates are generally comprised of long-term averages or comparatively less dynamic measures of location and access such as roads or proximity to urban centres. Bearing this in mind,

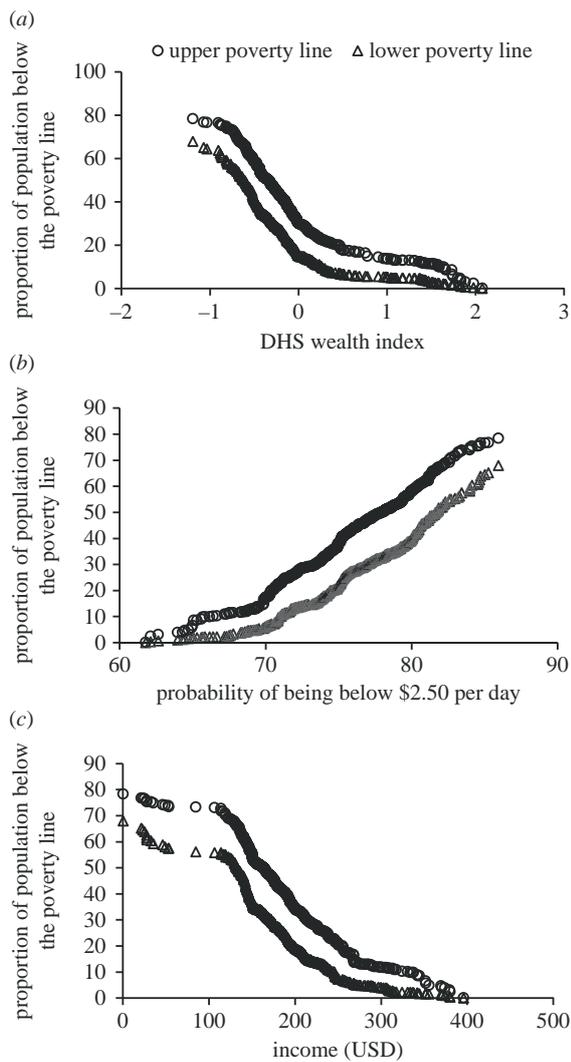

**Figure 5.** Comparison of the proportion of people falling below upper (circles) and lower (triangles) poverty lines estimated by Ahmad *et al.* [64] and (*a*) predicted mean WI using mobile phone and remote sensing data, (*b*) predicted probability of being below $2.50 per day using remote sensing data and (*c*) predicted income using mobile phone and remote sensing data. All models were predicted at the upazila scale (Admin unit 3). Pearson's $r$ correlations: $-0.91$ and $-0.86$ for the WI; 0.99 and 0.97 for the PPI; and $-0.96$ and $-0.94$ for income, respectively ($p < 0.001$ for all models).

we find CDR data useful for estimating poverty in the absence of ancillary datasets.

Our findings provide further support for correlations between socio-economic measures and night-time light intensity [36,48,49], access to roads and cities [50,66], entropy of contacts [37,40] and mobility features [39]. The universal coverage of cell towers across Bangladesh made it possible to predict poverty at high-resolution in both urban and rural areas. Within urban areas, the high correlation with maps of slums in Dhaka suggests we are capturing the poorest populations. Even if the poorest populations are not generating call data [36], and thus not included in the CDRs, we still see a clear difference in WI predictions between slum and non-slum areas using tower level CDR aggregates. This finding extends recent work which predicted wealth and poverty at the district level, but were unable to verify predictions at finer scales [36].

Using CDRs and RS data within BGMs to produce accurate, high-resolution poverty maps in LMICs offers a way to complement census-based methods and provide more regular updates. Regularly updated poverty estimates are necessary to enable subnational monitoring of the SDGs during intercensal years and are critical to ensure mobilization of resources to end poverty in all its dimensions as set out in SDG 1. Poverty estimates are time sensitive and become obsolete when factors such as migration rates, infrastructure, and market integration change [67]. Furthermore, the methods presented here offer a workaround to estimating poverty with household survey data, which can be time consuming and expensive to obtain.

To end poverty in all its dimensions, it is likely that methods that exploit information from, and correlations between, many different data sources will provide the greatest benefit in understanding the distribution of human living conditions. To leverage data from differing sample sizes, temporal and spatial scales, BGMs provide such a rigorous framework. This study further provides an example of how aggregated CDR data can be processed in such a way that detailed maps can be created without revealing sensitive user and commercial information. As insights from CDRs and other remote sensing data become more widely available, analysing these data at regular intervals could allow for dynamic poverty mapping and provide the means for operationally monitoring poverty. The combination of spatial detail and frequent, repeated measurements may distinguish the transitorily poor from the chronically poor, and allow for monitoring economic shocks [68]. This offers the potential for a fuller characterization of the spatial distribution of poverty and provides the foundation for evidence-based strategies to eradicate poverty. Researchers would do well to use the additional information and granularity afforded by CDR data with matched individual-based consumption data to further infer novel and useful information from mobile data.

Authors' contributions. J.E.S. was responsible for research design, production of RS covariates, data cleaning and processing, statistical analyses, interpretation, drafting and production of the final manuscript. C.P. was responsible for survey data management, cleaning and processing, and interpretation and drafting of the final manuscript. P.S., J.Bj. and K.E. were responsible for CDR data management, cleaning, and production of CDR data, and interpretation and drafting of the final manuscript. J.Bl. was responsible for interpretation, drafting, and production of the final manuscript. V.A., T.B., Y.M., X.L. and E.W. were responsible for interpretation and production of the final manuscript. A.I. and K.N.H. were responsible for income survey data collection, management and processing. A.J.T. and L.B. were responsible for overall scientific management, interpretation and production of the final manuscript. All authors gave final approval for publication.

Competing interests. We declare we have no competing interests.

Funding. J.E.S., E.W. and J.Bl. are supported by the Bill & Melinda Gates Foundation (OPP1106936). C.P. is supported by the Bill & Melinda Gates Foundation (OPP1106427). X.L. acknowledges the Natural Science Foundation of China under grant nos. 71301165 and 71522014. A.J.T. is supported by funding from NIH/NIAID (U19AI089674), the Bill & Melinda Gates Foundation (OPP1106427, 1032350, OPP1134076 and OPP1094793), the Clinton Health Access Initiative, National Institutes of Health, and a Wellcome Trust Sustaining Health grant (106866/Z/15/Z). L.B. acknowledges the Swedish Research Council, grant no. D0313701. This work forms part of the WorldPop Project (www.worldpop.org.uk) and Flowminder Foundation (www.flowminder.org).

Acknowledgements. The authors gratefully acknowledge Dr Elisabeth zu Erbach-Schoenberg, Dr Nick Ruktanonchai and Dr Alessandro Sorichetta for useful discussions. We also wish to thank two anonymous reviewers for providing useful comments. The funders had no role in study design, data collection and analysis, decision to publish or preparation of the manuscript.

# Mapping poverty using mobile phone and satellite data

J.E. Steele, P. Sundsoy, C. Pezzulo, V. Alegana, T. Bird, J. Blumenstock, J. Bjelland, K. Engo-Monsen, YA de Montjoye, A. Iqbal, K. Hadiuzzaman, X. Lu, E. Wetter, A.J. Tatem, and L. Bengtsson

## Supplementary Information (SI)

**Table of Contents**

Mapping poverty using mobile phone and satellite data
*J.E. Steele, P. Sundsoy, C. Pezzulo, V. Alegana, T. Bird, J. Blumenstock, J. Bjelland, K. Engo-Monsen, YA de Montjoye, A. Iqbal, K. Hadiuzzaman, X. Lu, E. Wetter, A.J. Tatem, and L. Bengtsson*

# A. Input data

## A.1. Geolocated survey data

The growing number of georeferenced household survey data from low and middle-income countries allows us to explore poverty metrics and comparisons between them while explicitly considering their geographic distribution. In Bangladesh, we utilised three geographically referenced datasets representing asset, consumption, and income-based measures of wellbeing (Figure S1).

### A.1.1. Demographic and Health Survey Wealth Index

The Demographic and Health Surveys (DHS) were designed primarily to collect household data on marriage, fertility, family planning, reproductive and child health, and HIV/AIDS in almost all lower income countries[1]. Through the assembling of indicators correlated with a household's economic status (e.g. ownership of television, telephone, radio as well as variables describing type of floor and ceiling material and other facilities), a wealth index is calculated for each country at each time[2] based on the idea that the possession of assets and access to services and amenities are related to the relative economic position of the household in the country[3]. By its construction, the wealth index is a relative measure of wealth within each survey; however, a new methodology has been developed in order to make it comparable across countries and through time[4]. Moreover, recent adjustments have been made to the methods of constructing the wealth index to overcome criticism that the original score was not adequately capturing the differences between urban and rural poverty or identifying the poorest of the poor[3].

The wealth index is constructed using a principal component analysis (PCA), which includes a long list of assets owned by households as well as other indicators. (The complete list of indicators included in the PCAs for each survey, as well as PCA analysis and results can be found at http://dhsprogram.com/topics/wealth-index/Wealth-Index-Construction.cfm). The first factor from the PCA, capturing the largest percentage of the variance within the dataset, is derived adjusting for urban and rural strata[3,5]. In practice, a national index and two area-specific indexes representing urban and rural strata are individually constructed using sets of assets/services specific to each in order to better capture differences between urban and rural areas, and compare the wealth index between them[3]. Subsequently, applying regression techniques described in Rutstein[3] and Rutstein[6], the three indexes are combined into a single wealth distribution and a composite national index is derived. This method ensures comparability between urban and rural areas.

Here we used the 2011 Bangladesh DHS[7] (Figure S1*A*), a nationally representative survey based on a two-stage stratified sample of households, where 600 enumeration areas (EA or cluster) were first selected with probability proportional to the EA size, (207 clusters in urban areas and 393 in rural areas). This first stage of selection provided a listing of households for the second stage, where a systematic sample of 30 households on average was selected per cluster, to create statistically reliable estimates of key demographic and health variables[7,8]. In recent DHS surveys where HIV/AIDS data are not collected, geolocations for each cluster are available. The survey cluster coordinates represent an estimated centre of the cluster and are collected in the field through GPS receivers. To maintain respondents' confidentiality, GPS positions for all clusters are randomly displaced by a maximum of five kilometres for rural clusters and a maximum of two kilometres for urban clusters[9–11].

Mapping poverty using mobile phone and satellite data
*J.E. Steele, P. Sundsoy, C. Pezzulo, V. Alegana, T. Bird, J. Blumenstock, J. Bjelland, K. Engo-Monsen, YA de Montjoye, A. Iqbal, K. Hadiuzzaman, X. Lu, E. Wetter, A.J. Tatem, and L. Bengtsson*

### A.1.2. Progress out of Poverty Index

The Progress out of Poverty Index (PPI) (Figure S1*B*) was designed to be easily collected, simple and cost-effective to implement and verify[12], while applying a rigorous methodology through selecting assets based on their statistical relationship with poverty[12,13]. In the case of Bangladesh, an easy-to-use poverty scorecard[13] of 10 questions was created in March 2013, based on data from the Bangladesh 2010 Household Income and Expenditure Survey (HIES). The questions selected are aggregated into a score highly correlated with poverty status as measured by the HIES. The scores included in the scorecard are then translated into likelihoods that the household has per-capita expenditure above or below a given poverty line[13].

A nationally representative survey of all adults in Bangladesh was undertaken by InterMedia Financial Inclusion Insight Project (www.finclusion.org) in 2014 (wave 2), where 6,000 Bangladeshi individuals aged 15 and above were interviewed[14] and geolocations for each individual were included in the survey. In Bangladesh, InterMedia adopted a stratified sample strategy, whereby divisions and subdivisions were first identified and interviews within each subdivision were distributed in proportion to population size. In order to select the individuals to interview, households were first randomly selected using electoral rolls to randomly assign starting points in each selected subdivision. After having identified the starting point, subsequent households were selected using the right-hand rule, and the Kish Grid method was applied to select an individual respondent from each household[14,15].

### A.1.3. Market research household surveys and income data

Two sequential large-scale market research household surveys were run by Telenor through its subsidiary, Grameenphone (GP), during 2 time periods between November and December 2013 (N=82,834, of which 55.3% GP subscribers) and February and March 2014 (N=87,509, of which 54.5% GP subscribers) (Figure S1*C*). The country was stratified in 226 sales territories by the phone company, and for every territory, an equal number of unions (in rural areas) and wards (in urban areas) were randomly selected. Four hundred households were surveyed in each territory, where a household was defined as a group of people sharing food from the same chula (fire/gas burner) or living under the same roof. Systematic sampling was then undertaken to select households by selecting every fourth household, starting from the selection of a random geographic point and direction within each ward or union. In the case of more than one household present in the complex or building, the fourth household was selected. In cases of non-response, the next household was then selected. Non-response rate was approximately 10% of households. Respondents within the household were selected via the Kish grid method[15] among those who were eligible. Eligibility was defined as individuals with their own phone, between 15 and 65 years of age. If a phone was shared between family members, usually the male head of household was interviewed. When the selected person was not home, the surveyors returned multiple times to try to reach the selected person. A very low non-response rate (less than 1-2%) was detected among respondents. The surveys were undertaken during working hours. To avoid that too many housewives were interviewed, given that men are more likely away for work, a ceiling on the number of housewives who could participate was also established. Sampling weights were applied to ensure national representativeness and correct for population sizes in urban and rural areas. Data quality control mechanisms were implemented and undertaken by the company; however, some sources of error were detected in matching household locations to phone number (approximately 20% of the cases).

A.2. Mobile phone call detail records (CDRs)

For the household income survey respondents described above, we collected 3-months of mobile phone metadata by subscriber consent. These metadata included call detail records (CDR) and top-up information, which were further processed into features. For each survey participant, 150 features from seven different feature families were constructed (Table S1). Household income was then linked to these metadata, resulting in three months of phone usage, matched with household income for each survey respondent. To preserve user anonymity, the local operator removes all personally identifying information from the data before analysis.

To be able to map poverty in other countries we focused on features that are easily reproducible, and easy to implement by local data warehouses. Most mobile operators generate similar features. CDR features range from metrics such as basic phone usage, top-up pattern, and social network to metrics of user mobility and handset usage. They include various parameters of the corresponding distributions such as weekly or monthly median, mean, and variance. In addition, we received pre-aggregated datasets of tower-level activity from 48,190,926 subscriber SIMs over a 4-month period. This includes monthly number of subscribers per home cell, where home cell corresponds to most frequent tower. These per-user features are not directly used, but further aggregated to the Voronoi polygons, and the aggregate features are used in covariate selection, model fitting, and prediction.

At the time of data acquisition, the mobile phone operator had an approximate 42% market share, and was the largest provider of mobile telecommunication services in Bangladesh. Multi-SIM activity is common in Bangladesh, but we believe that this should not create a systematic bias in poverty estimates because the geographic coverage of the operator is so extensive. In order to comply with national laws and regulations of Bangladesh, and the privacy policy of the Telenor group, the following measures were implemented in order to preserve the privacy rights of Grameenphone customers:

> 1) All customers are de-identified and only Telenor/Grameenphone employees have had access to any detailed CDR-/top-up data;
> 2) The processing of detailed CDR/top-up data resulted in aggregations of the data on a tower-level granularity; the tower-level aggregation makes re-identification impossible.

Hence, the resulting aggregated dataset is truly anonymous and involves no personal data.

Compared with other countries of comparable income levels, Bangladesh has a high mobile phone penetration, which includes rural areas. Fifty percent of the population above the age of fifteen has a mobile subscription[16]. The proportion of households with at least one mobile phone is increasing rapidly; between 2011 and 2014, household ownership across the whole of Bangladesh rose from 78% to 89%, with much of that growth concentrated among rural households[17]. The CDR data used in this study are available upon request for the replication of results only by contacting the corresponding author.

A.3. Remote Sensing-GIS covariates

Ancillary data layers used as remote sensing-GIS (hereafter RS) covariates were identified, assembled, and processed for the whole of Bangladesh at a 1-km spatial resolution. These data are described in Table S1 and include 25 raster and vector datasets obtained from existing sources or produced ad hoc for this study to include environmental and physical metrics likely to be associated with human welfare[18–22]. These data differed in spatial and temporal resolution, type, accuracy, and coverage. In order to align all data for model fitting and prediction, the following steps were taken:

1) Bangladesh was rasterized at a resolution of 30-arcsec (0.00833333 degree, corresponding to approximately 1-km at the equator);
2) Vector datasets were rasterized at a resolution of 30-arcsec;
3) When necessary, raster datasets were resampled to a resolution of 30-arcsec using an interpolation technique appropriate for the resolution and type of the original dataset;
4) All datasets were spatially aligned to make every pixel representing the same location coincident and match the rasterized study area.

Furthermore, for ad hoc datasets such as distance to roads and waterways, we used a customized Azimuthal Equidistant projection centred in the middle of the study area and clipped to a buffer extending 100 metres beyond its boundary to project the input data. This buffered area was rasterized to a resolution of approximately 927 metres, corresponding to 30-arcsec at the centre of the study area where distortion is smallest. Euclidean distance was calculated for each distance-to covariate within the customized projection. The resultant layers were then projected back to GCS WGS84, and made coincident with the rasterized study area. All datasets representing categorical variables (e.g. protected areas, global urban extent, etc.) were projected, rasterized, and/or resampled to 1-km resolution, spatially aligned to the rasterized study area, and converted into binary covariates, representing the presence or absence of a given feature. This resulted in twenty-five 1-km raster datasets, which were used to extract the mean, mode, or sum of each covariate for each Voronoi polygon, dependent on the type of dataset. These values were used for covariate selection, model fitting, and prediction.

A.3.1 GPS data displacement

In addition to the aforementioned processing, additional steps were undertaken to appropriately account for the displacement inherent in DHS data. When these data are collected, the latitude and longitude of the centre of each DHS cluster (representing numerous households) is collected in the field with a GPS receiver. To maintain respondents' confidentiality, GPS latitude/longitude positions for all DHS clusters are randomly displaced by a maximum of five kilometres for rural clusters and two kilometres for urban clusters. The displacement is restricted so the points stay within the country, within the DHS survey region, and within the second administrative level[9–11].

In order to account for the displacement in our analyses, we created buffers around each cluster centroid of 2 km and 5 km for urban and rural clusters, respectively, and subsequently extracted the RS covariate data for each buffer zone. For continuous covariates, the minimum, maximum, and mean values were calculated and extracted. For categorical covariates, the modal value was calculated and extracted.

# B. Statistical analyses and prediction mapping

## B.1. Covariate selection via generalized linear models

Stratifying models into urban and rural components produced the best fit models as measured by AIC. Top-up data produced the most important CDR feature family for all poverty measures and models. Within this feature family, significant covariates included recharge amounts and frequencies per tower, spending speeds and time between refills, and fractions of the lowest and highest available top-up amounts. Advanced phone usage was also an important CDR feature family, especially for PPI and income models. Sum, revenue, count, and volume of ingoing and outgoing multimedia messaging, Internet usage, and videos were prominent. Basic phone usage covariates measuring incoming and outgoing text counts were important for every model save for rural WI models. Mobility covariates including number and entropy of places and radius of gyration were also significant features in all three strata and poverty measures, as were social network features such as number and entropy of contacts.

Nighttime lights and covariates representing access - especially transport time to closest urban settlement and distance to roads - were the most important RS covariates for all three poverty measures and strata. Vegetation productivity, as measured by the Enhanced Vegetation Index (EVI), and elevation were also prominent RS features in all three strata, whereas climate variables featured prominently in rural models.

## B.2. Prediction mapping via Bayesian geostatistical models

Using the models selected and described in Tables S2*A-C*, we employed hierarchical Bayesian geostatistical models (BGMs) for prediction as described in our manuscript. All prediction maps not highlighted in our manuscript can be found in Figures S2-S6. Model performance was based on out-of-sample validation statistics calculated on a 20% test subset of poverty data input points (see Table 1 in manuscript). The performance of models built with CDR-only or RS-only data varied based on poverty measure and strata. RS-only models were more successful at predicting the WI for all three strata ($r^2$= 0.74, 0.71, 0.72 for national, urban, and rural models), as compared to CDR-only models. However, the CDR-only models performed nearly as well ($r^2$= 0.64, 0.70, 0.50 for national, urban, and rural models), and all urban WI models including CDRs outperformed national level models. The urban CDR-RS model exhibits the highest explained variance for any model ($r^2$=0.78), and the urban CDR-only model outperforms the national CDR-only model ($r^2$=0.70 versus $r^2$=0.64, respectively). For PPI and income measures of poverty, CDR data produced the best models in urban areas, whereas RS data produced the best models in rural areas. This highlights the compatibility of these two datasets for predicting different measures of poverty at different scales, as the best estimates and lowest error corresponded to the data with fine-scale spatial heterogeneity (CDRs in urban areas; RS data in rural areas). To that end, national poverty models generally performed best when utilising both CDRs and RS data.

To compare full model performance against a spatial interpolation model, we modelled the training data for all three poverty indicators using only the spatial random effect in the INLA model (see section 2.5 in manuscript). These results are shown in Table S3. We compared out-of-sample $r^2$ and RMSE values against results from the full models (see Table 1 in manuscript). The results show a spatial pattern in the WI data as the model built with only a spatial random effect yields an $r^2$=0.49, RMSE=0.578. When compared to the full model, the addition of covariate data increases the $r^2$ to 0.76, and the RMSE decreases to 0.394. Similarly for income, the data do show a slight spatial pattern, but the addition of covariate data to the model increases the predictive power and decreases the error. For the PPI, the covariates do not show a strong influence in the modelling results, and the model was driven by the spatial process, which suggests there's an underlying spatial covariate that we're not capturing in the model that could explain the data.

Model fit based on the spatial effect can also be considered using DIC, a hierarchical modelling generalization of the AIC and BIC, which can be useful in Bayesian modelling comparison. The BIC allows for comparing models using criterion based on the trade-off between the fit of the data to the model and the corresponding complexity of the model. Models with smaller DIC values are preferred over models with larger DIC values as the measure favours better fit and fewer parameters[23]. These results are shown in Table S4. For nearly every model with CDR data, DIC is greatly improved by accounting for the spatial covariance in the data structure. However, the income models see slight or no improvement from including the random spatial effect, likely due to the fact that they include and are thus penalised for many covariates.

# SI Tables and Figures

Table S1. Summary information for remote sensing-GIS and mobile phone call detail record datasets used for covariate selection and Bayesian geostatistical poverty mapping.

| Category | Description | Source | Resolution (Degrees) | Year |
|---|---|---|---|---|
| | | **RS-GIS** | | |
| Accessibility | Accessibility to populated places with more than 50k people | European Commission Joint Research Centre (http://forobs.jrc.ec.europa.eu/products/gam/) | 0.0083333 | 2000 |
| Population | Population count [per pixel] | WorldPop (http://www.worldpop.org.uk/) | 0.0008333 | 2010 |
| Population | Population count [per pixel] | CIESIN - Global Rural Urban Mapping Project (http://sedac.ciesin.columbia.edu/data/collection/grump-v1/sets/browse) | 0.0083333 | 2000 |
| Population | Population density [per sqkm] | CIESIN - Global Rural Urban Mapping Project (http://sedac.ciesin.columbia.edu/data/collection/grump-v1/sets/browse) | 0.0083333 | 2000 |
| Climate | Mean Aridity Index | CGIAR-CSI (http://www.cgiar-csi.org/data) | 0.0083333 | 1950-2000 |
| Climate | Average annual Potential Evapotranspiration [mm] | CGIAR-CSI (http://www.cgiar-csi.org/data) | 0.0083333 | 1950-2000 |
| Night-time lights | VIIRS night-time lights [W cm-2 sr-1] | NOAA VIIRS (http://ngdc.noaa.gov/eog/viirs.html) | 0.0041667 | 2014 |
| Elevation | Elevation [meter] | CGIAR-CSI (http://srtm.csi.cgiar.org/) | 0.0083316 | 2008 |
| Vegetation | Vegetation | MODIS MOD13A1 [Enhanced vegetation index] | 0.0041667 | 2010-2014 |
| Distance | Distance to roads [meter] | Input data from OSM (http://extract.bbbike.org/) | 0.0083333 | 2014 |
| Distance | Distance to waterways [meter] | Input data from OSM (http://extract.bbbike.org/) | 0.0083333 | 2014 |
| Urban/Rural | Urban/Rural | MODIS-based Global Urban extent | 0.0041670 | 2000-2001 |
| Urban/Rural | Urban/Rural | CIESIN - Global Rural Urban Mapping Project (http://sedac.ciesin.columbia.edu/data/collection/grump-v1/sets/browse) | 0.0083333 | 2000 |
| Urban/Rural | Global Human Settlement Layer | Global Land Cover Facility (www.landcover.org) | 0.002818 | 2014 |
| Protected Area | Protected areas | WDPA (http://www.protectedplanet.net/) | Vector | 2012 |
| Land Cover | Land cover | ESA GlobCover Project (http://due.esrin.esa.int/page_globcover.php) | 0.0027777 | 2009 |
| Land Cover | Land cover | IGBP MODIS MCD12Q1 (https://lpdaac.usgs.gov/dataset_discovery/modis/modis_products_table/mcd12q1) | 0.0041670 | 2012 |
| Land Cover | Land cover | ONRL DAAC Synergetic Land Cover Product (SYNMAP) (http://webmap.ornl.gov/wcsdown/wcsdown.jsp?dg_id=10024_1) | 0.0083333 | 2000-2001 |

Mapping poverty using mobile phone and satellite data
J.E. Steele, P. Sundsøy, C. Pezzulo, V. Alegana, T. Bird, J. Blumenstock, J. Bjelland, K. Engø-Monsen, YA de Montjoye, A. Iqbal, K. Hadiuzzaman, X. Lu, E. Wetter, A.J. Tatem, and L. Bengtsson

| | | | | |
|---|---|---|---|---|
| Demographic | Pregnancies | WorldPop (http://www.worldpop.org.uk/) | 0.0008333 | 2012 |
| Demographic | Births | WorldPop (http://www.worldpop.org.uk/) | 0.0008333 | 2012 |
| Ethnicity | Georeferenced ethnic groups | ETH Zurich (http://www.icr.ethz.ch/data/geoepr) | Vector | 2014 |
| Climate | Mean annual precipitation | WorldClim (http://www.worldclim.org/download) | 0.0083333 | 1950-2000 |
| Climate | Mean annual temperature | WorldClim (http://www.worldclim.org/download) | 0.0083333 | 1950-2000 |

**Call Detail Records**

| | | | | |
|---|---|---|---|---|
| Basic phone usage | Outgoing/incoming voice duration, SMS count, etc. | Telenor/Grameenphone | NA | 2013-2014 |
| Top-up transactions | Spending speed, recharge amount per transaction, fraction of lowest/highest recharge amount, coefficient of variation recharge amount, etc. | Telenor/Grameenphone | NA | 2013-2014 |
| Location/mobility | Home district/tower, radius of gyration, entropy of places, number of places, etc. | Telenor/Grameenphone | NA | 2013-2014 |
| Social Network | Interaction per contact, degree, entropy of contacts, etc. | Telenor/Grameenphone | NA | 2013-2014 |
| Handset type | Brand, manufacturer, camera enabled, smart/feature/basic phone, etc. | Telenor/Grameenphone | NA | 2013-2014 |
| Revenue | Charge of outgoing/incoming SMS, MMS, voice, video, value added services, roaming, internet, etc. | Telenor/Grameenphone | NA | 2013-2014 |
| Advanced phone usage | Internet volume/count, MMS count, video count/duration, value added services duration/count, etc. | Telenor/Grameenphone | NA | 2013-2014 |

Table S2A. Wealth Index models for RS-only, CDR-only, and CDR+RS data: national, urban, and rural strata.

| AIC | NATIONAL | URBAN | RURAL |
|---|---|---|---|
| RS-only | 690.61 | 333.44 | 161.42 |
| CDR-only | 907.78 | 373.97 | 164.14 |
| CDR-RS | 651.76 | 318.12 | 115.52 |
| **MODEL** | | | |
| RS-only | 1 + transport time to closest urban settlement + nighttime lights + EVI + elevation | 1 + distance to roads + distance to waterways + nighttime lights + elevation | 1 + transport time to closest urban settlement + annual temperature + annual precipitation + distance to roads + distance to waterways + nighttime lights |
| CDR-only | 1 + recharge average per tower + percent nocturnal calls + number of places + entropy of contacts + outgoing internet sessions + sum outgoing internet sessions + incoming voice duration + count incoming content management system + count sum incoming content management system + volume of incoming multimedia messages + recharge amount per transaction + count incoming multimedia messages + count incoming texts + weekly recharge amount | 1 + recharge average per tower + number of places + entropy of contacts + spending speed + average outgoing text count + sum count incoming content management system + weekly recharge amount | 1 + recharge average per tower + percent nocturnal calls + entropy of places + radius of gyration + interactions per contact + recharge amount (CV) + number of retailers visited weekly (CV) + sum incoming video duration + count incoming multimedia messages + weekly recharge frequency (CV) + sum incoming video count + recharge amount per transaction (CV) |
| CDR-RS | 1 + transport time to closest urban settlement + nighttime lights + EVI + elevation + recharge average per tower + percent nocturnal calls + outgoing internet sessions + count incoming content management system + recharge amount per transaction + count incoming texts + weekly recharge amount | 1 + distance to roads + distance to waterways + nighttime lights + elevation + recharge average per tower + spending speed + average outgoing text count + weekly recharge amount | 1 + transport time to closest urban settlement + annual temperature + distance to roads + distance to waterways + nighttime lights + recharge average per tower + percent nocturnal calls + entropy of places + radius of gyration + interactions per contact + recharge amount (CV) + number of retailers visited weekly (CV) + sum incoming video duration + weekly recharge frequency (CV) + sum incoming video count + recharge amount per transaction (CV) |

*CV=coefficient of variation

Table S2*B*. Progress out of Poverty Index models for RS-only, CDR-only, and CDR+RS data: national, urban, and rural strata.

| AIC | NATIONAL | URBAN | RURAL |
|---|---|---|---|
| RS-only | 41676 | 14099 | 27465 |
| CDR-only | 41562 | 14044 | 27421 |
| CDR-RS | 41502 | 14043 | 27382 |
| **MODEL** | | | |
| RS-only | 1 + annual precipitation + annual temperature + transport time to closest urban settlement + distance to roads + EVI + nighttime lights | 1 + annual precipitation + annual temperature + transport time to closest urban settlement + elevation | 1 + annual precipitation + annual temperature + transport time to closest urban settlement + distance to water + EVI + nighttime lights |
| CDR-only | 1 + subscribers per tower + recharge average per tower + entropy of places + entropy of contacts + average outgoing text count + sum outgoing multimedia messages + fraction of 10 Thaka top-ups (min amount) + outgoing internet sessions + sum outgoing internet sessions + sum count incoming content management system + number of retailers visited weekly (CV) + sum revenue outgoing multimedia messages + spending speed variance + count incoming multimedia messages + sum count incoming texts + weekly recharge frequency (CV) + median time between refills + incoming video count + outgoing internet volume + time variable (CV) | 1 + subscribers per tower + recharge average per tower + sum outgoing multimedia messages + count outgoing internet sessions + sum count outgoing internet sessions + number of retailers visited weekly (CV) + volume of incoming multimedia messages + outgoing text charges + sum revenue outgoing multimedia messages + incoming video duration + sum incoming video duration + spending speed variance + sum count incoming multimedia messages + weekly recharge amount + incoming video count + sum incoming video count + number of retailers visited weekly | 1 + subscribers per tower + recharge average per tower + number of places + entropy of places + sum duration outgoing value added services + count outgoing texts + sum count outgoing texts + sum volume of outgoing multimedia messaging + fraction of 300 Thaka top-ups + count outgoing internet sessions + sum count outgoing internet sessions + incoming voice duration + number of retailers visited weekly (CV) + volume of incoming multimedia messages + outgoing text charges + sum outgoing text charges + sum revenue outgoing multimedia messages + spending speed variance + count incoming texts + sum count incoming texts + weekly recharge frequency (CV) + median time between refills + outgoing internet volume + recharge amount per transaction (CV) + time variable (CV) |
| CDR-RS | 1 + annual precipitation + annual temperature + transport time to closest urban settlement + distance to road + elevation + subscribers per tower + recharge average per tower + entropy of places + entropy of contacts + average outgoing text count + sum outgoing multimedia messages + outgoing internet sessions + sum outgoing internet sessions + number of retailers visited weekly (CV) + sum revenue outgoing multimedia messages + spending speed variance + sum count incoming multimedia messages + count incoming multimedia messages + sum count incoming texts + weekly recharge frequency (CV) + median time between refills + incoming video count + outgoing internet volume + time variable (CV) | 1 + annual precipitation + annual temperature + EVI + elevation + subscribers per tower + recharge average per tower + sum outgoing multimedia messages + number of retailers visited weekly (CV) + volume of incoming multimedia messages + sum volume of incoming multimedia messages + incoming video duration + sum incoming video duration + sum revenue outgoing multimedia messages + outgoing text charges + sum count incoming multimedia messages + weekly recharge amount + incoming video count + sum incoming video count + number of retailers visited weekly | 1 + annual precipitation + annual temperature + transport time to closest urban settlement + distance to water + EVI + nighttime lights + subscribers per tower + recharge average per tower + outgoing multimedia messages + fraction of 300 Thaka top-ups + number of retailers visited weekly (CV) + volume of incoming multimedia messages + outgoing text charges + sum revenue outgoing multimedia messages + spending speed variance + median time between refills + time variable (CV) |

*CV=coefficient of variation

Table S2C. Income models for RS-only, CDR-only, and CDR+RS data: national, urban, and rural strata.

| AIC | NATIONAL | URBAN | RURAL |
|---|---|---|---|
| RS-only | 83194 | 17295 | 64569 |
| CDR-only | 83109 | 17180 | 64413 |
| CDR+RS | 82895 | 17129 | 64330 |
| **MODEL** | | | |
| RS-only | 1 + nighttime lights + transport time to closest urban settlement + annual temperature + EVI + distance to water + annual precipitation + elevation | 1 + nighttime lights + transport time to closest urban settlement + distance to roads | 1 + nighttime lights + annual temperature + distance to roads + annual precipitation + elevation |
| CDR-only | 1 + percent nocturnal calls + number of places + entropy of places + entropy of contacts + radius of gyration + interactions per contact + spending speed + outgoing video duration + sum outgoing video duration + average outgoing text count + sum outgoing text count + fraction of 300 Thaka top-ups + fraction of 10 Thaka top-ups + recharge amount (CV) + number of retailers visited weekly (CV) + recharge amount per transaction + spending speed variance + sum spending speed variance + sum count incoming texts + handset weight + outgoing voice duration + sum outgoing voice duration + sum outgoing internet volume + time variable (CV) + fraction of 200 Thaka top-ups | 1 + percent nocturnal calls + number of places + entropy of places + entropy of contacts + radius of gyration + spending speed + outgoing video duration + average outgoing text count + sum outgoing video duration + fraction of 300 Thaka top-ups + volume of incoming multimedia messages + sum volume of incoming multimedia messages + sum recharge amount per transaction + count incoming multimedia messages + sum count incoming multimedia messages + sum outgoing voice duration + outgoing internet volume + recharge amount per transaction (CV) + time variable (CV) | 1 + number of places + entropy of places + entropy of contacts + spending speed + duration outgoing value added services + sum duration outgoing value added services + sum outgoing video duration + handset weight + software OS version + fraction of 300 Thaka top-ups + fraction of 10 Thaka top-ups + coefficient of variation: recharge amount + number of retailers visited weekly (CV) + volume of incoming multimedia messages + sum volume of incoming multimedia messages + spending speed variance + sum spending speed variance + sum count incoming multimedia messages + count incoming texts + sum count incoming texts + weekly recharge amount + outgoing voice duration + sum outgoing voice duration + outgoing internet volume + time variable (CV) |
| CDR-RS | 1 + nighttime lights + transport time to closest urban settlement + EVI + distance to road + percent nocturnal calls + number of places + entropy of contacts + radius of gyration + spending speed + outgoing video duration + average outgoing text count + sum outgoing text count + recharge amount (CV) + number of retailers visited weekly (CV) + recharge amount per transaction + spending speed variance + sum spending speed variance + sum count incoming texts + handset weight + outgoing voice duration + sum outgoing voice duration + sum outgoing internet volume + time variable (CV) + fraction of 200 Thaka top-ups | 1 + nighttime lights + transport time to closest urban settlement + annual temperature + distance to roads + annual temperature + percent nocturnal calls + number of places + radius of gyration + spending speed + outgoing video duration + sum outgoing video duration + average outgoing text count + sum outgoing text count + fraction of 300 Thaka top-ups + number of retailers visited weekly (CV) + volume of incoming multimedia messages + recharge amount per transaction + count incoming multimedia messages + sum count incoming multimedia messages + sum outgoing voice duration + outgoing internet volume + recharge amount per transaction (CV) + time variable (CV) | 1 + nighttime lights + annual precipitation + number of places + entropy of places + entropy of contacts + spending speed + sum outgoing video duration + handset weight + software OS version + fraction of 300 Thaka top-ups + fraction of 10 Thaka top-ups + coefficient of variation: recharge amount + number of retailers visited weekly (CV) + weekly recharge amount + outgoing voice duration + sum outgoing voice duration + outgoing internet volume + time variable (CV) |

*CV=coefficient of variation

Table S3. Comparison of $r^2$ and RMSE for INLA models run with only a structured spatial random effect (Spatial interpolation) and the full model (Spatial model + covariates).

| Poverty Metric | Spatial interpolation | Spatial model + covariates (from Table 1) |
|---|---|---|
| | $R^2$, RMSE | $R^2$, RMSE |
| DHS WI | 0.49, 0.578 | 0.76, 0.394 |
| PPI | 0.31, 58.727 | 0.32, 57.439 |
| Income | 0.10, 123.963 | 0.27, 105.465 |

Mapping poverty using mobile phone and satellite data
*J.E. Steele, P. Sundsoy, C. Pezzulo, V. Alegana, T. Bird, J. Blumenstock, J. Bjelland, K. Engo-Monsen, YA de Montjoye, A. Iqbal, K. Hadiuzzaman, X. Lu, E. Wetter, A.J. Tatem, and L. Bengtsson*

Table S4. Comparison of deviance information criterion (DIC) model fit for Bayesian geostatistical models run with a structured spatial random effect (Spatial model) and without (Non-spatial model).

| Poverty Metric | Model | Spatial model | Non-spatial model |
|---|---|---|---|
| **WHOLE COUNTRY** | | | |
| | | DIC | DIC |
| DHS WI | CDR - RS | 463.7 | 574.6 |
| | CDR | 272.5 | 862.2 |
| | RS | 465.7 | 581.5 |
| PPI | CDR - RS | 1361.3 | 1439.8 |
| | CDR | 1349.7 | 1473.1 |
| | RS | 1358.6 | 1421.4 |
| Income | CDR - RS | 66142.5 | 66143.0 |
| | CDR | 66314.6 | 66314.3 |
| | RS | 66482.6 | 66480.4 |
| **URBAN** | | | |
| | | DIC | DIC |
| DHS WI | CDR - RS | 449.4 | 576.0 |
| | CDR | 239.2 | 873.2 |
| | RS | 454.1 | 582.0 |
| PPI | CDR - RS | 1371.6 | 1432.3 |
| | CDR | 1365.7 | 1470.1 |
| | RS | 1358.3 | 1417.7 |
| Income | CDR - RS | 66180.6 | 66179.8 |
| | CDR | 66363.3 | 66365.7 |
| | RS | 66693.0 | 66690.3 |
| **RURAL** | | | |
| | | DIC | DIC |
| DHS WI | CDR - RS | 458.0 | 574.5 |
| | CDR | 63.1 | 873.5 |
| | RS | 451.5 | 595.5 |
| PPI | CDR - RS | 1376.9 | 1444.6 |
| | CDR | 1342.9 | 1475.9 |
| | RS | 1357.7 | 1419.4 |
| Income | CDR - RS | 66262.5 | 66260.6 |
| | CDR | 66395.0 | 66392.1 |
| | RS | 65548.9 | 66503.8 |

Mapping poverty using mobile phone and satellite data
*J.E. Steele, P. Sundsoy, C. Pezzulo, V. Alegana, T. Bird, J. Blumenstock, J. Bjelland, YA de Montjoye, K. Engo-Monsen, A. Iqbal, K. Hadiuzzaman, X. Lu, E. Wetter, L. Bengtsson, and A.J. Tatem*

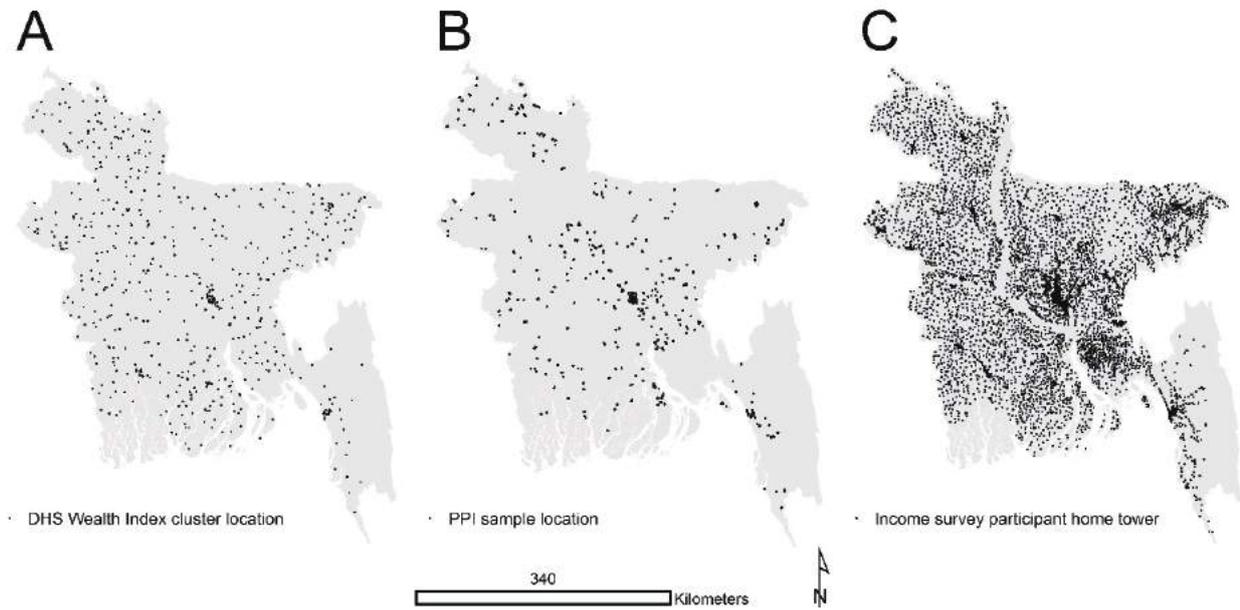

Figure S1. Survey sample locations for DHS wealth index (A), Progress out of Poverty Index (B), and income survey respondents (C).

Mapping poverty using mobile phone and satellite data
*J.E. Steele, P. Sundsoy, C. Pezzulo, V. Alegana, T. Bird, J. Blumenstock, J. Bjelland, YA de Montjoye, K. Engo-Monsen, A. Iqbal, K. Hadiuzzaman, X. Lu, E. Wetter, L. Bengtsson, and A.J. Tatem*

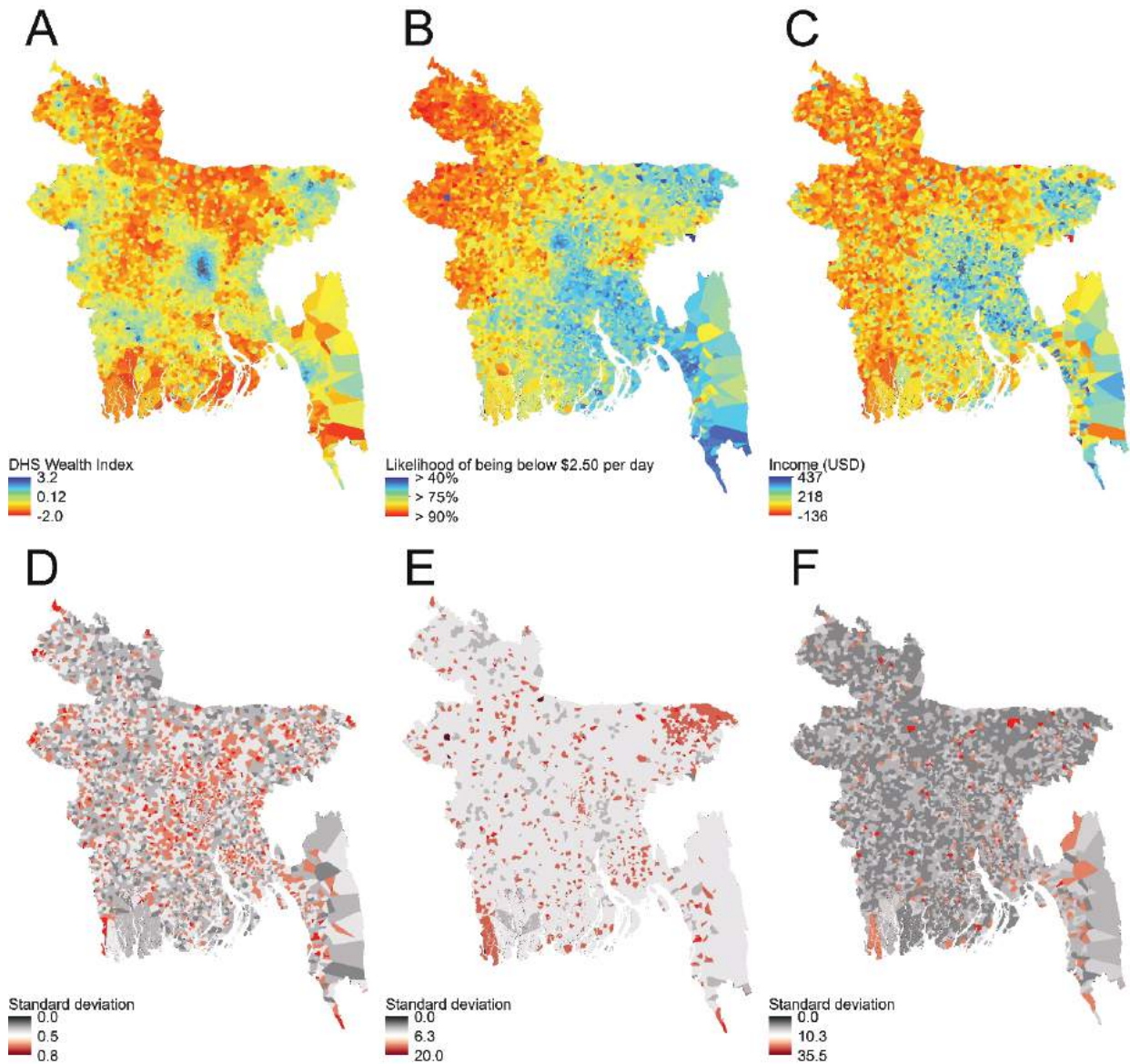

Figure S2. National level prediction maps for mean wealth index (A) with uncertainty (D); mean probability of households being below $2.50/day (B) with uncertainty (E); and mean USD income (C) with uncertainty (F). Maps were generated using call detail record features only and Bayesian geostatistical models. Red indicates poorer areas in prediction maps, and higher error in uncertainty maps.

Mapping poverty using mobile phone and satellite data
*J.E. Steele, P. Sundsoy, C. Pezzulo, V. Alegana, T. Bird, J. Blumenstock, J. Bjelland, YA de Montjoye, K. Engo-Monsen, A. Iqbal, K. Hadiuzzaman, X. Lu, E. Wetter, L. Bengtsson, and A.J. Tatem*

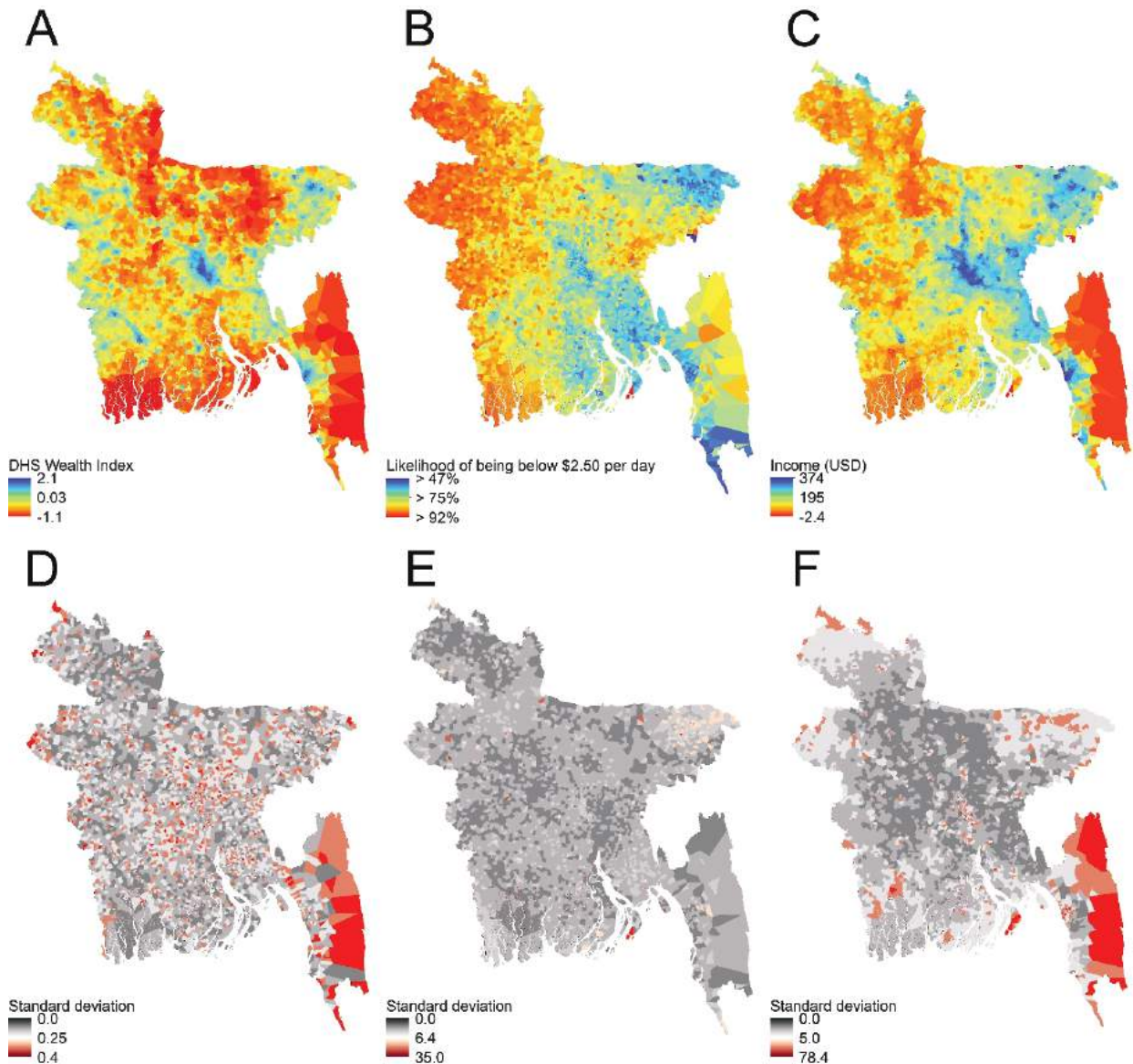

Figure S3. National level prediction maps for mean wealth index (A) with uncertainty (D); mean probability of households being below $2.50/day (B) with uncertainty (E); and mean USD income (C) with uncertainty (F). Wealth index and income maps were generated using remote sensing data only; PPI maps were generated using call detail record features and remote sensing data. All maps were generated using Bayesian geostatistical models. Red indicates poorer areas in prediction maps, and higher error in uncertainty maps.

Mapping poverty using mobile phone and satellite data
*J.E. Steele, P. Sundsoy, C. Pezzulo, V. Alegana, T. Bird, J. Blumenstock, J. Bjelland, YA de Montjoye, K. Engo-Monsen, A. Iqbal, K. Hadiuzzaman, X. Lu, E. Wetter, L. Bengtsson, and A.J. Tatem*

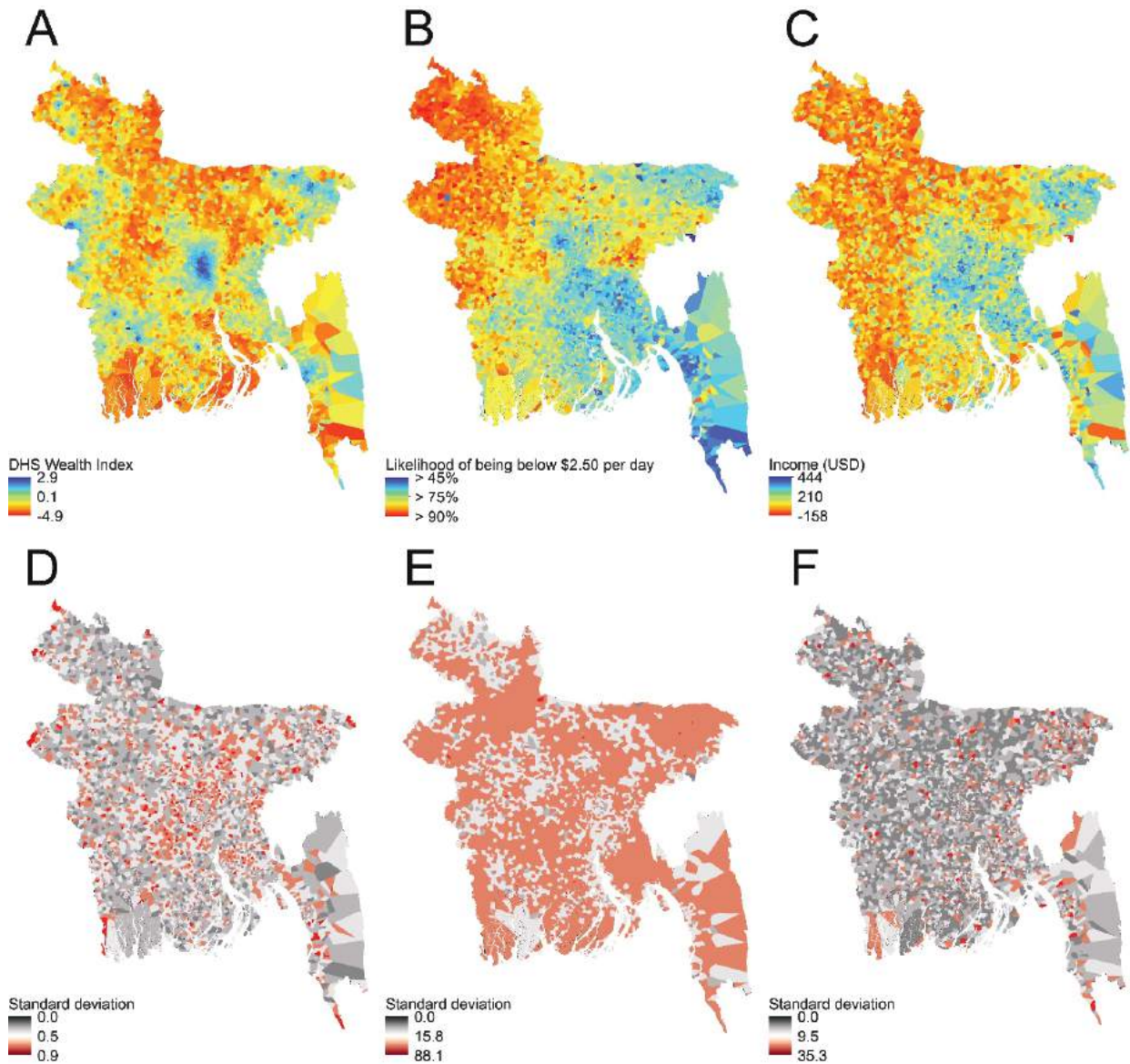

Figure S4. Stratified urban/rural prediction maps for mean wealth index (A) with uncertainty (D); mean probability of households being below $2.50/day (B) with uncertainty (E); and mean USD income (C) with uncertainty (F). Maps were generated using call detail record features only and Bayesian geostatistical models. Red indicates poorer areas in prediction maps, and higher error in uncertainty maps.

Mapping poverty using mobile phone and satellite data
*J.E. Steele, P. Sundsoy, C. Pezzulo, V. Alegana, T. Bird, J. Blumenstock, J. Bjelland, YA de Montjoye, K. Engo-Monsen, A. Iqbal, K. Hadiuzzaman, X. Lu, E. Wetter, L. Bengtsson, and A.J. Tatem*

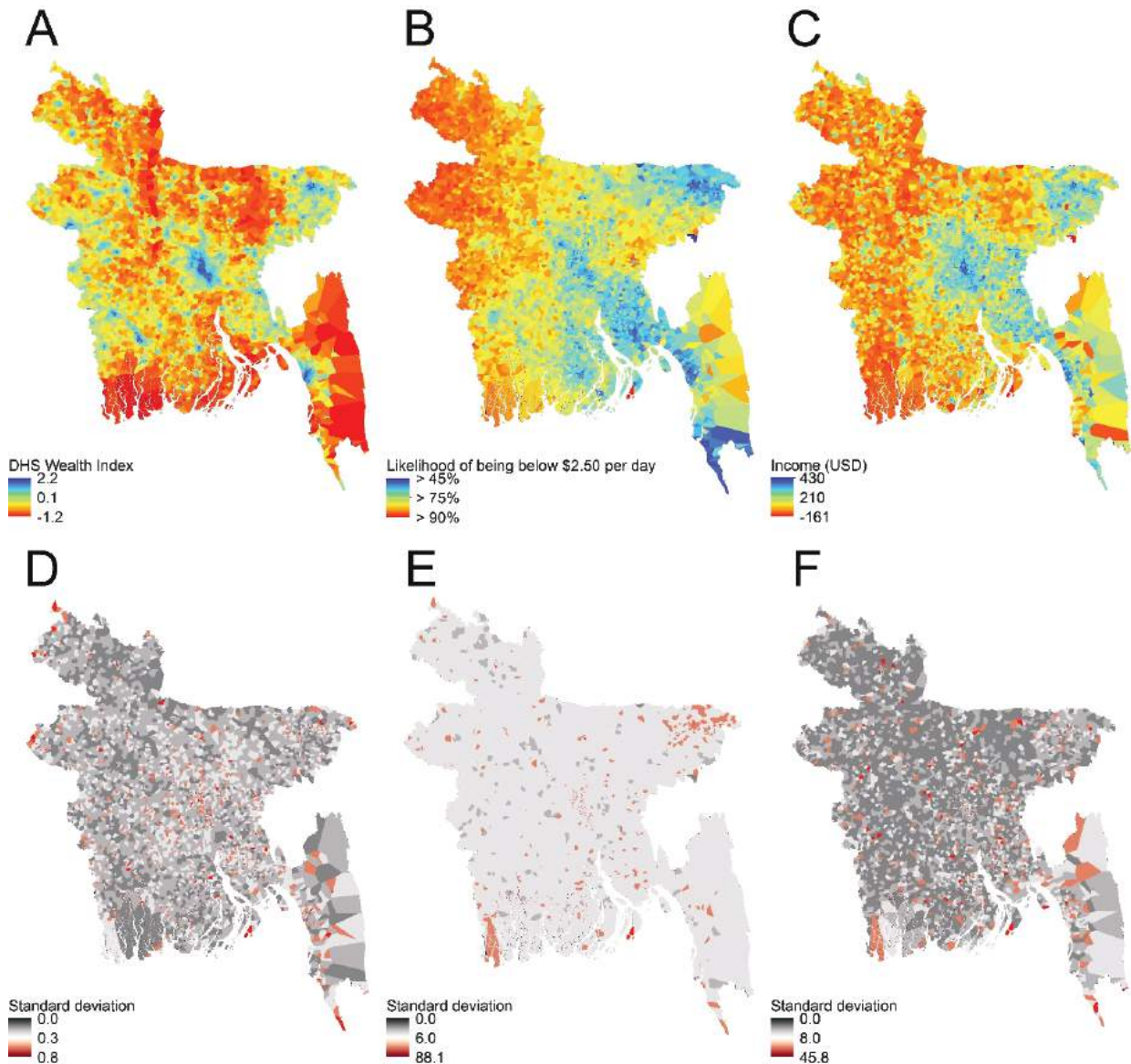

Figure S5. Stratified urban/rural prediction maps for mean wealth index (A) with uncertainty (D); mean probability of households being below $2.50/day (B) with uncertainty (E); and mean USD income (C) with uncertainty (F). Maps were generated using call detail record features, remote sensing data, and Bayesian geostatistical models. Red indicates poorer areas in prediction maps, and higher error in uncertainty maps.

Mapping poverty using mobile phone and satellite data
*J.E. Steele, P. Sundsoy, C. Pezzulo, V. Alegana, T. Bird, J. Blumenstock, J. Bjelland, YA de Montjoye, K. Engo-Monsen, A. Iqbal, K. Hadiuzzaman, X. Lu, E. Wetter, L. Bengtsson, and A.J. Tatem*

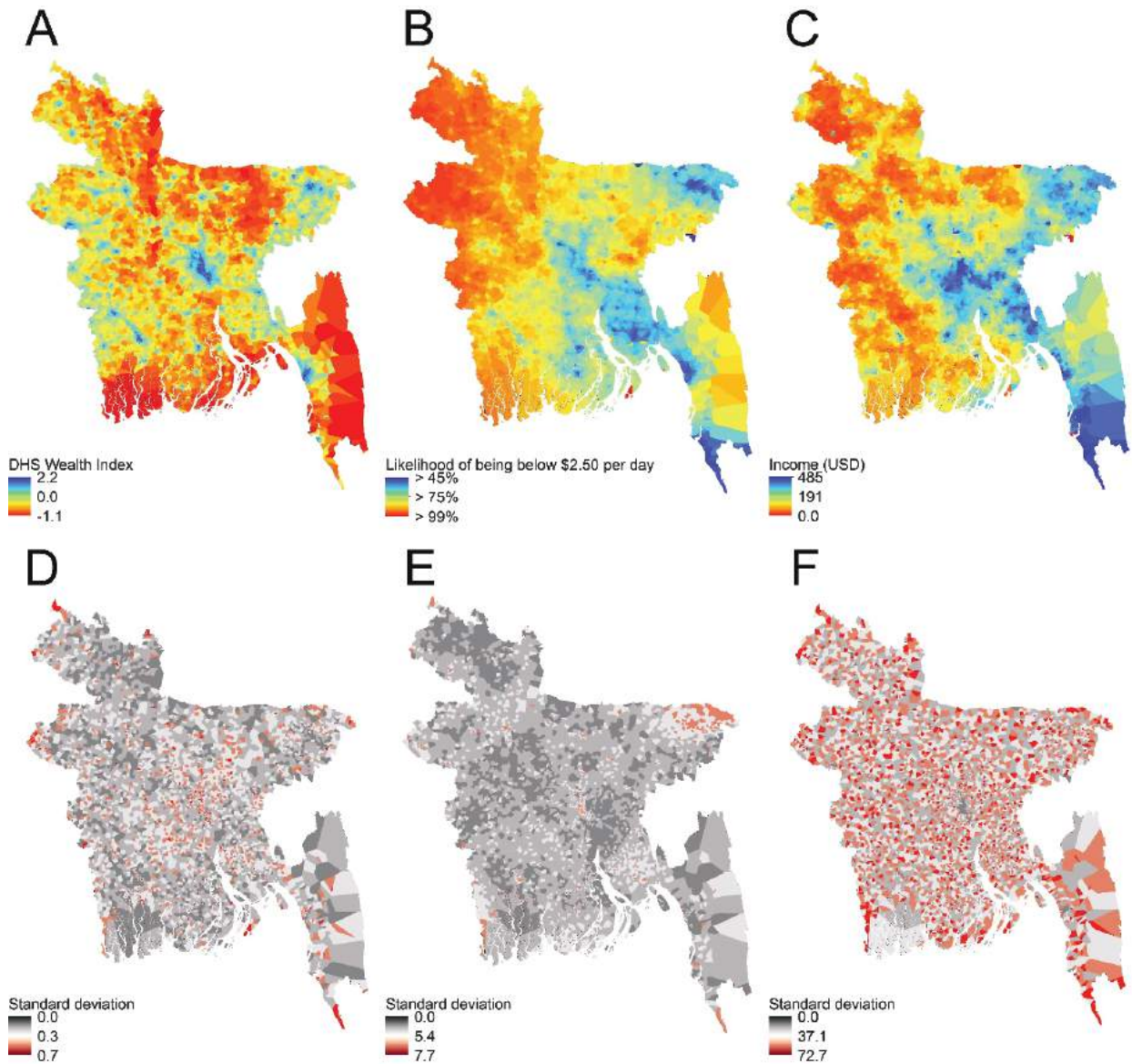

Figure S6. Stratified urban/rural prediction maps for mean wealth index (A) with uncertainty (D); mean probability of households being below $2.50/day (B) with uncertainty (E); and mean USD income (C) with uncertainty (F). Maps were generated using remote sensing data only and Bayesian geostatistical models. Red indicates poorer areas in prediction maps, and higher error in uncertainty maps.

# ◆ Paper 4

# The activation of core social networks in the wake of the 22 July Oslo bombing

Pål Roe Sundsøy, Johannes Bjelland, Geoffrey Canright, Kenth Engø-Monsen
Research & Future Studies
Telenor Group
Oslo, Norway
pal-roe.sundsoy@telenor.com

Rich Ling
IT University/ Telenor Group
Copenhagen Denmark/Oslo, Norway
rili@itu.dk

*Abstract*—This paper examines how core social networks were activated in the wake of the bombing in Oslo on July 22, 2011. Empirical mobile traffic data illuminate exceptional behavior, just after the bombing in Oslo. We find that in the minutes after the bombing people called ties that were (1) close socially and (2) perceived to be in danger; that is, people who were close to the bombing point. Our main findings: (1) individuals first focus on their single closest contact ('best friend'); but, soon after, switch to spending more mobile communication resources than average on contacts ranked 2—5. (2) we see clearly a large increase (over typical) in traffic to and from, and not least within, the affected area (Oslo)—in some cases more than a 300% increase, right after the bombing. Interestingly, we also find a marked increase in traffic for relationships where both persons were outside of Oslo. All these results illustrate the importance of social contact in this highly unusual situation. This paper underscores how the mobile phone is an instrument of the intimate sphere. The situation on the 22nd of July in Oslo is a prime example of an unexpected situation, where individuals use the mobile to get critical information on their loved ones in their core network.

*large-scale social network analysis; telecommunications;22 July; mobile communication in emergencies; social cohesion; terrorist networks; terror;SNA; mobile communication networks; call detail records; CDR*

## I. Introduction

On Friday 22 July at 15:26 a powerful bomb exploded in *Regjeringskvartalet* (the center of national administration) in central Oslo, Norway. It killed 8 people and seriously injured almost 100 others. Following the bombing, Anders Behring Breivik (a right-wing, white extremist) drove approximately 1.5 hours and took a ferry to the small island Utøya where the Norwegian Labor party has a summer camp for its youth members. There he shot 69 of the attendees and wounded 60. This is the deadliest attack on Norway since the Second World War.

For reasons we will discuss below, this paper will focus on the immediate reaction of the general population in Oslo (and also Norway in general) to the bombing event alone. We do this by examining their calling behavior, as assessed using anonymized traffic data from the mobile phone network.

In earlier but related work, Dutton [1] and Katz and Rice [2] presented qualitative results on the 9/11 catastrophe, emphasizing the need to reach out to the closest tie. Cohen and Lemish [4] examined the geographical distribution of traffic after a terrorist bombing in Israel. Studies of other types of disasters include Erikson's [6] discussion of communication needs (absent mobile telephony) during and after flooding from a breached coal sludge dam, Weick's [7] study of verbal communication during a firefighting disaster, and Figley and Jones' [8] qualitative results on the use of mobile telephones around the Virginia Tech shootings.

Our work is distinct from these others, in that we use large scale phone logs to understand communication patterns around the disaster of July 22. Such studies, utilizing unique and valuable mobile communication data for research, have begun to appear in recent years; for examples, see [9]—[16].

It is clear that, in events such as 22 July, we feel the need to reach out to near family and friends—to check on one another's wellbeing, possibly to organize assistance, and to also make sense of the situation. In this paper, we will examine behavior—specifically, mobile phone usage—which can help illuminate the ways these needs are met in the face of a disaster.

## II. Definition of the event and the data used in the analysis

As noted above, the terror attack on July 22 consisted of two deadly events: the bombing in Oslo at 15:26, and the shootings on the island Utøya starting around 18:00. In this paper we will focus almost entirely on the response to the bombing in Oslo. There were three main factors which guided this choice of focus:

*1) Privacy issues.* Even though we use anonymized call and sms data in all of our research, we observe that the Utøya event involved relatively few individuals (compared to the Oslo event) so that there is a small chance that results for Utøya might be traced to individuals. This is not possible for the highly aggregated data describing the response to the Oslo bombing.

*2) Temporal dimensions.* In addition, the bombing was a temporally well-defined event. By contrast, the shootings on Utøya were not witnessed by nearly as many people, and the breadth of the tragedy was not immediately obvious. Due to these considerations, the bombings were obvious in the network traffic picture, while the shootings were much less distinct.

*3) Technical difficulty.* Anonymized call data can be connected to locations by using demographic data. We assume that, statistically, many anonymous subscribers with a home address in Oslo are either physically in Oslo at the time of the bombing—or are suspected to be there by their close contacts—or return promptly to Oslo after the attack. In short: we can use postal address as a simple proxy for location. This idea however cannot work for the Utøya massacre.

Thus, we focus on communication behavior that is localized in time around the bombing event at 15:26. We also study the geographic dependence of this behavior, focusing on Oslo (the site of the bombing) and comparing communication behavior there to that in other areas. For simplicity, we have restricted most of our time-based results to voice communication only—we find that voice is a good proxy for a more complete definition of a mobile communication relationship (such as voice+sms). Our geographic results however include both voice and sms.

### III. SPIKE IN AGGREGATED VOICE TRAFFIC

Before presenting our social-network results, we present here the overall response of voice traffic—aggregated over all customers in the Telenor subscriber base, i.e. around 3 million Norwegian customers—to the bombing.

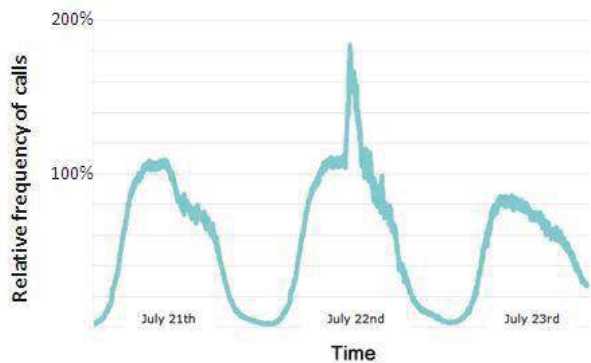

**Figure 1** Voice calls per minute, aggregated across the whole country. Here we count only those calls that were successfully terminated.

Figure 1 shows the general voice traffic, displayed minute by minute for July $21^{st}$, $22^{nd}$, and $23^{rd}$, and aggregated across the whole country. We see that the Friday (July 22) behavior is much like the Thursday behavior—except for a sharp peak rising up immediately after the bombing. In this peak we see a strong increase in successfully originated voice calls. We see that the number of calls/minute rises to about twice that we would expect from normal traffic in this period. The corresponding SMS analysis shows a much less distinct peak. The preference for voice communication in this situation shows the importance of immediacy in this situation.

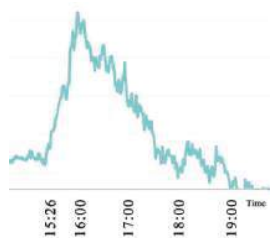

**Figure 2** Voice calls per minute (zoomed from Figure 1).

In Fig 2 we give an expanded view of the time-dependent voice traffic during the spike. We see that the rapid increase in traffic reaches a peak around 30 minutes after the explosion (around 4 pm). The traffic then gradually returns to a normal calling pattern. An exception to this is seen in two small peaks between 18:00 and 19:00. These two peaks correspond to the time of the Utøya massacre. Compared to the peak caused by the explosion, these smaller peaks show that this was a much more isolated event. Next we want to further investigate the calling patterns in this period.

### IV. DEFINITION OF THE CORE NETWORK

We want to better understand how the closest relationships in a social network are affected by the emergency situation which the bombing attack represents. To do this, we will, for each node, rank the node's relationships according to the number of calls sent to the node's various contacts.

We draw on earlier work [17] and define the *closest relation* using a simple definition: the closest relation is defined as the person that an individual called the most over a 3 month period prior to the bombing. Figure 3 illustrates the idea: node A has had 37 calls to node B during a 3 month interval, which defines node B as A's closest relation. Node C will be his second closest relation, and so on. Based on these definitions, we can generate a list, for all subscribers, giving closest relation, second closest relation, third closest relation, etc.

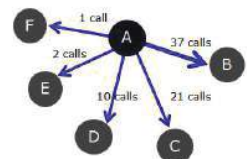

**Figure 3** Closest relation definition. B is A's closest relation, C is A's second closest relation, etc.

### V. EFFECTS OF THE BOMBING ON THE CORE NETWORK

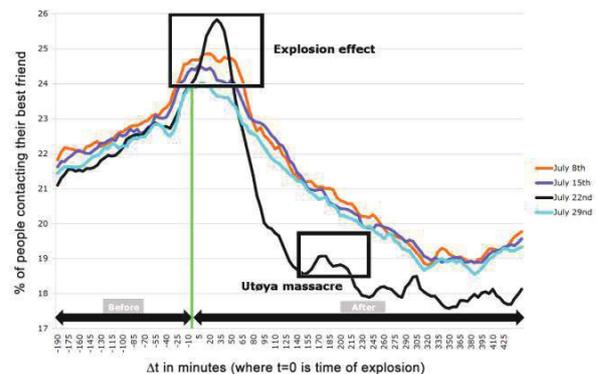

**Figure 4** Percent of active subscribers contacting their closest relation as a function of time. The percentage is shown for Friday $8^{th}$, $15^{th}$, $22^{nd}$ and $29^{th}$ of July 2011.

Using the definitions of Section IV, we now look at (i) what relations are contacted, and (ii) when they are contacted—in the period just before and after the bombing. We focus on an

11 hour time interval (12 noon to 11pm), and on all subscribers that were active in Telenor's network during July 22$^{nd}$.

The black curve in Figure 4 represents the percentage of active subscribers contacting their closest relation (in each time interval of 5 minutes) as a function of time, where t=0 is equal to the time of explosion.

We notice that the black curve reaches a peak around 25 minutes after the explosion. At this time, around 26% of all active subscribers contacted their closest relation. At about the same time, the explosion was reported in the largest online newspapers. We also notice a smaller peak when the Utøya massacre starts.

This curve can be compared to a normal Friday in order to see when and how often we usually contact our closest relations. Therefore, as a comparison, we have included in Figure 4 curves from the two Fridays before the explosion (red and blue color) and the Friday after (light blue color).

By looking at the rise of the normal curves before the explosion, we see that it is more common to contact the closest relation later in the day (we notice a peak around 4 pm). On normal days, this is often a type of "coordination" call between partners to organize the transition from work to the week-end.

When compared to the other Fridays, we observe a 7% increase in people contacting their closest relation on the 22$^{nd}$. We also observe that there is a 14% *decrease* in these contacts *after* the explosion. The reasons for this decrease are not obvious. One hypothesis is that, after reaching our closest contact, we then cycle through our other contacts more than normal. This is perhaps to check on their wellbeing. This interaction with weaker social ties is at the expense of contact with our closest relation.

To further examine this hypothesis, we have normalized the black curve in Fig 4, by dividing it by the average of the closest-contact traffic on the normal Fridays. This gives the fraction of people contacting their closest relation, as compared to a normal Friday. The result is shown in Fig 5.

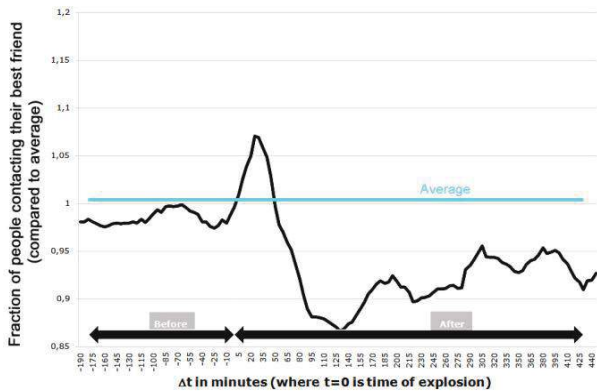

**Figure 5** Fraction of active subscribers contacting their closest relation compared to a normal Friday (based on all active subscribers in the 11 hours period). This plot shows the same as Fig 4, but normalized with the average from the other three Fridays.

We see that this doubly-normalized curve roughly follows the average (a little beneath) before the bombing, then rises by about 7%, and subsequently falls by about 14% when compared to a normal Friday afternoon.

In Figure 6 we show doubly-normalized curves (percentage of contacts for number x contact, compared to average, over time) for x = 1—5. This enables us to compare communication with the five first contacts on a common scale.

We see in Figure 6 a clear picture. Before the explosion, we see that all curves roughly follow the average. When the explosion hits, the closest relation is contacted first (peak at 25 min, 7% higher). This is followed by contacting the second closest relation (peak at 40 min, 15% more than average), then the third, fourth and fifth closest relations (peak at 45 min, 20% higher than average). We notice that relatively more time than average is spent subsequently on 'medium-distant' relations (3 to 5). Of course, the third, fourth and fifth closest relations are relatively seldom called. Our data shows that they normally command about 4.5%, 2%, and 1%, respectively, of all calls. In contrast, link one commands 69% of the traffic, and link 2 receives 13%. As shown in other work, these core links are extremely important social contacts [17].

Figure 6 shows that, in the wake of the explosion, there was an increase in the need to reach these more remote contacts (3 to 5).

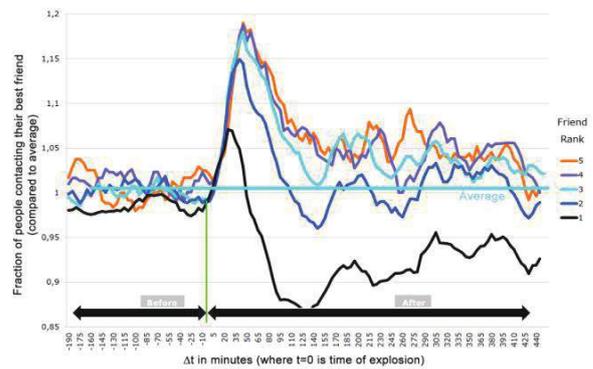

**Figure 6** Fraction of active subscribers contacting their 1-5 closest relations vs time for all active subscribers. All curves are normalized by average communication for the same contact number (1—5).

Thus we see that, while the 'absolute' amount of traffic generated in calls to links 3, 4, and 5 is small, the proportional increase for them is greater than that for the nearer contacts 1 and 2. Indeed, contact with link number 1 in fact decreases proportionally, after the initial peak. It is difficult to completely specify here the reason for the relatively early fall in contact with the best relationship. It may be simply that after mobile contact is achieved with the closest contact, there is an urgency to check on other links that are called less often, and so we do not dwell as much as normal with link number one. It may also be that one quickly arranges to meet physically with link one, thus obviating the need for further contact over the mobile phone.

## VI. GEOGRAPHICAL SPREADING

In section III we saw an approximate doubling of voice calls (on a nationwide basis) in the wake of the explosion. In this section, we will look at how this traffic was distributed between different regions in the country. To place each customer geographically, we use their registered postal address. Of course, there is considerable uncertainty about where these "Oslo residents" were to be found physically at the time of the event. We believe however that the results obtained by this method support the argument that this error source is not too large. We also note in passing that close relations to Oslo residents might also very well not know where their Oslo friends are at the time of the attack—and so are motivated to contact them, in the face of this same uncertainty.

significantly more in the aftermath of the bombing. The people calling to and from Oslo have a motivation to call which is based on possible proximity of at least one of them to the bomb. Here the need is to hear if the contact in Oslo safe. This motivation was not necessarily in place for the people neither living in Oslo nor calling to others in Oslo. Rather, the strong diagonal in Figs 7(b)—7(d) speaks to another motivation for coming into contact. Perhaps it speaks to the need to process the situation, and to consider the events with someone who is socially and physically close. We note further that this strong diagonal also mimics the normal geographical pattern of telephonic communication [6].

## VII. SUMMARY AND FUTURE WORK

There have been several analyses of how people have used

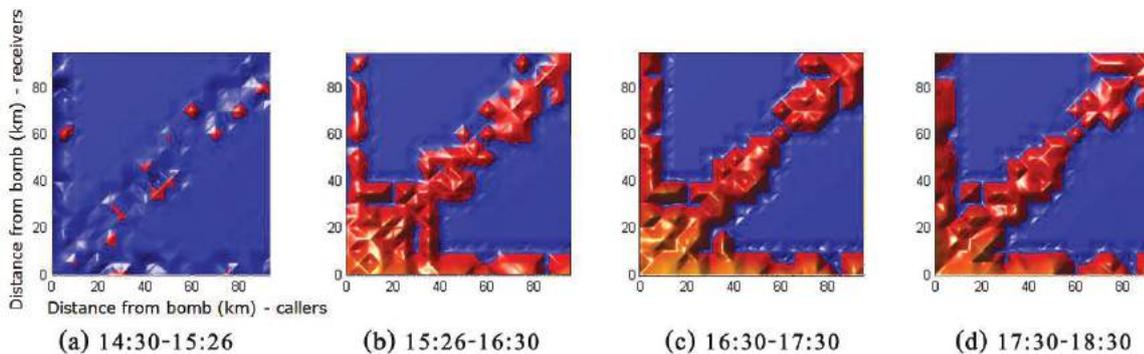

**Figure 7** Change in activation of social ties between different geographical regions, for 4 different time intervals. Blue=Average or beneath, red=2 times average, yellow=4 times average.

Figure 7 shows the activation of social links (defined by either SMS or voice) before and after the bomb, and how this activation varies with distance from the centre of explosion.

The x-axis shows the distance from the bomb (in km) for the people making calls, while y-axis shows the distance from the bomb (in km) for the people receiving calls. The colors on the z-axis show the number of social relations activated compared to an average Friday. A blue color shows average, red is about twice normal and yellow indicates four times the normal telephone traffic.

Fig 7(a) shows that the activation of geographically based social ties via the mobile phone during the hour previous to the explosion is fairly typical. Fig 7(b) shows the geographic distribution of links in the hour just after the explosion. The red line along the y-axis (x = 0 means the caller is in Oslo) shows that the outgoing link activations from Oslo are twice as large as the normal, while the incoming link activations to Oslo (following the x-axis) are two to four times the average. This result is even stronger in the following hour (up to 2 hours after the explosion—Fig 7(c)), and still quite strong up to three hours after (Fig 7(d)).

The area of Figure 7 located around (x=0,y=0) shows communication which is internal to Oslo. We see that the number of activated relations increases with a factor 4 in this area. Interestingly, we can also see along the diagonal that people living near to each other, and outside Oslo, talked mobile telephony during wide-scale disasters. Several authors have written on mobile telephony use in the case of 11 September in the US ([1][2] Dutton and Katz and Rice). In the work by Dutton and Nainoa [5], there was an analysis of press coverage of mobile phones related to the 11 September events. The work by Katz and Rice [2] is a qualitative analysis of peoples' immediate reaction to the disaster. Randall Collins [3] also examines the social reaction to the 11 September bombings, though not through the lens of mobile communication. Cohen and Lemish [4] examined the use of mobile phones in the case of terror bombings in Israel. They had access to the telephone traffic in the aftermath of the bombings and showed, as we do here, that there was an immediate peak in traffic generated near the site of the bombing. In many of these analyses, the material used qualitative (interview) data to show that people used the mobile phone in order to be in contact with those who were emotionally close to them.

None of these analyses have taken the step to examine the material from a social network perspective, using quantitative communication log data. Indeed, we feel that this is the unique contribution of this paper. We are able to parse out the way that we use the mobile phone to indeed call those people who are closest to us in the wake of this untoward event.

The analysis here is of a general emergency in a broad social context. We see that time was a critical element in the mobilization of different people in the social networks. People called the most important link first, and then moved on to the secondary, tertiary, etc. links in order to inform and to mobilize one another. These findings can be scaled down to smaller scale emergencies within the immediate social sphere. When, for example, a family member is hurt to the degree that they need medical attention, there is also the mobilization of the immediate network of family and friends via the mobile phone. The "one-hop" people need to be alerted, and need to reshuffle their meetings and tasks so as to take care of the injured person or perhaps carry out the necessary tasks (e.g. pick up children at daycare) that that the injured person cannot carry out.

The behavior that we see here is not surprising—it confirms in fact our expectations: we reach out to our most important contacts first. We believe that this unsurprising, 'common-sense' aspect of our results does not at all detract from their value, for two reasons. First, it is never guaranteed that one will get one's common-sense expectations fulfilled—surprises can happen. Hence measurements which test our expectations are of value, whether or not they confirm them. Here, in response to the occurrence of a rare event, we have taken the rare opportunity to make such measurements and test our expectations. These measurements show how mobile telephony is becoming increasingly embedded in society.

Our results give value in another way: they strengthen the claim that the use of *frequency* of contact as a measure of *closeness* of the relationship is justified. The clear (and expected) time ordering of the communication with contacts leaves little doubt that the definition of closeness given in Section IV is in fact appropriate here; otherwise our results make little sense.

Thus, we see that the mobile phone has gained a central position in the coordination of action within the core social sphere. This can operate when there is a major social event such as the bombing on 22 July, or when the family faces an emergency with regard to one of its members.

This study has several limitations. These include the fact that the data are taken only from one operator in Norway: Telenor. Thus the data is not necessarily generalizable to all of Norway. Also, as noted, the location based data use the postal code of the subscriber, and not the actual location of the phone at the time of the bombing. This is a result of privacy issues and is not to be avoided.

When thinking of further work, we have several items that we wish to develop. These include examining the "cascades" of calls between individuals as the events developed. This can include analysis of the sizes of the cascade groups, the timing and the geographical dimensions of these call sequences. In this paper we have examined the strong ties. It is also interesting to examine the calls to the weak ties, and to look at the people who were *not* called in the period after the bombing. What are the characteristics of these people? In addition we have access to the streaming of different types of music in the aftermath of the bombing and the shootings. The data show interesting preferences for songs that celebrate Norwegian national symbols. This material suggests that people were active in the reconstruction of the sense of being Norwegian through these activities. Finally, we are interested in providing a better theoretical foundation for the material. The analysis presented here is very empirically driven. We are considering theoretical approaches that might help to clarify the observations.

# ◆ Paper 5

# Detecting climate adaptation with mobile network data in Bangladesh: anomalies in communication, mobility and consumption patterns during cyclone Mahasen

Xin Lu[1,2,3] · David J. Wrathall[4] · Pål Roe Sundsøy[5] ·
Md. Nadiruzzaman[6,7] · Erik Wetter[2,8] · Asif Iqbal[5] ·
Taimur Qureshi[5] · Andrew J. Tatem[2,9] ·
Geoffrey S. Canright[5] · Kenth Engø-Monsen[5] ·
Linus Bengtsson[1,2]

**Abstract** Large-scale data from digital infrastructure, like mobile phone networks, provides rich information on the behavior of millions of people in areas affected by climate stress. Using anonymized data on mobility and calling behavior from 5.1 million Grameenphone users in Barisal Division and Chittagong District, Bangladesh, we investigate the effect of Cyclone Mahasen, which struck Barisal and Chittagong in May 2013. We characterize spatiotemporal patterns and anomalies in calling frequency, mobile recharges, and population movements before, during and after the cyclone. While it was originally anticipated that the analysis might detect mass evacuations and displacement from coastal areas in the weeks following the storm, no evidence was found to suggest any permanent changes in population distributions. We detect anomalous patterns of mobility both around the time of early warning messages and the storm's landfall, showing where and when mobility occurred as well as its characteristics. We find that anomalous patterns of mobility and calling frequency correlate with rainfall intensity ($r = .75$, $p < 0.05$) and use calling frequency to construct a spatiotemporal distribution of cyclone impact as the storm moves across the affected region. Likewise, from mobile recharge purchases we show the spatiotemporal patterns in people's preparation for the storm in vulnerable areas. In addition to demonstrating how anomaly detection can be useful for modeling human adaptation to climate extremes, we also identify several promising avenues for future improvement of disaster planning and response activities.

**Keywords** Climate change adaptation · Migration · Resilience · Mobile network data · Anomaly detection · Disaster risk

## 1 Introduction

The increasingly robust evidence base in climate sciences relies on the measurement of normal trends and analysis of deviations (Bindoff et al. 2013). Techniques for detecting anomalies have produced key findings on changing atmospheric and surface temperature (Jones et al. 1999; Mann et al. 1998), oceanic circulation (Hurrell 1995; Thompson and Wallace 1998), arctic temperatures and ice cover (Serreze et al. 2000; Stroeve et al. 2007; Vinje 2001), intensity of tropical rainfall and cyclones (Knutson et al. 2010; Trenberth 2011); and seasonal variability and extremes (Seneviratne et al. 2012). Anomaly detection principles have also shown how earth's ecosystems (Lucht et al. 2002; Stenseth et al. 2002) and biota including agriculture (Lenoir et al. 2008) have responded to climate change. However, human behavior in response to disasters also deviates from normal behavioral patterns. In this paper, we aim to use anomaly detection to investigate behavioral responses in a human population exposed to an extreme weather event.

Vulnerable people in low- and middle-income countries respond to weather extremes associated with climate change, such as tropical cyclones and flooding, with a variety of behaviors that appear anomalous against a baseline (here termed "adaptations") such as moving animals to safety, harvesting crops early, reinforcing and repairing flood embankments, and changing household spending behaviors. In more extreme cases, short-term adaptive responses include evacuation and displacement. Weather extremes can, in the long-term, undermine livelihoods, push people into poverty, and elicit an extraordinary adaptive response in these circumstances: permanent migration (Black et al. 2011; Brouwer et al. 2007), a subject of rich academic debate (summarized in (Black et al. 2013) featured centrally in 5th Assessment of the IPCC (Adger et al. 2014; Olsson et al. 2014). Unfortunately, our ability to detect anomalous human behaviors is not on par with our large-scale measurements of biophysical systems at relevant temporal and spatial scales.

Climate science has seen rapid progress in the measurement, and prediction of changes and extremes in biophysical systems in high resolution across geographic and temporal scales. To understand the impacts of climate change on human society it is imperative to measure anomalous behavioral responses as they coincide with hazards at the common spatiotemporal scales in which they occur (Palmer and Smith 2014). This is especially crucial where people are dependent on stable environmental conditions for livelihoods, and where both climate change and the burden of adaptation threaten human security and development (Adger et al. 2014; Field et al. 2014). Methodologies that focus on large-scale spatial indicators of both human behavioral and environmental change, and make use of temporally adjusted longitudinal data are required to establish baselines and link short-term responses and long-term outcomes (Palmer and Smith 2014).

As of the end of 2014, mobile networks served a total of 3.6 billion unique mobile subscribers, roughly half of the global population (GSMA Intelligence 2015). Mobile operator data are updated in close to real-time and have a vast geographic reach. The data generated from mobile operators enable measurement of some characteristics of social networks, migration, and patterns of household economic behavior at a previously unprecedented scale (Bagrow et al. 2011; Palmer and Smith 2014; Zolli 2012). Operator data has been used during relief operations after the Haiti 2010 earthquake (Bengtsson et al. 2011; Lu et al. 2012) and cholera outbreaks (Bengtsson et al. 2015) and the Nepal 2015 earthquake (Wilson et al. 2016), making them a very promising proxy indicator for measuring impacts of climate change, and weather extremes. In Rwanda, retrospective analyses of network activity was used to estimate the epicenter of an earthquake and to infer humanitarian needs in the weeks after the earthquake (Kapoor et al. 2010). Likewise, Blumenstock and colleagues identified unusual patterns of person-to-person transfers of airtime credits through social networks to identify a geographical pattern of earthquake impact (Blumenstock et al. 2011). Anomaly detection methods have previously been applied to mobile network data to identify unusual calling patterns after floods (Pastor-Escuredo et al. 2014), and in the interest of improving the normal operation of mobile networks (Karatepe and Zeydan 2014). They have been used for anomaly detection for detecting and classifying social disturbances, like conflict and violence in data-poor circumstances (Dobra et al. 2014; Young et al. 2014). One study showed the diffusion of anomalous calling patterns through intimate social networks in the wake of a terrorist bombing in Oslo (Sundsøy et al. 2012). Various studies have concluded that in the wake of disasters anomaly detection could reduce the cost, increase timeliness and improve the geographic focus of emergency response activities (Candia et al. 2008; Pawling et al. 2007).

The extreme South of coastal Bangladesh, with its low elevation and routine exposure to intense tropical cyclones, exemplifies an area with high climate pressure and is a fitting location to explore mobile network data before and after climate shocks. We searched for anomalous patterns of phone usage that could provide insight into adaptive preparations and responses (Martin et al. 2014; McGranahan et al. 2007; Penning-Rowsell et al. 2013), and examined how spatial and temporal patterns in large sets of operator data from the Grameenphone mobile network in Bangladesh around tropical cyclone Mahasen could inform impact assessment and adaptation in cyclone affected areas. We investigated three hypotheses. First, anomalous patterns of calling frequency represent the affected populations' physical contact with the storm in the most affected areas during landfall. Second, as communication is an important tool during an environmental crisis, we hypothesized that anomalous mobile recharges represent behaviors of people preparing for impacts in the most vulnerable areas. Finally, we hypothesized that cyclones drive anomalous flows of users between towers, indicating evacuation, displacement and migration.

## 2 Cyclone Mahasen

Cyclone Mahasen struck Bangladesh on 16 May 2013. Before landfall it moved northward along the Bay of Bengal. Forecasts estimated a landfall in the heavily populated Chittagong

District, and the government's Comprehensive Disaster Management Programme concentrated early warnings there. However, in the final hours of 15 May, the storm veered to the north, making landfall over the rural Barisal Division, at approximately 3:00 a.m. on 16 May (Fig. S1a). During the course of 16 May, Mahasen moved eastward along the coast into Chittagong, and northward into India, where rainfall and wind speed rapidly diminished (Gutro and Pierce 2013).

Mahasen was a relatively weak storm compared to earlier cyclones in Bangladesh, such as Aila and Sidr (REACH Initiative 2013). While it affected an estimated 1.3 million people (REACH Initiative 2013) and impacts on crops and homes were extensive, the death toll was relatively small. Seventeen perished in the storm, mostly from falling trees, and unlike previous storms, no fishermen were lost (Associated Press 2013). The minimal loss of life was regarded as a major victory for the Comprehensive Disaster Management Programme's early warning system (UNDP 2013).

## 3 Cyclone impacts and population-level adaptation

### 3.1 Mobile phone data

We used a de-identified set of call detail records (CDRs) from 5.1 million Grameenphone users collected between 1 April and 30 June of 2013 in the Barisal Division and Chittagong District of Bangladesh. The dataset began six weeks before the landfall of Cyclone Mahasen (16 May 2013) and continued for six weeks after landfall (1 April to 30 June 2013) (Fig. S1). CDRs are compiled by network operators principally for the purposes of billing customers for their use of the network. De-identified data entries include information on the time of the call, the mobile phone tower used and the duration of call, and can thus be used to indicate the geographical position and movements of users. To limit potential biases resulting from subscriber churn, and new users entering the dataset due to impacts of the storm, we limited the study to SIM cards that had placed at least one call before the cyclone landfall (16 May); and also made at least one call in the last ten days of the data collection period (21–30 June).

Since Mahasen was a relatively weak cyclone, the performance of the Grameenphone network remained virtually undisturbed during and after landfall, guaranteeing continuous relay of CDR throughout the study period. An analysis of tower function anomalies appears in the *Supporting Information* (S2), along with a general discussion on the Grameenphone network, the dataset, and the representativeness of data for the general population (S1).

### 3.2 Calling frequency and rainfall measurements

During "normal" circumstances, which we defined as the average calls per hour for any given hour across the data set, regular daily and weekly cycles of calls were apparent (Fig. 1a). Users concentrated phone usage in the daytime hours, with a spike occurring toward late evening. A small shift in the temporal distribution of calls occurred on Fridays (the first day of the weekend) when calls began later in the day. Increases in calling frequency coincided with several events within the data set, most notably on 25 June, when a major religious festival, Shab-e-Barat was celebrated. Likewise smaller increases coincided with the Bengali New Year in early April and a series of protests in early May.

However in the early hours of Thursday 16 May 2013, as Mahasen made landfall across Barisal, we observed a dramatic increase in call frequency relative to "normal," which we defined

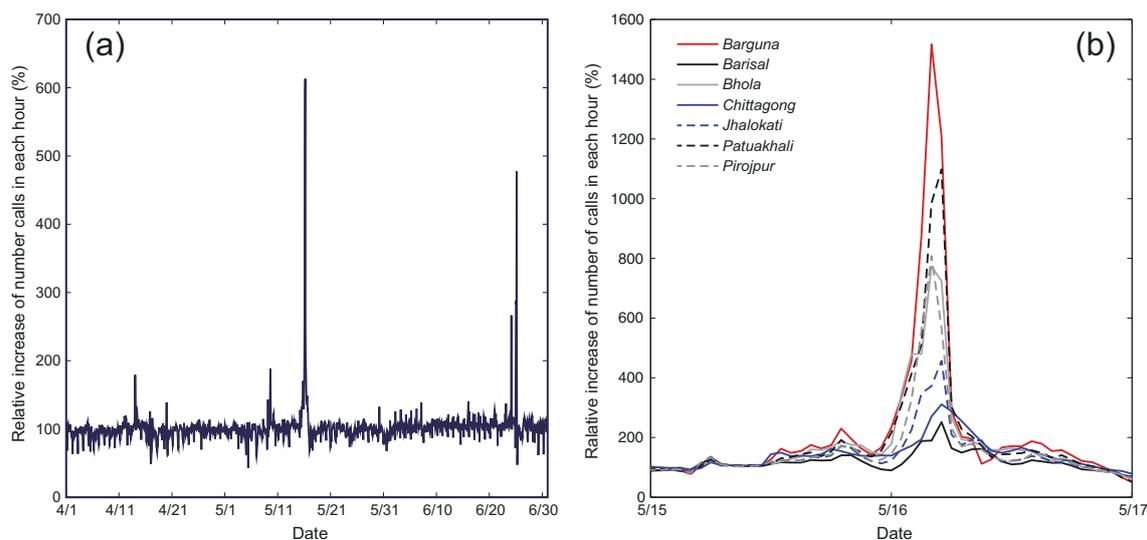

**Fig. 1** Change in call frequency. **a** For each hour, the number of calls is compared to the average number of calls made during that same hour across the whole period. Relative calling frequency spikes on 16 May as Mahasen makes landfall. **b** Calling frequency during cyclone landfall at the district level. For each of the seven districts, the change of calling frequency in the morning of 16 May is shown

as calls per hour compared with the same hour on all Thursdays in the dataset (Fig. 1a). Deconstructing calling frequency among the more vulnerable coastal districts (Barguna, Bhola, Patuakhali, Pirojpur), we saw calling volumes increase by at least seven times the average level (Fig. 1b). In Barguna, calling frequency increased by a factor of 15. Throughout the evening and early hours, a spatiotemporal pattern emerged in peak calling frequency. In Barguna and Pirojpur, in the extreme south and west of Barisal, peaks occurred between 3:00 and 4:00 a.m., while in the northern districts of Jhalokati and Barisal, the peak occurred between 4:00 and 5:00 a.m. This suggests that call frequency corresponded with the physical manifestations of the cyclone as it moved over Barisal Division from the South West. One alternative explanation is that people might call friends and relatives when a cyclone is approaching in order to communicate concerns for wellbeing, encourage evacuation plans, and coordinate preparations. However, between 00:00 and 6:00 a.m. at the time Mahasen was making landfall in Barisal, in the areas where the cyclone was predicted to make landfall (the Chittagong district), calling frequency was close to normal levels. These differences provide support for the hypothesis that calling frequency represented a behavioral response to sensory experience of the storm.

To further investigate the relationship between calling frequency and physical manifestations of the storm, we conducted a spatiotemporal comparison of calling frequency with rainfall data from NASA's Tropical Rainfall Measurement Mission (TRMM) satellite. The TRMM satellite passed over Bangladesh at 3:32 a.m., measuring rainfall during the cyclone's landfall, reaching 67 mm per hour in some areas (Gutro and Pierce 2013). Locations of maximum rainfall were clearly correlated with locations of maximum increase in calling frequency (Fig. 2). Even areas with moderate rainfall, for example a narrow band of rainfall to the east of Chittagong (Fig. 2a) also exhibited an increase in calling frequency (Fig. 2b). This adds supporting evidence that clusters of high calling frequency represented contact with the cyclone's most severe physical effects.

Past research has shown that rainfall data alone is often too low resolution and intermittent to make any inferences about cyclone damage (Auffhammer et al. 2013). Detailed spatiotemporal data on call frequency may improve inferences about the effect of weather extremes on vulnerable people, and is identified here as an area for future research.

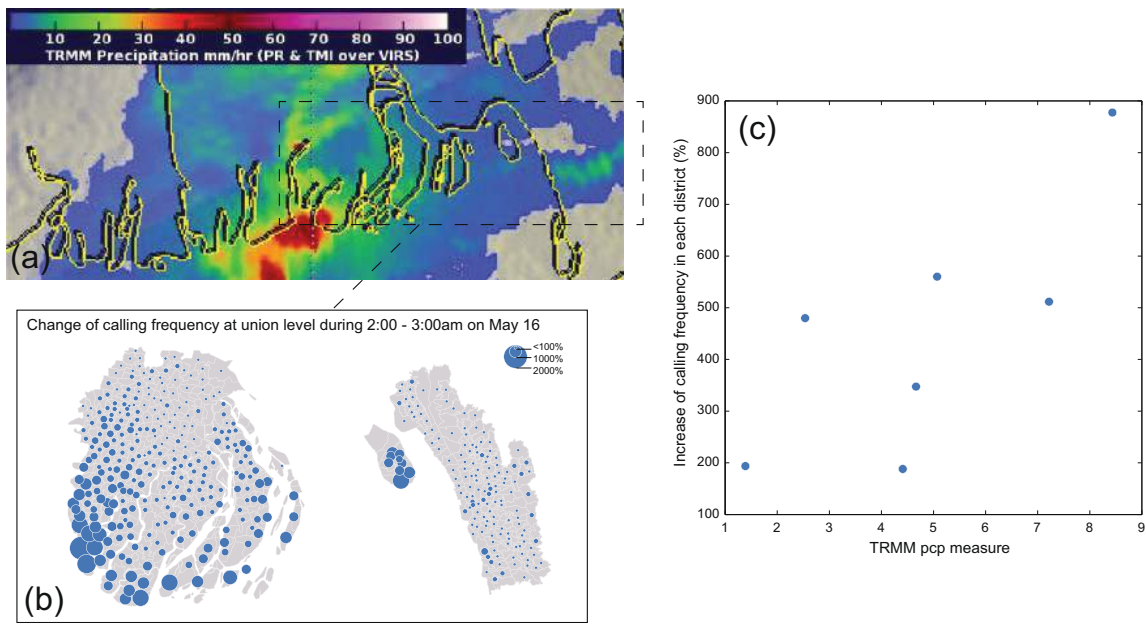

**Fig. 2** Call anomalies and rainfall. **a** Precipitation measurements from NASA's Tropical Rainfall Measurement Mission captured at 3:32 a.m. show the distribution of rainfall in the study area, reprinted from Gutro and Pierce 2013, with permission from the authors. **b** The geographical distribution of call frequency at 3:00 a.m. on 16 May. **c** Rainfall is plotted with calling frequency at the district level. Correlation coefficient = 0.75, $p = 0.05$

### 3.3 Recharge behaviors

Next, we investigated how mobile recharges or top-ups can complement call frequency to provide insight on how vulnerable people prepare for climate impacts. To accomplish this, we relied on a second data set, consisting of mobile recharge purchases from 892 retailers in Barisal and Chittagong Divisions during the original three-month timeline, 1 April to 30 June of 2013 (Fig. 3). Recharges are the amount of money that users credit to their SIM card to access the network. They allowed an investigation of the geographic distribution of changes in expenditures before and after the cyclone. In Bangladesh, mobile credits represent a small but significant proportion (~3 %) of the household budget (Lucini and Hatt 2014), and disasters increase demand for private communication (Samarajiva 2005). We hypothesized that spikes in recharges represented knowledge of the cyclone and preparations for its impacts.

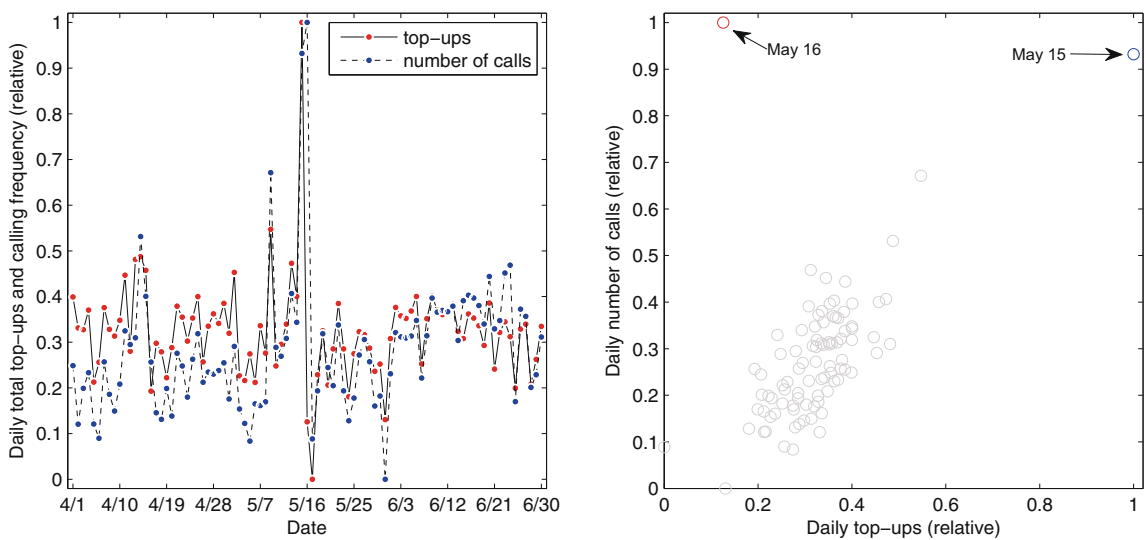

**Fig. 3** Comparison of daily recharges and calling frequency, both features are presented with min-max normalization. **a** During the cyclone, while calling frequency is high for both 15 May and 16 May, recharges are high only on 15 May and then drop to low levels on 16 May. **b** Excluding 16 May, there is a strong linear relationship between recharges and calling frequency (corr without 16 May = 0.798, $p < 0.000$, corr with 16 May = 0.576, $p < 0.000$), implying that recharges indicate users preparing for a potential disaster

In the second half of 15 May, as forecasts and early warnings were transmitted across radio and television, a large increase in recharges is evident, coinciding with a high volume of calls placed on the same day (Fig. 3). However even as calling frequency remained high on 16 May, recharges dropped below the predicted level, and continued at low levels during the following day. This suggests that users recharged their phones as part of their storm preparation and awareness of vulnerability, planning for the need to communicate with family and friends during and after the cyclone.

### 3.4 Estimating evacuation, displacement and migration

Usage patterns in the data also enabled us to analyse short-term features of evacuation, displacement and migration, which would be extremely hard to quantify using standard survey-based research but were readily apparent in CDRs. Using CDRs and tower locations to identify moving SIM cards, we created a series of mobility networks, which quantify the direction, volume and distance of flows between locations at specified time intervals before and after Cyclone Mahasen (Figs. 4 and 5). Note that the mobility networks during normal periods were almost perfectly symmetrical, meaning that the numbers of users entering an area are roughly equal to the number of users leaving an area (Fig. 4). In contrast, anomalies appeared as larger than normal flows in one or both directions (Fig. 5), and indicated spatiotemporally explicit patterns of movement, such as evacuation, displacement and permanent migration that took place at specific moments coinciding with the storm. Because asymmetrical flows might also represent, for example, the onset of migration season, a calendar festival or a popular protest, it is important to be cautious in assigning causation.

Prior to the storm, large changes in the flow network were notable in Chittagong City, as people evacuated in response to the forecasts that Mahasen would make landfall over

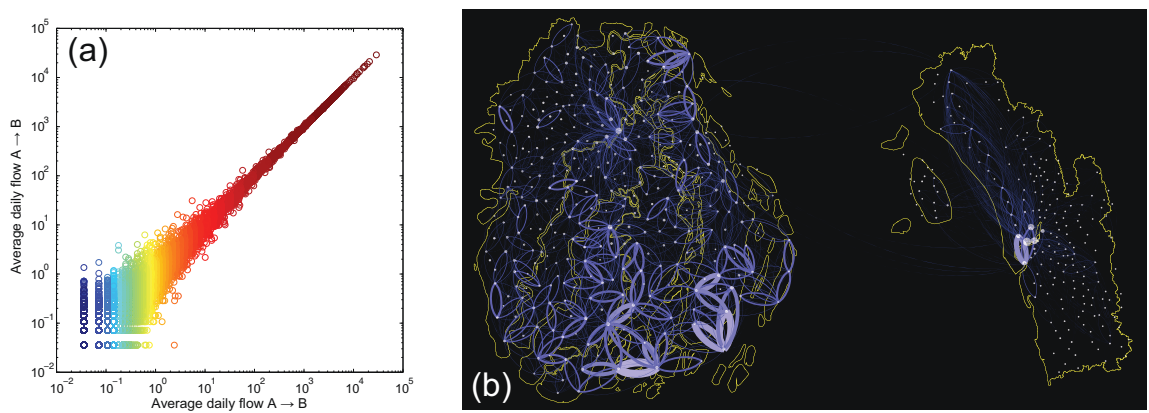

**Fig. 4** Symmetry in "normal" flow networks. **a** The average daily flow between each union in both directions between April 1, to April 28 (i.e., for four "normal" weeks prior to the storm) are shown. The correlation is extremely high (corr = 0.999, $p < 0.000$), indicating a high level of symmetry in "normal" mobility patterns, i.e. each day, roughly as many people leave an area as those who enter it. **b** This symmetry of "normal" flow can be represented geospatially. SIM cards are included only if they accessed more than one tower in a day. Links indicate areas where 10 or more movements were observed at distances greater than 10 km. Direction of flow is clockwise from the point of origin

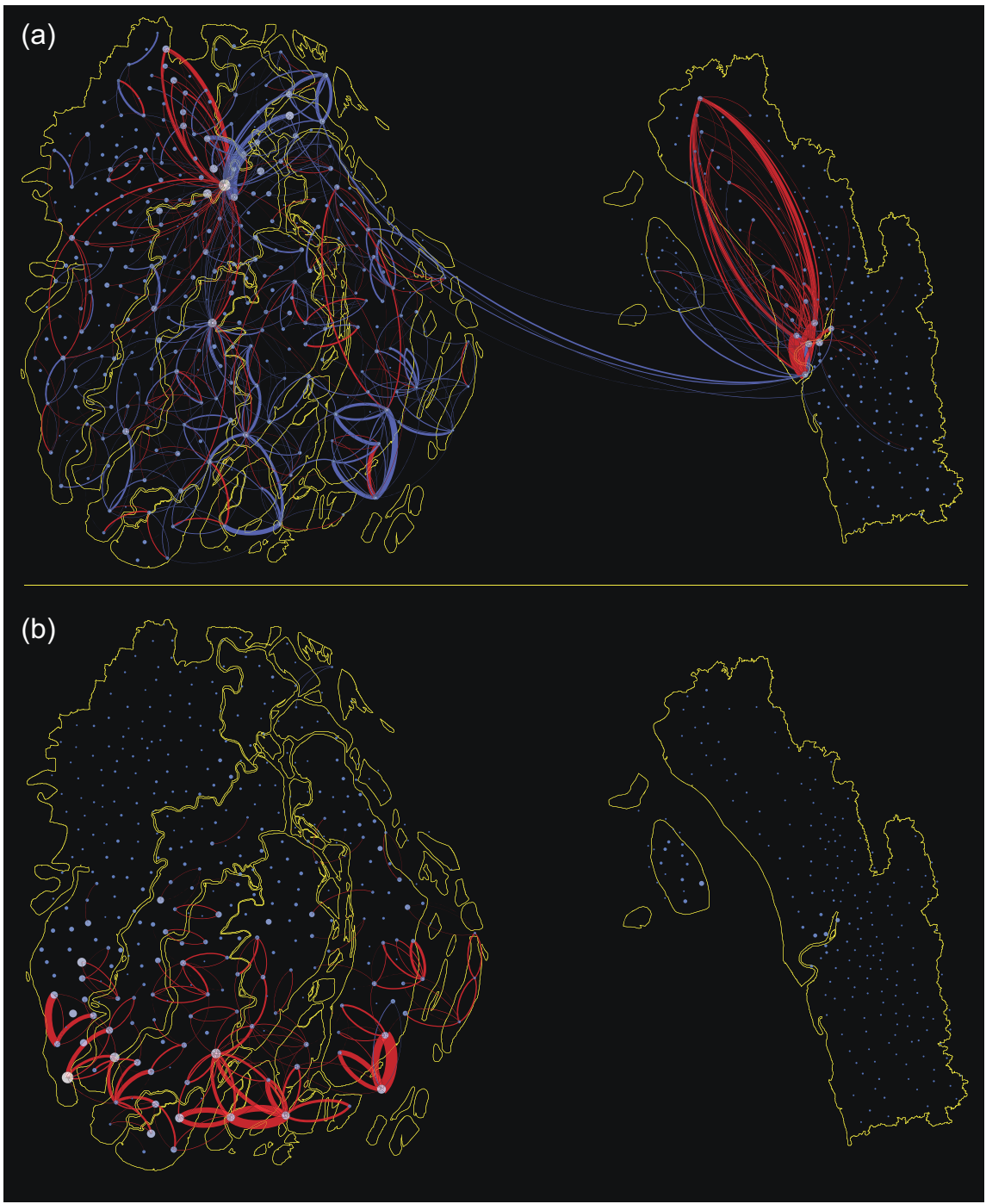

**Fig. 5** Evacuation and landfall flow networks. **a** The mobility network on 15 May (the day prior to landfall) is compared with 24 April (3 weeks before the storm during the same hourly period). Positive flows are shown in *red*, indicating increased flow on 15 May, while negative flows are shown in *blue*, indicating decreased flow on 15 May. Thickness of link represents relative volume of flow. To appear in the flow network, a user had to make at least two calls. Each SIM contributed only one movement (the first and last observed location). Links indicate areas where 10 or more movements were observed, at distances greater than 10 km. **b** The mobility network during landfall on 16 May, 00:00–6:00 a.m., is compared with 25 April (3 weeks prior during the same hourly period). Unusual mobility is observed in the affected area, where warnings were not concentrated

Chittagong (Fig. 5a) Meanwhile, there were less than normal flows in Barisal at the same time, suggesting people were not evacuating to other areas in large numbers, but rather suspending regular trips.

In the early hours of May 16, during cyclone landfall, at the time when people should have been in shelters, above normal flows of SIMs were evident in the margin of sub-districts in Barisal nearest to the coast, indicating that people were moving about at night, during the storm (Fig. 5b). This suggests that people evacuated too late, and would have been in danger if the storm intensity had been greater. Mobility patterns in Barisal during landfall contrasted sharply with mobility in Chittagong during the same time, where patterns were virtually unchanged from normal flows for that day and hour. Although the exact explanation for these differences is unknown, officials in the Ministry of Disaster Planning and Response indicated that early warnings were not made in Barisal until too late because all forecasting indicated that the storm would make landfall in Chittagong (Nadiruzzaman 2013). Other possible explanations for delayed evacuation in rural areas were that men commonly stay behind to look after livestock and protect homes and assets from thieves.

In sum, the mobility patterns evident in mobile network data allow researchers to perform an audit of early warning program effectiveness on the basis of early and mid-storm population movements. In this case, the early warning system in Chittagong apparently accomplished the aim of motivating evacuation during appropriate times.

## 4 Quantifying impacts and behavioral responses using anomaly detection techniques

To automatically detect human behavioral changes in our study, we used a sigma-model to evaluate the stability of the observed sequence of activities extracted from customers' usage data in the mobile network. Specifically, for each time series of a quantified activity, $E = \{e_1, e_2, e_3, ..., e_t\}$, in which $e_i \in R (1 \leq i \leq t)$ is the measure of the activity (e.g., number of calls at time $i$, and $t$ the length of evaluated time window) we highlight time points $I = \{1 \leq i_1, i_2, ..., i_M \leq t\}$ in which each observation $e_{i_m}$ at time point $i_m$ exceeds three standard deviations from E's average during the time period.

As the studied activity reached the predetermined thresholds, three standard deviations from the mean, it was flagged as anomalous. In this way, any unusual patterns of network usage could be identified, and further analysis would determine what these anomalies represented about cyclone impacts. When the number of anomalous cases is very large, the procedure may result in considerable false positives (type I error) (Candia et al. 2008). To avoid this limitation, we also calculated the total number of anomalies detected (see Fig. 6a).

### 4.1 Anomaly detection for calling frequency

Unusual calling patterns provide a measure of behavioral response to storm severity. In the timeline, several clusters of calling frequency anomalies were observed (Fig. 6). The first occurred on 14 April, the Bengali New Year, followed by a drop the following day. A second cluster occurred on 9 May around an infamous series of national protests, dubbed "the Siege of Dhaka" in which several dozen people were killed across the nation in a series of violent protests. The next two clusters coincided with Mahasen, which made landfall on 15 and 16 May, and a cold front, which flooded the southern coast between 30 May and 1 June. Finally a large spike on 25 June coincided with Shab-e-Barat, an important religious festival, when people commonly call their relatives.

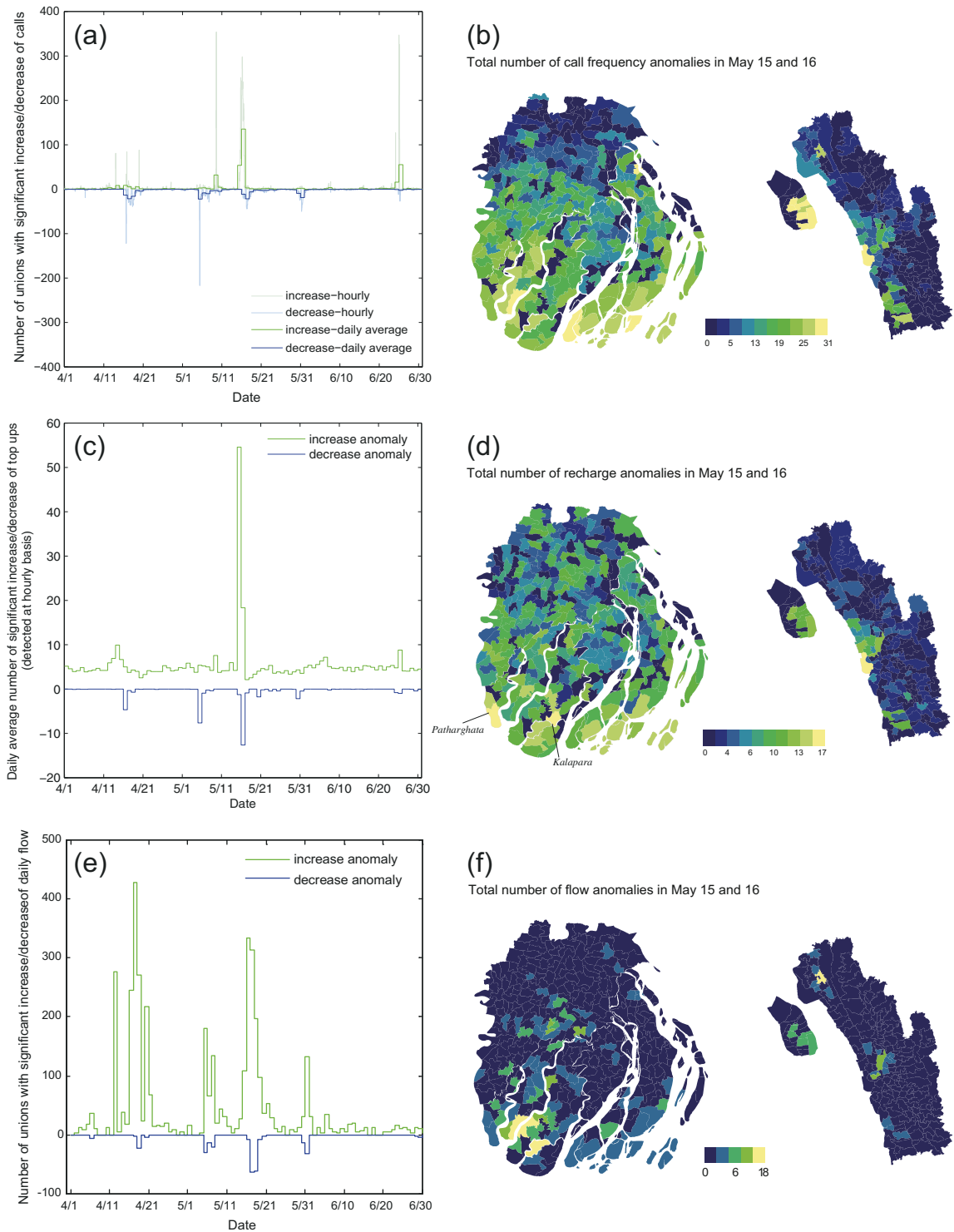

**Fig. 6** The temporal and spatial distribution of anomalies in calling frequency (**a**, **b**), recharge anomalies (**c**, **d**), and flow anomalies (**e**, **f**). The threshold for detection was set at three standard deviations from the mean of baseline. In the flow anomaly analysis, the restriction for including flows in the investigation was that they were positive for each day in the timeline

The most profound and longest lasting set of anomalies coincided with Mahasen. With few exceptions, calling frequency anomalies were concentrated along the vulnerable coastline, and in areas where the storm made landfall (Fig. 6b). These anomalies spatiotemporally coincided with cyclone landfall, and indicated when and where behavioral response to the physical cyclone were strongest.

Typically, post-cyclone damage assessments, which determine the form that disaster relief should take, rely on the reporting of damage by local officials. These rapid reports typically form the basis for selection of areas in which more detailed information on impacts and needs may be collected through household surveys. However systematic biases and delays can be introduced at various stages of the assessment process (Hallegatte and Przyluski 2010) due to limited capacity of responding agencies in many resource constrained settings. This analysis indicates how mobile network data could be used to overcome potential biases in the site selection portion of post-cyclone damage assessments by indicating when and where impacts have occurred.

### 4.2 Anomaly detection relating to recharge behavior

Mobile recharge anomalies differ from other anomalous behaviors detected in the timeline in two ways: they are almost exclusively concentrated around the cyclone impact zone, and occured before the event, indicating foreknowledge of the coming cyclone and preparation for its impacts. As with calling frequency, recharge anomalies were concentrated along the vulnerable coastline, where the cyclone first made landfall (Fig. 6). But whereas calling frequency anomalies were detected widely across the region, mobile recharge anomalies were concentrated in Kalapara and Patharghata (Fig. 6d), areas noted as pockets of exceptional vulnerability within this already vulnerable landscape (Ahamed et al. 2012), where impacts of Cyclones Sidr (2007) and Aila (2009) were most severe. This suggests that anomalous preparation behaviors can reveal the areas where people perceive themselves to be vulnerable (Fig. 6d).

Alternative explanations include that these were areas where people had greater access to recharge vendors and spending money, however if this were the case, similar anomalies would have been observed in other urban centers of similar size. It is plausible that people undertook other anticipatory actions in areas where recharge anomalies were detected, but further study is required to link mobile recharges to overall disposition toward cyclone preparedness.

### 4.3 Anomaly detection relating to population movements

Forms of human mobility that deviate from normal patterns can also be detected from mobile network data (Fig. 6). To identify anomalous flows, we investigated daily flow between each pair of locations (unions) during the whole period. If the daily flow between unions A and B exceeded three standard deviations from the mean for that weekday during the study period, a signal was generated for both unions A and B. Since a substantial number of location pairs normally have low flows, false alarms could result from small absolute increases. While anomalous decreases in daily flow were less likely to produce false alarms, there is potentially more noise in increases due to the small flows which normally pass between some locations. For simplicity, to decrease noise and to obtain a conservative measure of flow anomalies, we therefore focused on pairs of locations having non-zero flow during all days during the whole period.

As with call frequency anomalies, the flow anomalies detected in this analysis corresponded with the significant events in the time line: Bengali New Year (14 April), nation-wide protests (8 May), Cyclone Mahasen (16 May), and heavy rainstorms (30 May) (Fig. 6e). The largest cluster of anomalous flow increases coincided with the Bengali New Year, where very few anomalous flow decreases were simultaneously observed. Both anomalous flow increases and decreases were apparent between 15 and 19 May, before and after Mahasen struck, nevertheless the frequency of anomalous increases was eight times that of decreases. Note that the areas with most anomalous flow events coincided very well with the area in which rainfall intensity was the highest during cyclone landfall (Fig. 2).

## 5 Conclusion

In this paper, we show how data from mobile networks provides insights into behavioral responses to Cyclone Mahasen and its impacts. We show that anomalous patterns of calling frequency are correlated with rainfall intensity at local scales, likely providing a defined spatiotemporal measure of users' physical exposure to the storm. We show that mobile recharge purchases increase in vulnerable impact zones before landfall, representing preparations for potential environmental hazards. We also identify anomalous patterns of mobility during evacuation and storm landfall, indicating how people respond to storm forecasts and early warnings. The analysis is in agreement with the official joint needs assessment, which saw little evidence of mass displacement. We also show how, in future applications, anomalous flows of SIM cards between mobile towers can provide a much needed audit of the effectiveness of forecasting and early warnings systems, and indicate the new locations of displaced people. Rapid, cost-effective and accurate tools for assessing the effectiveness of early warning systems, and indicating the location of displaced people are currently in short supply.

Based on comparisons with rainfall measurements at landfall, and considering the substantial weakening of cyclones as they pass over land, calling frequency and population movement anomalies seemed to provide the best proxy indicators for cyclone impacts among those evaluated. Traditional methods for assessing cyclone impacts and human behavioral responses have well known limitations (Hallegatte and Przyluski 2010), and the anomaly detection technique applied to mobile network data presented here (building on work of Blumenstock et al. 2011; Candia et al. 2008; Dobra et al. 2014; Kapoor et al. 2010; Pawling et al. 2007; Sundsøy et al. 2012, and Young et al. 2014), may overcome some of these challenges, and demonstrates the potential value of mobile network data as a complement to current cyclone impact assessment tools. Specifically, the spatiotemporal distributions of anomalous usage activity could be used to improve the timeliness and cost-effectiveness of cyclone impact assessments. Data from mobile networks may be very useful as a tool to prioritize locations in which rapid needs assessments are performed after cyclone landfall, with the potential to drastically reduce the time to reach those most in need.

While the study provided a robust analysis of the behavior of Grameenphone subscribers, the primary limitations of the study involved the representativeness of data for the general population. However, the general features of behavior change that we found to be most useful, i.e. sharp increases in calling frequency and changes in mobility, may well result independently of mobile operator and are likely to reflect natural human responses to shocks. Likewise, the study concentrated on Mahasen, a relatively Cyclone, which despite maximum rainfall of 68 mm/h dissipated quickly. Findings cannot be generalized about larger, more energetic cyclones, where storm surges and flooding can cause greater destruction. Finally, other causes of increased calling frequency and mobility than those indicating a need for post-disaster assistance may exist after a disaster, and thus network data should, at our present level of understanding, be used as a complement to, not a replacement for, other information sources.

To overcome these limitations and to better understand the effects of multiple types of environmental disruption, future work should incorporate mobile network data covering longer time spans. Longitudinal household measures of storm impacts and improved environmental impact models can provide external validation of the spatiotemporal patterns of anomalous usage that are apparent in the mobile network data. Additionally, as we illustrate in the Supporting Information (S2), analysis of other aspects of network function, such as service interruptions, which do not convey information on human behavior, may still provide a proxy for spatiotemporal damage to infrastructure.

Detecting anomalous usage patterns from mobile network data is a promising avenue for researching human behavioral responses to impacts associated with climate change across large spatiotemporal scales. Data from mobile networks may become an important tool for prioritizing areas for rapid needs assessments following cyclones.

**Acknowledgments** Grameenphone provided data, analysis and dedicated support to the project. The Bangladesh Telecommunications Regulatory Commission provided guidance and assured that the project adhered to regulatory standards. Rockefeller Foundation financed the work. DW and MN acknowledge Munich Re Foundation, United Nations University-Institute for Environment and Human Security, and the International Centre for Climate Change and Development for salary support. XL acknowledges the Natural Science Foundation of China under Grant Nos. 71301165 and 71522014. AJT is supported by Wellcome Trust Sustaining Health Grant (106866/Z/15/Z) and funding from the Bill and Melinda Gates Foundation (OPP1106427, 1032350, OPP1134076, OPP1117016). LB was funded by the Swedish Research Council. Stephen Roddick helped to assemble the project team.

# Supporting Information:

# Detecting climate adaptation with mobile network data in Bangladesh: Anomalies in communication, mobility and consumption patterns during Cyclone Mahasen

*Xin Lu*[a,b,c], *David J. Wrathall*[d], *Pål Roe Sundsøy*[e], *Md. Nadiruzzaman*[f,g],
*Erik Wetter*[b,h], *Asif Iqbal*[e], *Taimur Qureshi*[e], *Andrew Tatem*[b,i], *Geoffrey S. Canright*[e],
*Kenth Engø-Monsen*[e], *Linus Bengtsson*[a,b]*

[a]*Department of Public Health Sciences, Karolinska Institutet, Stockholm, Sweden;*
[b]*Flowminder Foundation, Stockholm, Sweden;*
[c]*College of Information System and Management, National University of Defense Technology, Changsha, China;*
[d]*Oregon State University, College of Earth, Ocean and Atmospheric Sciences, Corvallis, Oregon, USA;*
[e]*Telenor Research, Oslo, Norway;*
[f]*Department of Geography, University of Exeter, Exeter, UK;*
[g]*International Centre for Climate Change and Development, Dhaka, Bangladesh;*
[h]*Stockholm School of Economics, Stockholm, Sweden;*
[i]*WorldPop, Department of Geography and Environment, University of Southampton, Southampton, UK.*

*Linus Bengtsson: linus.bengtsson@flowminder.org

## S1. Dataset Characteristics

De-identified call detail records (CDRs) from 5.1 million Grameenphone users were collected between 1 April and 30 June of 2013 in the Barisal Division and Chittagong District of Bangladesh. The data begins six weeks before the landfall of Cyclone Mahasen (16 May 2013) and continues for six weeks after landfall (1 April to 30 June 2013) (Fig. S1a). CDRs are compiled by network operators principally for the purposes of billing customers for their use of the network. De-identified data entries include information on the time of the call, the mobile phone tower used and the duration of call, and can thus be used to indicate the geographical position and movements of users. To limit potential biases resulting from subscriber churn, and new users entering the dataset due to impacts of the storm, we limited the study to SIM cards that had placed at least one call before the cyclone landfall (16 May); and also made at least one call in the last ten days of the data collection period (21-30 June).

A distribution of 986 towers across the Barisal Division and Chittagong District forms the basis for our spatiotemporal analysis of calling frequency, mobility and top-up behaviors. We assigned users a position within the network based on the tower through which his or her most recent call was routed (Fig. S1a). The coverage area of a mobile tower (BTS, base transceiver station) can range from a few hundred meters in a city, up to tens of kilometers in rural areas. The location of tower positions were further scrambled within 200 meters. The average and maximum distance of each tower and its nearest neighbor tower, were 2.1 km and 16.8 km, respectively.

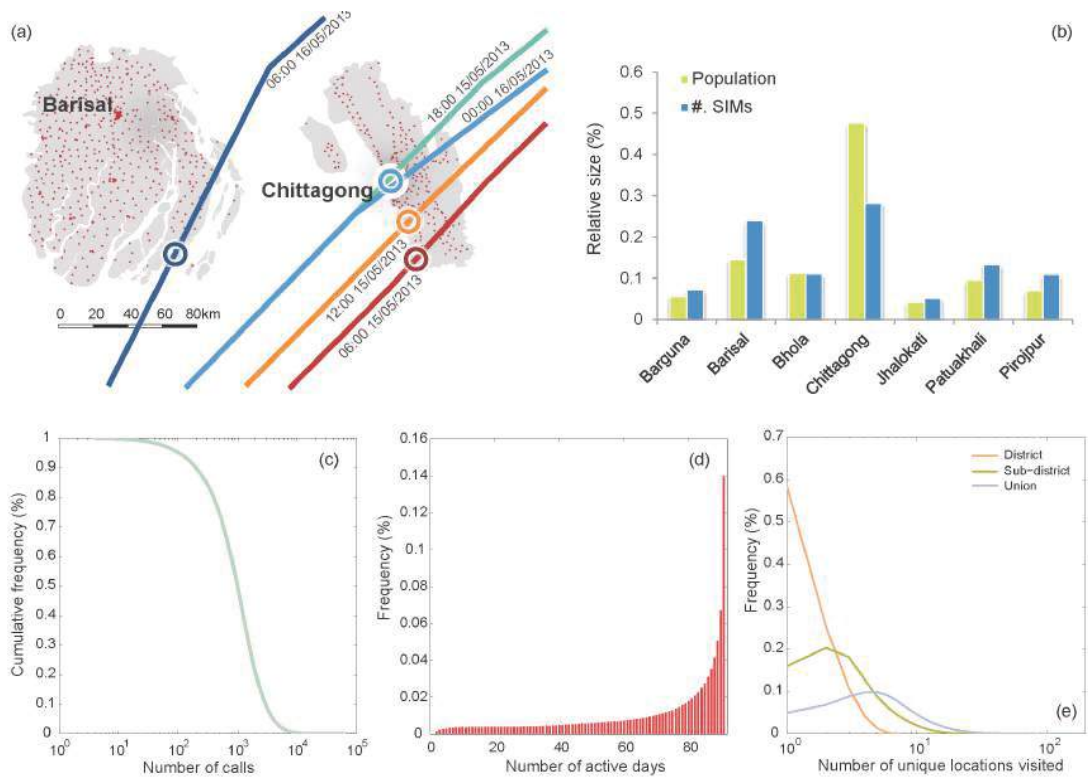

*Fig. S1 Study area, distribution of users, distribution of calls, number of active days, and unique locations visited. a) Map of study area, including tower map and cyclone path. Colored lines represent the forecasted versus actual cyclone paths. Forecasts predicted landfall over Chittagong, but on its final approach, Mahasen passed over Barisal. Actual versus predicted storm paths allow us to distinguish anticipatory and actual behaviors. b) Distribution of population and number of SIM cards at district level. Population data are taken from the 2011 census, and SIM cards' locations are determined by the district in which they appeared most frequently during evening (10:00pm-6:00am) during the study period. c) A distribution of the number of calls per user shows that 90% of users made at least 200 calls during the three months period, and 50% of users made more than 1000 calls during the period. d) The number of active days on the network shows that 89% of users were active on 30 days, and 50% of users were active on at least 80 days. e) A distribution of unique locations visited shows that 95% of users visited more than one union.*

A comparison of number of SIM cards to population shows high agreement between the number of subscribers and the census figures (Fig. S1b). Deviations are seen in Chittagong district. In Bangladesh, household possession of a mobile phone grew from 78% of all households in 2011 to 89% in 2014, exceeding usage in other countries of similar socioeconomic profiles (National Institute of Population Research and Training (NIPORT) 2015). Users from the Grameenphone network, the largest in Bangladesh, numbered at 42 million in 2013, and comprised 61% of all mobile users nationwide (Telenor 2013). In this dataset, 90% of users made more than 200 calls during the study period and were active 80 days or more out of the 90 day study period (Fig. S1c, d). Likewise, SIM mobility is high: 95% of users appeared in more than one union, 84% in more than one sub-district, and 42% in more than one district (Fig. S1e). For the purposes of this study, we assumed that behaviors exhibited in CDRs are representative for the entire population. While earlier research support this notion, more studies are needed to adequately understand the representativeness of mobile phone users in Bangladesh.

## S2. Quantifying impacts and infrastructure resilience

The functioning of mobile network towers potentially provides an additional proxy for cyclone damage. We undertook an analysis of network function during the study period, investigating towers that were inactive during Mahasen. Because of the potential for damaged equipment and interruption of services, mobile operators invest heavily in network resilience. Therefore in the most powerful cyclones, disabled towers could indicate where damages are concentrated, and could provide a proxy for other infrastructural damages sustained.

In the analysis of towers with no calling activity, we see the same four events that appear in other sections. Focusing on Mahasen, we evaluated the number of inactive towers in Mahasen's impact zone, i.e., towers that registered zero calls, calculated on an hourly basis (Fig. S2). The Grameenphone network held up very well. During Cyclone Mahasen, only 60 towers went offline during landfall followed by 120 towers going offline during the course of the following day. It is likely that rather than suffering damages, towers went offline the following day as power sources failed and reserve batteries became depleted. Notably, patterns of "offline" towers between Mahasen and a heavy downpour that occurred between 30 May and 2 June were remarkably similar (Fig. S2).

In contrast to anomalies in calling frequency, recharges and mobility, tower function does not represent behavioral information, but may provide a potential indicator on the distribution of environmental impact to infrastructure. Mahasen was a relatively weak storm, and as such registered only minor disruption, however during major events, data showing inactive towers or loss of tower function may also indicate the locations where damages are concentrated, helping focus impact assessment on the most damaged sites. Further research is required to make meaningful inferences about the relationship between network function and infrastructural damage.

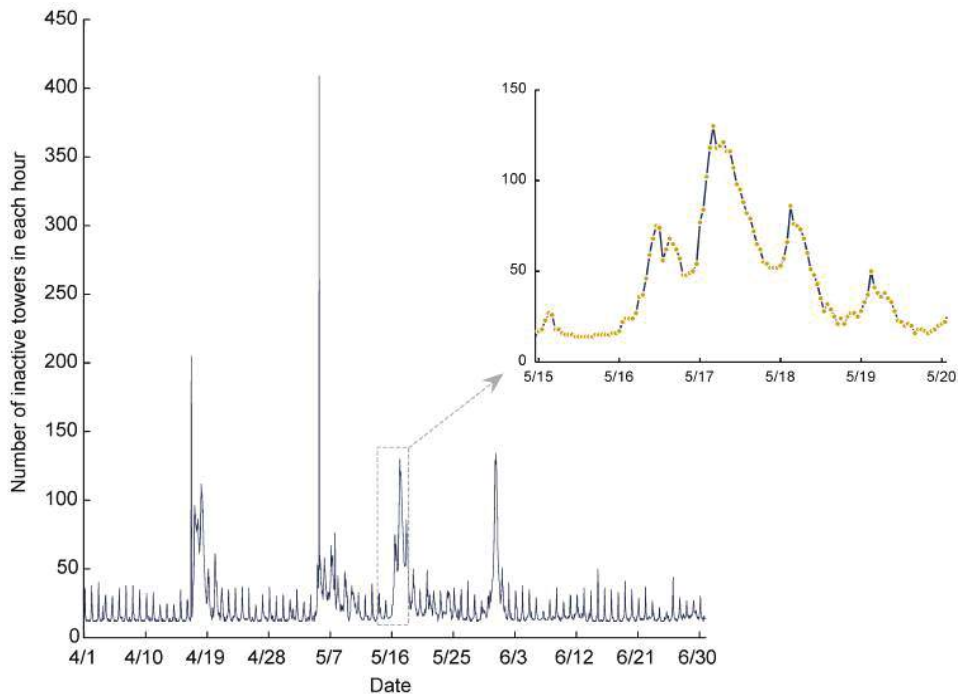

*Fig. S2* *Number of towers with zero calls over the study period. For all filtered SIM cards, the number of calls received by each tower in each hour is calculated. Towers are classified as inactive during any particular hour if they received zero calls. There are four major periods when the number of inactive towers increased: 1) the Bengali New Year, 2) a national demonstration, 3) Cyclone Mahasen, and 4) a severe rainstorm.*

# ◆ Paper 6

# Comparing and visualizing the social spreading of products on a large social network

Pål Roe Sundsøy, Johannes Bjelland, Geoffrey Canright, Kenth Engø-Monsen
Corporate Development, Markets
Telenor ASA
Oslo, Norway
pal-roe.sundsoy@telenor.com

Rich Ling
IT University/Corporate Development, Markets
Telenor ASA
Copenhagen Denmark/Oslo, Norway
rili@itu.dk

*Abstract*— By combining mobile traffic data and product adoption history from one of the markets of the telecom provider Telenor, we define and measure an adoption network—roughly, the social network among adopters. We study and compare the evolution of this adoption network over time for several products – the iPhone handset, the Doro handset, the iPad 3G and videotelephony. We show how the structure of the adoption network changes over time, and how it can be used to study the social effects of product diffusion. Specifically, we show that the evolution of the Largest Connected Component (LCC) and the size distribution of the other components vary strongly with different products. We also introduce simple tests for quantifying the social spreading effect by comparing actual product diffusion on the network to random based spreading models. As videotelephony is adopted pairwise, we suggest two types of tests: transactional- and node based adoption test. These tests indicate strong social network dependencies in adoption for all products except the Doro handset. People who talk together, are also likely to adopt together. Supporting this, we also find that adoption probability increases with the number of adopting friends for all the products in this study. We believe that the strongest spreading of adoption takes place in the dense core of the underlying network, and gives rise to a dominant LCC in the adoption network, which we call "the social network monster". This is supported by measuring the eigenvector centrality of the adopters. We believe that the size of the monster is a good indicator for whether or not a product is going to "take off".

*Social network analysis, Product diffusion, graph theory, eigenvector centrality, telecom, social network structure, viral spreading, data mining, network visualization, iPhone, iPad, Doro, LCC.*

*Revised extended version: 2011.03.25*

## I. Introduction

This paper is an extended version of [29], where new methodology is developed and applied to existing datasets. A new longitudinal dataset is also added. The new methodology quantifies the social spreading effects for transactional products, and is applied on the video telephony data. In addition we enrich the paper by adding spreading studies of the recent iPad 3G tablet.

The current study is motivated by the question of how people adopt new products and services, and what role the underlying social network structure plays in this process. The effect of the social network on product adoption and diffusion has been well documented in early market research, see e.g [9] for an overview of this research. Most of the early studies have suffered from limited network data availability, since social networks have traditionally been difficult to measure. Several theoretical network models have been developed. Some are less realistic due to the evolutionary nature and power law degree distributions [20][21]. Analyses of more realistic models can be found in [14] [22][23].

In recent years massive social network data have been made available to researchers through electronic phone logs [3][6][5][7] and online social network services [2][8]. These studies have confirmed that "the network matters" when customers decide to churn [4][6] and when purchase decisions are made [3][12]. Most of the existing research on product diffusion on networks has been focused on a single product, with a static snapshot of the social network. For an overview of analyses of large networks, the reader is referred to [13]-[19]. A few papers study the evolution of real-world networks [24]-[28]. In this paper we will present an empirical study of how the social network among adopters of telecom-products develops over time. In addition we will show how the product diffusion depends on the underlying social network.

We know that, for many products, a person's adoption probability increases with the number of that person's friends or contacts that have adopted the same product [3][6]. This can be interpreted as inter-personal or social influence, and can be measured empirically. These measurements do not typically say anything about the large-scale structure of the social network. In telecommunications it is possible to obtain detailed anonymized mobile traffic data for a large connected network of users. One can then use this telephony network as a proxy for the underlying social network. Studies show that a telephony network is a very good proxy for the real social network [30]. Furthermore, by combining telephone network data over time with the adoption history for a product of interest, it is possible to observe how different products spread over the social network.

Using anonymized datasets from one of Telenor's markets, we will show how two different handsets have spread over the social network. The cases being used in this study are the highly buzzed iPhone, and the less fancy, but user-friendly,

Doro type handset, which is more common among elderly people.

Both handsets are tracked from their early introduction and followed for a period of two years. We also present the tracking of a *transactional* product, mobile video telephony, a potentially useful product which allows users to talk to each other, while simultaneously viewing one another (or one another's surroundings)—given certain technological preconditions.

At the end we will show how a computer tablet, specifically iPad 3G, is spreading over the network. We include this as a preview on ongoing research due to the recent introduction in the market.

We start by introducing the *adoption network*, a construction which is readily visualized and which gives insight into the spreading of a product or service. Our figures will include many visualizations, which, we believe, are useful in understanding the product diffusion process on the underlying social network.

II. THE ADOPTION NETWORK

We will in this section define what we mean by the adoption network, followed by an empirical example.

*A. Definition of adoption network*

We define an adoption network as follows. Given a measured telephony network, the node set of the adoption network is the set of subscribers that have adopted a given product, and the links are the communication links belonging to this subset.

Mathematically, an adoption network is thus a subgraph of the whole mobile communication network $C = (c_{ij})$, where $C$ represents (most generally) a weighted, directed, and possibly disconnected graph.

The mobile communication matrix $C$ places a link between each pair of communicating subscribers, so that each nonzero element $c_{ij}$ represents communication. The communication can be based for example on a weighted sum of SMS and voice duration (in which case we call these weighted links *W-links*), or other transactional data like video telephony traffic. All the results in this paper depend only on whether the communication link exists or not, without consideration to weight or direction. We consider traffic between Telenor subscribers in one market, which implies that the matrix will be $n \times n$ large, where $n$ is the size of the customer base (several million subscribers).

The adoption network is then simply the subgraph of $C$ formed by including only the adopting nodes and their common links. As we will see, for a *transactional* product (video telephony), there are two distinct useful choices for the communication links to be used in defining the adoption graph: (i) the standard (voice + SMS) links, or (ii) the links representing the use of the transactional service.

*B. Introducing the "social network monster" by example*

Figure 1 shows the empirical iPhone adoption network from Q4 2007. (This was measured before the iPhone had been introduced into the Telenor net; hence these users have presumably bought their iPhones in the US and "cracked" them for use on the Telenor net). The data show that 42% of the iPhone users communicated with at least one other iPhone-user, which speaks to the social nature of technology consumption, while 58% did not have any iPhone contacts. We call the latter *isolates*. We do not include isolates in any of our visualizations of adoption networks, but do include them in all results counting number of users. We also study the connected components of the adoption network, where the connected components are subgraphs in which any two nodes are connected to each other by paths. Using this convention, we find for example that the largest connected component (LCC) in the adoption network of Figure 1 includes 24.7% of the total number of adopters (while representing over half of the nodes visible in Figure 1). When the LCC in the adoption network is much bigger than all other connected components, and also represents a large fraction of all adopters, we will call the LCC a "social network monster". We note that this is not a precise definition; but we find that such monsters are typically found in adoption networks, and hence believe that the concept is useful.

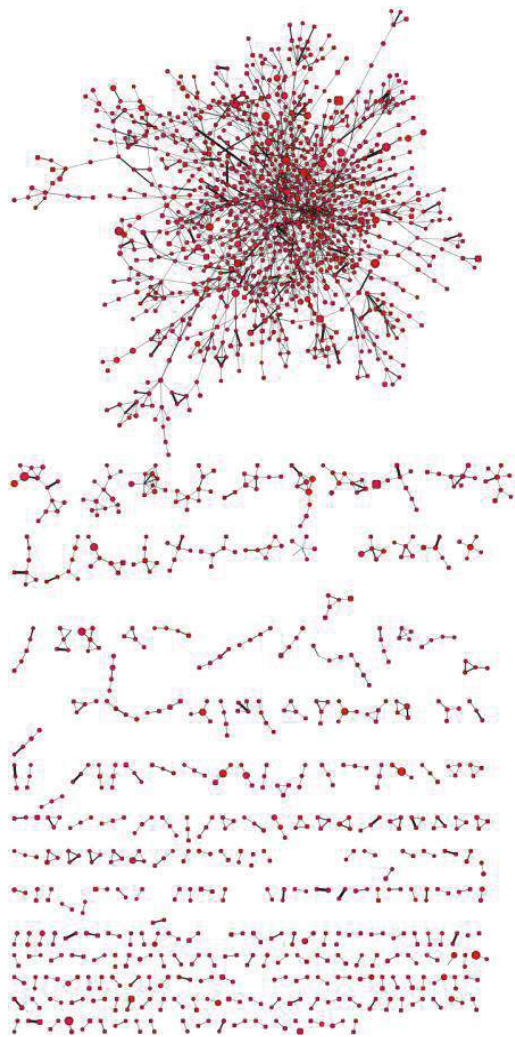

Figure 1. iPhone Q4 2007 adoption network. One node represents one subscriber. Node size represents downloaded internet volume. Link width represents a weighted sum of SMS+voice. Isolates—adopters who are not connected to other adopters—are not shown in the picture.

III. TIME EVOLUTION OF ADOPTION NETWORKS

By studying the time evolution of an adoption network, we can get some insight into how the product which defines the adoption network is diffusing over the underlying social network. In particular we will often focus on the time evolution of the LCC of the adoption network – which may or may not form a social network monster. We recall from Figure 1 that the other components are often rather small compared to the LCC. Hence we argue that studying the evolution of the LCC itself gives useful insight into the strength of the network spreading mechanisms in operation. It also gives insight into the broader context of adoption. As described in [12], two friends adopting together does not necessarily imply social influence – there might also be external factors that control the adoption. In this paper we will not try to separate the 'influence-effects' from external effects such as network homophily. Instead, when we observe a tendency that people who talk together also adopt together, we will use the term 'social spreading'—without making any implicit claim as to the underlying mechanism.

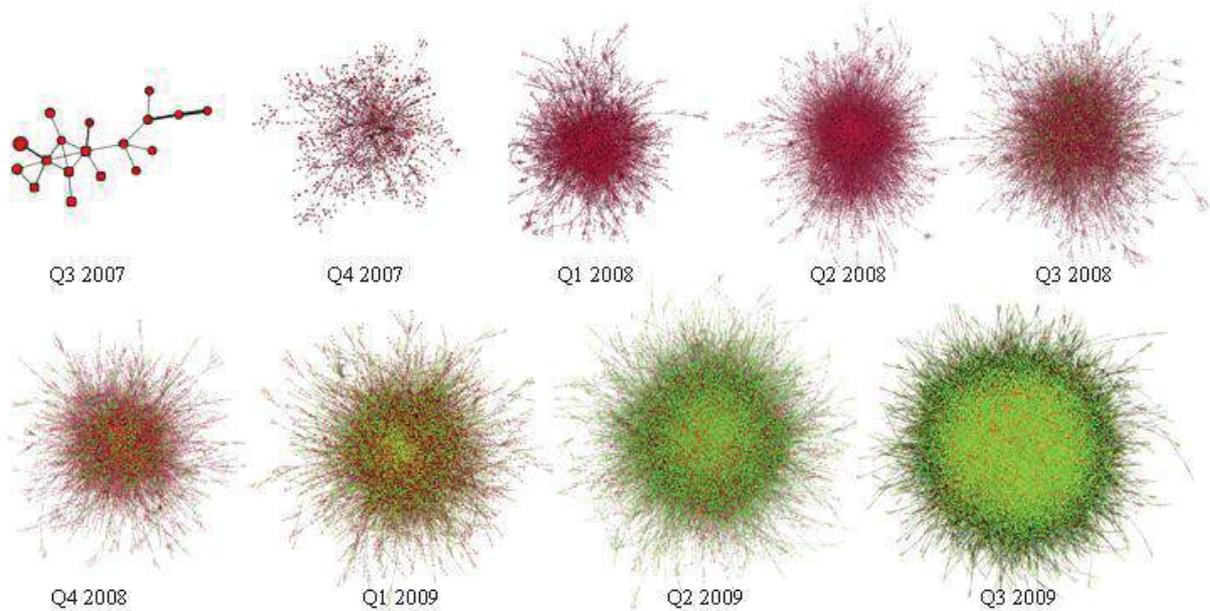

Figure 2. Time evolution of the iPhone adoption network. One node represents one subscriber. Node color: represents iPhone model: red=2G, green=iPhone 3G, yellow=3GS. Node size, link width, and node shape (attributes which are visible in Q3 2007) represent, respectively, internet volume, weighted sum of SMS and voice traffic, and subscription type. Round node shape represents business users, while square represents consumers.

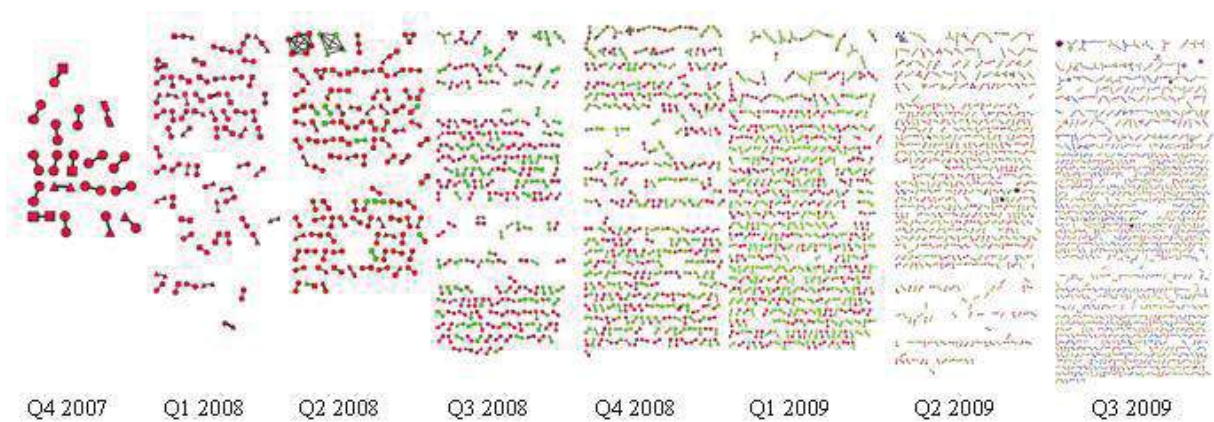

Figure 3. Time evolution of Doro adoption network. One node represent one subscriber. Node color represents Doro Model: red=HandleEasy 326,328, green=HandleEasy 330, blue=PhoneEasy 410, Purple=Other Doro models. Node shape represents age of user: A circle means that user is older than 70 year. Link width represents weighted sum of SMS and Voice traffic.

*A. The iPhone case*

The iPhone 2G was officially released in the US in late Q2 2007 followed by 3G in early Q3 2008 and 3GS late Q2 2009. It was released on the Telenor net in 2009. Despite the existence of various models, we have chosen to look at the iPhone as one distinct product, since (as we will see) the older models are naturally substituted in our network. Figure 2 shows the development of the iPhone monster in one particular market. We observe how the 2G phone is gradually substituted by 3G (red to green), followed by 3GS in Q3 2009 (yellow nodes). The 2G model falls from 100% to 10% with respect to all the iPhone subscribers in the adoption network. In Q1 2009 we observe the same amount of 2G as 3G models.

We show only the LCC in Figure 2 because the other components are visually very much like those seen in Figure 1; the main change over time is that the non-LCC components increase greatly in number, but not in size. That is, essentially all significant growth in component size occurs (in the iPhone case) in the LCC. We regard this growth as a sign that the iPhone is spreading strongly ("taking off") over the social network. It is worth noting that there is a significant marketing "buzz" and external social pressure associated with the iPhone that is perhaps unique. We will offer in later Sections other kinds of measurements which support this conclusion.

*B. The DORO case*

The next example is the Doro. As with the iPhone, there are several different models that are considered collectively. It is a handset which is easy to use, and mainly targeted towards elderly people [10]. Since Doro has a relatively low number of non-isolated users in all quarters studied, we present in Figure 3 visualizations of the whole adoption network (minus isolates) over the entire time period, from introduction (in Q4 2007) to Q3 2009. Figure 3 shows that most Doro users that are not isolates appear in pairs in the adoption network. The social network monster never appears—the contrast with the iPhone case is striking. We believe that the kind of adoption network evolution seen in Figure 3 is indicative of a product where "buzz" effects—social influence in the spreading of adoption— are weak or absent, whereas what we see in Figure 2 indicates strong buzz effects. It is possible to argue that the adoption of the Doro is more of an individual choice, or perhaps even the choice of the user's children who wish to be in contact with their elderly parents. We note finally that the tiny "monster" (LCC) seen in Q3 2009 of Figure 3 consists entirely of enterprise subscribers. Hence we speculate that these users are not the elderly of the target segment, but rather users with some other interest in the product.

*C. Mobile Video Telephony case*

Compared to iPhone and Doro, video telephony has no value for an isolated user; thus users will always appear in pairs. A similar (pairs-only) constraint may be seen in [1], where the connections are based on romantic relations. In the video telephony case we actually have two distinct link sets which may be used to define an adoption network: W-links (voice + SMS), and video links. Thus for mobile video telephony we create and study two distinct adoption networks:

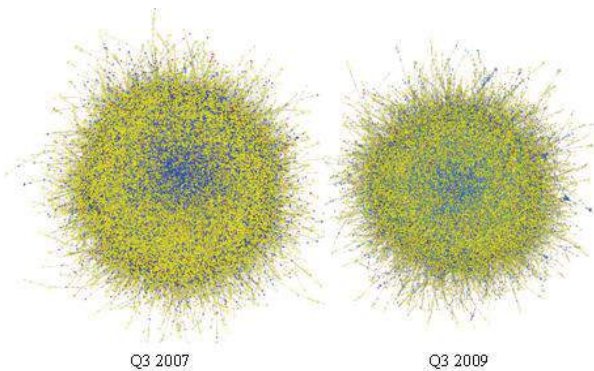

Figure 4. Time evolution of Mobile Video Telephony adoption network (WVAN—where the social links include all communication). Only two quarters are shown due to the fairly stable LCC. Blue node color represents

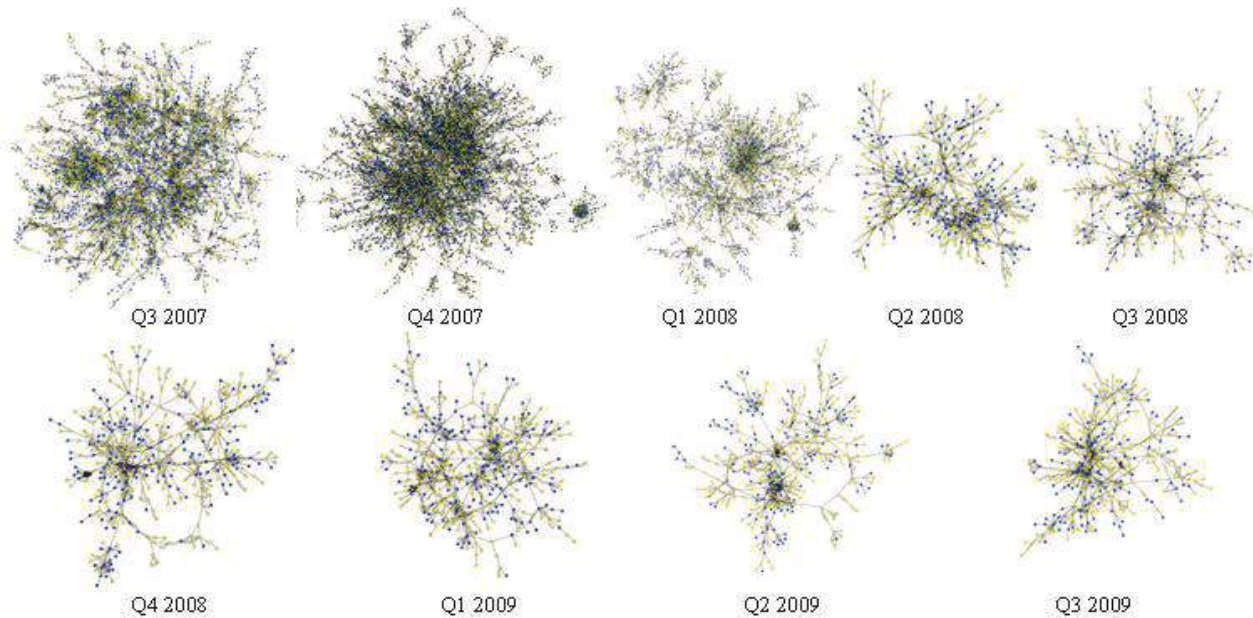

Figure 5. Time evolution of the Mobile Video Telephony adoption network. Links are real video links, where width represents duration of video conversations. Enterprise adopters have blue node color, while consumer adopters are tagged yellow.

- The video-link set gives rise to the video adoption network (VAN)
- The W-linkset (voice+SMS) gives rise to the W-Video adoption network (WVAN).

We find that these two networks for video telephony have quite different behavior. We consider first the WVAN. This network connects users who (i) have a communication link (voice and/or SMS) and (ii) both use video telephony—not necessarily with each other. For the WVAN we find consistently a large social monster, much like the one seen in Figure 2 that starts to form. However, differently from Figure 2, the monster in the WVAN actually diminishes in size over time—both in absolute number of users, and in the percentage of users in the LCC. Figure 4 illustrates this by showing two WVAN video-monsters which are two years apart.

We gain even more insight by looking at the time evolution of the VAN. Figure 5 shows the time evolution of the VAN-LCC. We see growth in the monster from Q3 2007 to Q4 2007, followed by a rather dramatic breaking down of the LCC after that time. Hence we see indications that the service itself had the potential to form a real social monster and take off, but some change in the service and user conditions killed that takeoff – in this case, we have found that a new pricing model was introduced.

*D. Comparison of the social network monsters over time*

Figure 6 sums up much of what we have seen in the visualizations of the last subsections. The figure shows the fraction of adopters in various components of the adoption network. Subscribers in the blue area are adopters which have no connection to other adopters. These users (termed isolates here, and referred to as singletons in [2]) have not been visible in our visualizations. The users in the green area correspond to the adopters in the social network monster (there is in every case only one component with >1000 users).

We first consider Figure 6(a), the figure describing iPhones. Here we see that the growth of the monster (green), as a percentage of the total number of users, has not been monotonic. The monster has however grown monotonically in the absolute number of users—see again Figure 2. We conclude from this that the number of isolated subscribers grew more rapidly than did the core. This implies that some change in the offering has induced a large growth in the number of new users in this period (Q2 2008—Q3 2008). One candidate explanation is the appearance of 3G handsets in this time period. Another likely explanation for many new users is the fact that "legitimate" iPhones were first available on the Telenor net at this time.

Figure 6(b) (Doro) simply confirms the picture seen in Figure 3: no monster, essentially no large LCCs. At the same time we see an enormous dominance of isolates. This is consistent with the hypothesis that Doro users are elderly (which we can confirm), and that they speak mostly with other generations, ie, non-Doro users. Again, it suggests that the adoption of this phone is not based on network influence, but on more ego-based considerations.

Figure 6(c) shows the VAN, while 6(d) shows the WVAN for the video product. Again we confirm the qualitative picture obtained from Figs. 4 and 5: the WVAN-monster decays slowly, while the VAN-monster collapses. In the case of the video service, the collapse corresponds to the initiation of payment for the system. Using the lingo of the iPhone example, this would be the same as turning of the "buzz". We also see in Figure 6(c) a dominance of two-node components—not surprising for a transactional service—and a complete absence of isolates. The latter result, while not surprising, is not in fact guaranteed (for WVAN) by our definitions: we will see two isolates in WVAN every time two subscribers use video transactions, but have no other (W) communication, and have no friends using video transactions. We see that this simply

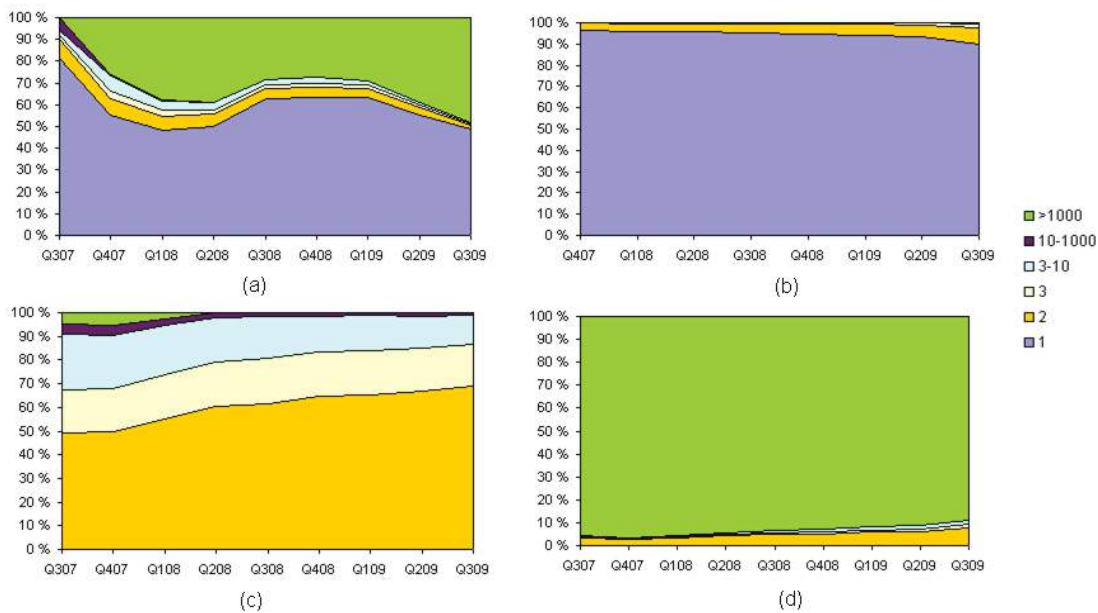

Figure 6. Fraction of subscribers in components of various sizes in the adoption networks for: (a) iPhone (b) Doro (c) Video (video-links) (d) Video (W-links)

does not happen—primarily because every pair that uses video also uses voice, SMS, or both.

We offer some quantitative details illustrating the dynamics seen in Figs. 6(c) and 6(d). We observe that 95.8% of users are in the core of WVAN in Q3 2007, while only 5.7% are in the VAN core. Two years later, the corresponding numbers are 88.7% for WVAN and 0.76% for VAN.

## IV. CENTRALITY OF ADOPTERS

We have seen how the iPhone and video adopters form a giant monster with respect to the W-links, while the Doro adopters do not. Motivated by this, we calculate the social centrality for all the adopters in each group, comparing it to the centrality of the whole customer base. Our expectation is that the involvement of highly central users is essential to the development of a large monster. To measure centrality we use the well-known eigenvector centrality (EVC). We believe high EVC will be strongly correlated with presence in the social monster, as it is already known to be correlated with strong spreading [11].

Figure 7 shows the EVC distributions for all the adopters. We see that video users are the most central, with iPhone just behind. We also see that the Doro adopters are rather peripheral socially, with their distribution falling well below that for the entire customer base. This supports our expectation that people in the giant monsters tend to be more central than the rest of the customer base. We believe that one may find, among these customers, the influential early adopters—those that adopt new products and services fairly early, and stimulate (or perhaps demand) others to do the same.

## V. KAPPA-TEST

In all our results so far we see indirect evidence for social spreading effects (or their apparent absence, in the Doro case). As another test for social spreading, we introduce a simple statistical test, the kappa-test or κ-test.

### A. Definition of the κ-test

We consider again the entire social network (as proxied by our communication graph) and define two types of links:

- A-links: links where neither, or only one, of the two connected nodes have adopted the product.
- B-links: links where the two connected nodes have both adopted the product.

We regard B-links to be the links which can indicate (but not confirm) social influence. We also recognize however that B-links can arise by other mechanisms, and even by chance. In order to evaluate the significance of the B-links that we observe in the empirical adoption data, then, we compare the empirical number of B-links (call it $n_{B,emp}$) with the number found by distributing at random the same number of adopters over the same social network, and then counting the resulting

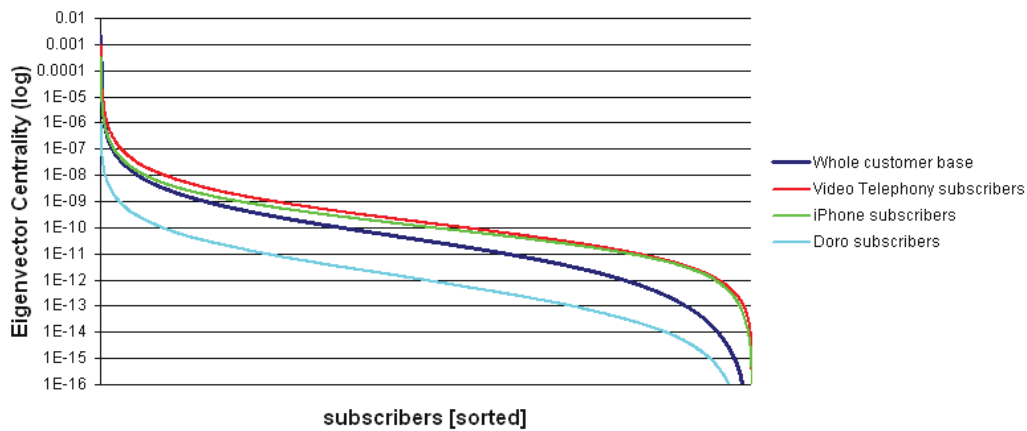

Figure 7. Eigenvector centrality distributions for the whole customer base, Doro users, iPhone users, and video telephony users. Distributions are from Q3 2009. For ease of comparison, the x-scale is normalized so as to run from 0 to 100% for all displayed distributions.

number of B-links $n_{B,rand}$.

We then define $\kappa \equiv n_{B,emp}/n_{B,rand}$. Clearly, if κ is significantly larger than 1, we have strong evidence for social spreading effects. More precisely, κ>1 implies that people who communicate with each other tend to adopt together.

Figure 8 illustrates what happens when we scatter the iPhone adopters in Q1 2009 randomly over the empirical social network. The monster is still there, but it is smaller (by more than a factor 3) than the empirical monster seen in Figure 2. Comparing the corresponding whole adoption networks of Figs. 2 and 8 by using the κ-test, we find that there are over twice as many links in the empirical adoption network compared to the random reference model—that is, κ is 2.18. We take this to be evidence that social spreading has occurred—more precisely, that people who talk together adopt together much more often than chance would predict.

Figure 8 illustrates an important point which gives insight into both monsters and social network structure. The point is that monsters arise even in the complete absence of social spreading effects. The monster seen in Figure 8 is thus telling us something about the structure of the social network itself—that it has a "dense core" in which a dominating LCC arises even in the case of random adoption. At the same time, this dense core must include the set of users who give rise to the empirical monsters that we observe. The empirical LCC is simply larger than the random one (for the same number of adopters), due to social spreading (arising from mechanisms such as social influence or homophily).

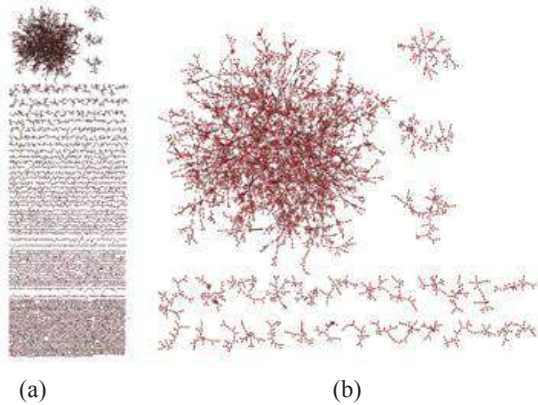

(a)　　　　　　　　　　(b)

Figure 8.　a) iPhone network from random reference model, Q1 2009. b) LCC zoomed in

Figure 9 shows the evolution of κ over time, both for the iPhone and for Doro. We notice that κ is very large in the early stages of product adoption (eg, around 28.7 for the iPhone in Q3 2007). We find this to be typical: the first adopters are not randomly distributed, but rather tend to lie in a few small connected social groups. The large value for κ tells us that this observed distribution of the early adopters on a social network is extremely unlikely to have occurred by chance.

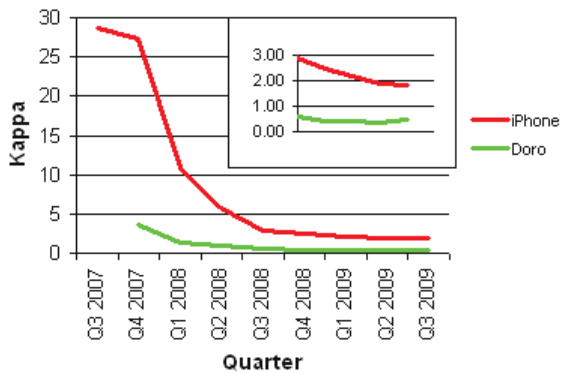

Figure 9.　Kappa for Doro and iPhone

We notice also that κ is consistently less than 1 for Doro, after the early phase of adoption. While we argue that κ>1 is evidence for social spreading effects, we do not believe that κ<1 proves that such effects are not occurring. What κ<1 does say is that adopting friends are found less often than a random model would predict. Our explanation for this is that the random model hits the dense core more often than the actual empirical adopters do. In other words, the empirical adopters are socially peripheral. This idea is in agreement with the EVC distribution seen in Fig 6.

Finally we note that our κ test is not performed for the video adoption network in this chapter. The reason is that (as discussed in Section III) the transactional nature of the video service constrains both VAN (exactly) and WVAN (empirically) such that there are no isolates. This constraint is not captured by the random reference model of the κ test. The purpose now is to introduce a link-based version of the node-based kappa-test.

VI.　LINK BASED KAPPA-TEST.

The kappa-test defined in V considers products which are adopted nodewise, like e.g. the adoption of handset. In the case of transactional products like videotelephony, product adoption is defined in a pairwise manner: the nature of the product is *to activate links on the social network*. We consider videotelephony as a *link-based* product. The spreading process of a link-based product will be different compared to a node-based product, and hence the random spreading model in the kappa-test has to reflect this. In the node-based kappa-test the *nodes* are spread randomly and then the resulting links are compared to the empirical number of links between adopters. For link-based products, we suggest a random model where links in the social network are randomly activated.

We suggest the following for a link-based kappa-test: We start from the underlying social network, and track the time-evolution (adoption) of a link-based service; e.g. MMS or video telephony. In other words; a link-based service is a service in which the adoption requires both end-points of a link to adopt the service. Both nodes at the end-point of a link are mutually depending on each other in starting using the link-based service—communicating with each over the link. The link adoption of the network-based service, also gives rise to an adoption network.

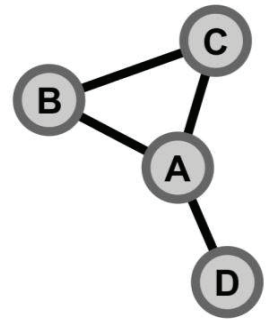

Figure 10. llustration of open and closed triangles. A-C and A-D are adjacent links since they share one common node and make an open triangle. A-B, B-C and A-C together make a closed triangle.

Again, we are interested in measuring 'adopting together', so for links this is interpreted as meaning that two links have adopted together if they have a common node as end-point. In other words; two links are said to adopt together when the two links are adjacent. Fig 10 illustrates how we count adjacent links, two links A-C and A-D define an adjacent pair of links, since the node A is a common node for both links. Counting the number of open triangles in the adoption network then measures the occurrence of links adopting together. The random adoption network is generated through randomly choosing the links to adopt the link-based service from the set of possible links in the underlying social network, and the number of open triangles is then counted in the random case. We suggest then to define the link-based kappa test as;

$$\kappa = \frac{\#\ adjacent\ links\ in\ empirical\ network}{\#\ adjacent\ links\ in\ random\ network} \quad (1)$$

The above framework can now be used for generating kappa tests for comparing the empirical and random occurrence of links and adjacent links in the true adoption network and a randomly simulated adoption network, respectively. These adoption networks are considered relative to the underlying 'true' social network that can be measured. Summarizing, we have the following two kappa tests:

- Node based adoption, counting links: κ = empirical number of links / random number of links

- Link based adoption, counting open triangles: κ = empirical number of open triangles / random number of open triangles.

We count the total number of open links in the network using the binomial coefficient and sum over all nodes $i$:

$$\sum_{i=1}^{n} \binom{k_i}{2} = \sum_{i=1}^{n} \frac{k_i!}{2!\,(k_i - 2)!}$$
$$= \sum_{i=1}^{n} \frac{1}{2} k_i (k_i - 1)$$
$$= \frac{1}{2}\sum_{i=1}^{n} k_i^2 - \frac{1}{2}\sum_{i=1}^{n} k_i \quad (2)$$
$$= \frac{1}{2}\sum_{i=1}^{n} k_i^2 - L$$

where $k_i$ is node degree, $n$ is the total number of nodes, and L is the number of links – all measured relative to the adoption network.

From the above discussion, it is also possible to consider other types of kappa-tests. For example, one can envision kappa-tests that compares;

- The number of closed triangles to the number of random triangles.
- The number of closed loops of some order to the number of random closed loops.

The idea is to consider local network structures, and how these local structures potentially 'adopt' together.

*A. Quarterly development of the link kappa.*

From figure 11 we see that the link kappa value defined in (1) is steadily increasing from around 15 to about 20. A kappa value of 15 means that there is 15 times as many adjacent links in the empirical data than in the random reference case! We can then conclude that the distribution of video usage on the social network is far from a random process where social relations are randomly 'activated' to become video links.

The number of adjacent links is related to the node degree according to (2). Figure 12 shows the degree distribution for Q3 2009 – both for the random 'link activation' model and for the real empirical video adoption network (VAN). As expected from the high link kappa value, the degree of the nodes in the empirical adoption graph is higher than in the corresponding random model (higher number of adopting neighbors means higher number of adjacent links). In the random case around 90% of the nodes have degree 1 – the least degree possible since the unit we use in this experiment is *links*. An isolated *link* will give both end nodes degree 1. This is expected, since the random model does not take into account that the videotelephony users, once they have started using the service, will use the service to communicate with more than one network neighbor. A more sophisticated random reference model could also take this into account. The simple random link activation model acts as basic reference and assigns the same activation probabilities to all links on the network. We reserve testing more advanced reference spreading models for future work.

Another way to measure clustering among adopters is to use the global clustering coefficient defined as total number of closed triplets divided by the total number of adjacent links. See e.g. reference [13] for a detailed discussion of the global clustering coefficient. Using the same 'kappa' logic as before, a completely random adoption process will also show some clustering. To compensate for this effect, we use the kappa test logic on the global clustering coefficient and define a new link kappa measure as $C_{emp}/C_{rnd}$, where $C_{emp}$ is the clustering coefficient calculated from the empirical network and $C_{rnd}$ is calculated from the random video adoption network. Figure 13 shows the development of the ratio $C_{emp}/C_{rnd}$.

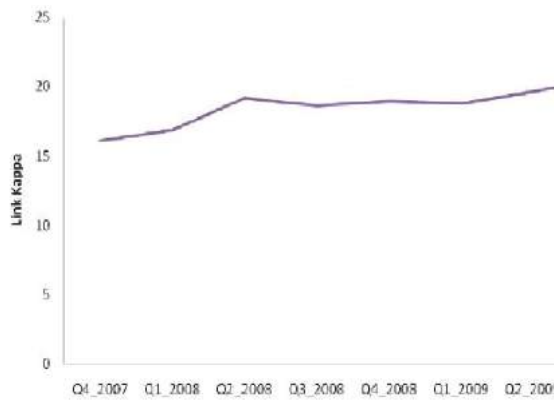

Figure 11 Historical development of link kappa for video telephony. Link kappa is defined as the number of total adjacent links in the empirical data to the number of adjacent links in the random reference case. The random reference network is constructed by choosing an equal number of links randomly from the underlying social network and then count the number of adjacent links.

In summary – measuring the social network dependence on video telephony is more difficult than measuring effect on node attributes like choice of handset. Video telephony is a link attribute involving choices of both end nodes. A simple reshuffling of individual adopter status will not take this into account. Our simple link kappa test can be used for quantifying network dependent adoption for services of a transactional nature. In the case of video telephony we find that the empirical relations are clustered on the social network compared to a random link activation model.

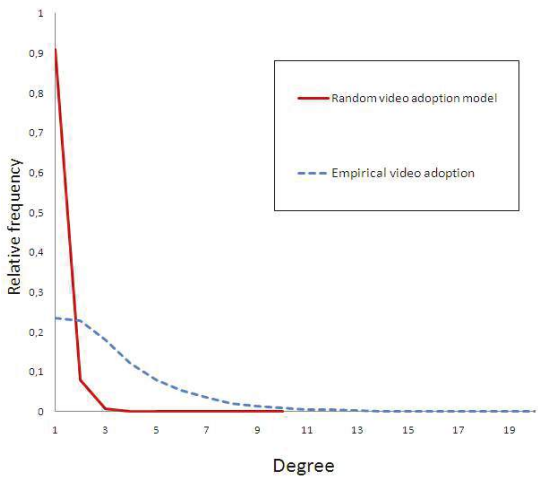

Figure 12 Node degree distribution for the 'random link activation' reference model and the empirical video adoption network (VAN), Q3 2009.

VII. CORRELATED ADOPTION PROBABILITY

Our final test for social spreading effects is to measure the probability $p_k$ that a subscriber has adopted a product, given that $k$ of the subscriber's friends have adopted the product. This conditional probability does not indicate causation, because it makes no reference to time order—it simply measures (again) how strongly those that communicate together tend to adopt together. We measure $p_k$ simply by first finding all subscribers with $k$ adopting friends, and then finding the fraction of these that have themselves adopted.

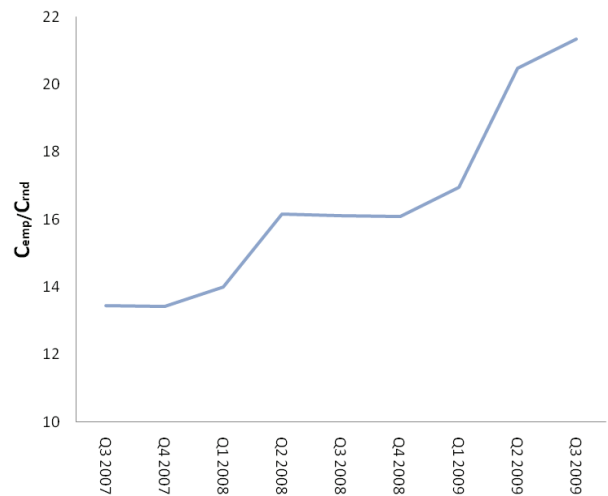

Figure 13 Clustering coefficient kappa test defined as the empirical global clustering coefficient divided by the global clustering coefficient for the random network. The global clustering coefficient is defined as total number of closed triangles divided by the total number of adjacent links in the network.

Figure 14 shows $p_k$ vs $k$ for the three products, for $0 \leq k \leq 3$. For higher $k$, the Doro data are too dominated by noise (a low $n$) to be useful. (The results for iPhone and video have better adoption profiles, and so are meaningful at least out to $k$=10; but their qualitative behavior—monotonic increase, at roughly constant slope, with increasing $k$—is like that seen in Figure 14.) Figure 14 supports our claim that there are some social spreading effects operating on Doro adoption—since we see a steady increase of $p_k$ with $k$. Such effects are not visible from our κ test, for reasons given above; yet we see for example that, if we know that a subscriber has one friend using a Doro phone, that subscriber's probability of using one him- or herself is roughly *twice* the adoption probability for a subscriber with no adopting friends.

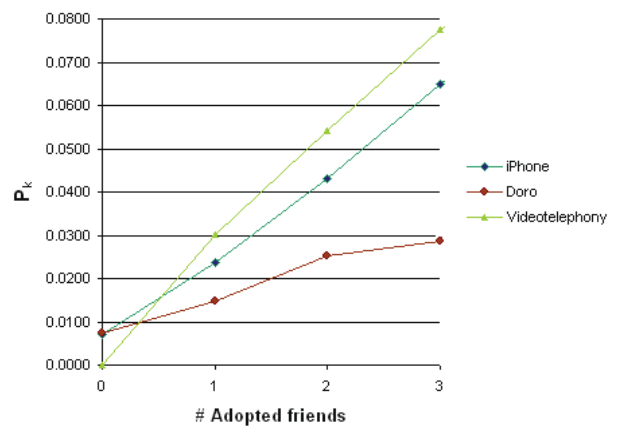

Figure 14. Adoption probability $p_k$ vs the number $k$ of adopting friends, for three products. In each case we see a monotonic growth of $p_k$ with $k$— indicating that some kind of social spreading is occurring.

The iPhone and video $p_k$ results lie, not surprisingly, considerably higher than the Doro curve. This is consistent with our claim that social spreading effects are much stronger for these products. For example, knowing that a person has one friend using an iPhone roughly triples the probability (compared to someone with no adopting friends) that this person also uses an iPhone. This comparison is however not possible to make for the video service, because the probability of using video telephony and having no (W-)friends who use video telephony is (as discussed above) empirically zero. Otherwise the video results are qualitatively the same as the iPhone results.

We also note that the $p_k$ for video-telephony is consistent with the degree distribution for video adopters in figure 12, where we see that the nodes in the empirical VAN generally have a higher degree than expected from the simulated random link activation model. Since the adoption probability, $p_k$, increases with number of adopting neighbors, one will expect a higher number of activated neighbors in the adoption graph compared to a random model. A higher number of neighbors will again result in a higher number of adjacent links in the empirical data and thus the high link kappa value we see in figure 11.

## VIII. TIME EVOLUTION OF COMPUTER TABLET ADOPTION NETWORKS

Until now we have looked into the social interaction between users of different handsets, as well as video telephony. Recently, so called computer tablets, such as Apple iPad 3G and Samsung Galaxy Tab, have received increased attention. These tablet computers are particularly marketed as a platform for audio and visual media such as books, periodicals, movies, music and games, as well as web content. As a preview of current research we will show the social network of iPad 3G-adopters, and its time-evolution from its release date, as well as some early statistical results. We are not able to study uptake of WiFi-only tablets, since we depend on a SIM to map tablets to the social network. Computer tablet networks address some fundamental differences from the adoption networks already mentioned. An important difference from traditional handsets is that the SIM card placed in a tablet is not necessary the same SIM card which is being used for social interaction, It is typically a "twin SIM" solution, where a customer gets two SIMs on the same subscription. We use the SIM in the traditional handset as the node and map the social network. This will give us a picture of the social interaction among iPad 3G users. We will show that the 'non-pad' SIM cards are most often placed in an iPhone.

## IX. iPAD 3G ADOPTION NETWORK

As with the iPhone, the release date of iPad 3G varies with region - it was officially released in the US Apr 3$^{rd}$ 2010. In Telenor net it was released in November 2010. As the visualizations will show, thousands of users bought their iPad in the US and used it on Telenor net before they actually were officially released by Telenor in November 2010. Since such tablets are quite new in the given market, we have chosen to study the time evolution with a more fine-grained granularity - month by month.

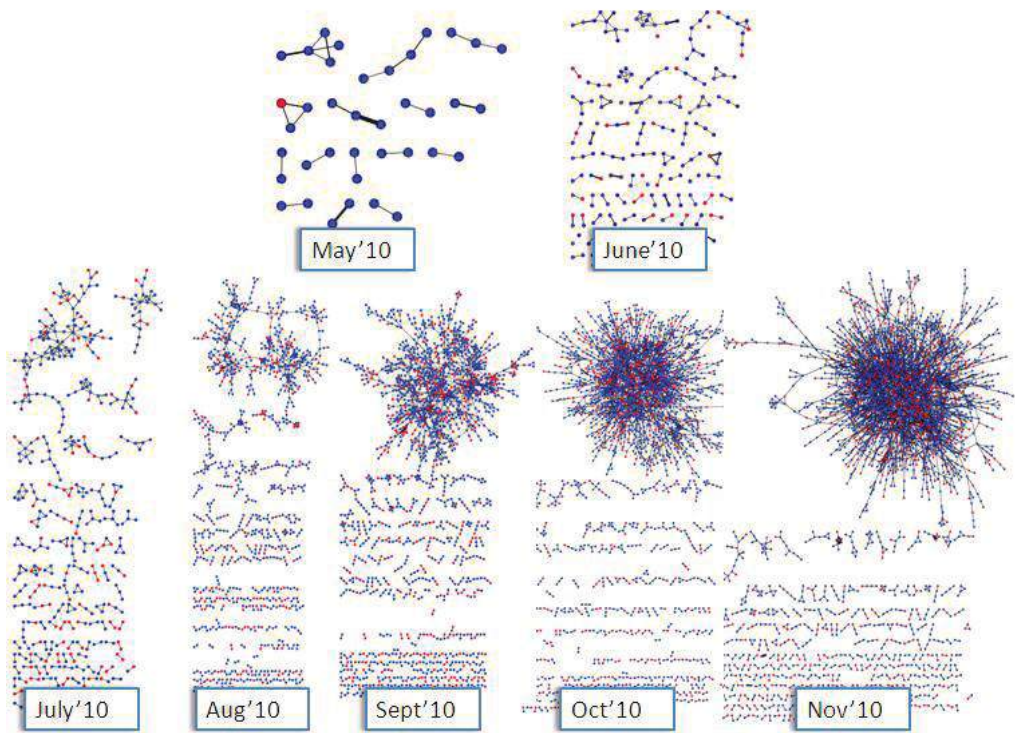

Figure 15. Time evolution of iPad 3G adoption network. One node represent one subscriber. Blue node color represent an iPad 3G user which also uses an iPhone, while red color represent an iPad 3G user with another handset. Link width represents weighted sum of SMS and Voice traffic. Isolates are not included in the visualization.

Fig 15 shows the time evolution of the iPad 3G adoption network for the first seven months. The structure reminds us of the iPhone evolution. We have here also included smaller social components, but for visualization purposes we have not included the isolates (iPad 3G-adopters with no iPad 3G friends), but we remind the reader that these are also part of the adoption network. From the visualization we see how the largest connected component gradually turns into a "monster" within a couple of months. When a specific iPad-user also uses an iPhone, the node is colored blue. Visually it is seen that the connected iPad-users have a high percentage of iPhone's. In general we find that 53.4% of the users also are using iPhone for social interaction. We observe that in the cases where the iPad-user has at least one iPad-friend (adopters shown in Fig 15) there are as many as 72.1% also having an iPhone. The remaining users, which represent the disconnected users in the adoption network (isolates), there are only 38.6% which have adopted an iPhone. This shows that iPhone is much more common phenomena in socially connected groups of iPad users. This also supports the idea of an Apple "tribe" of users.

## X. KAPPA-TEST FOR IPAD 3G

As stated in section V, $\kappa>1$ implies that people who communicate together also tend to adopt together. Figure 16 shows the time evolution of $\kappa$ for iPad 3G. We notice that it is significantly above 1, which indicate strong evidence for social network effects. $\kappa$ varies between 16.9 and 72, which means that at most there are 72 times more empirical iPad-relations compared to the relations we find when spreading the adopters randomly on the whole communication network. We notice that the numbers are significantly higher than for

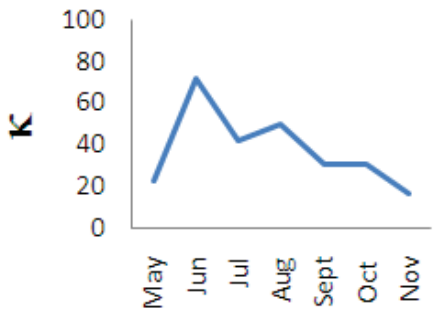

Figure 16.  Kappa for iPad 3G

iPhone – and from the perspective of $\kappa$, this trend supports that the spreading is stronger compared to iPhone, which is also reflected through the number of sales [31].

The rapid increase of kappa from May to June can be explained by the time it takes from the first adopters got their pads until friends are influenced and are able to get one. It is a product of how the individuals got their iPads since it involved travel to the U.S. or knowing a friend who was traveling there.

The decrease from june can be due to increased exposure of the product in the media which can lead to less dependence on social spreading, in combination with the increased number of sales. Due to the nature of $\kappa$, it will approach 1 if we have complete saturation of iPads in the market.

## XI. IPAD 3G ADOPTION PROBABILITY

Fig 17 corresponds to the correlation adoption probability which we measured in section VI. We measure the probability $p_k$ that a subscriber has adopted iPad 3G, given that $k$ of the subscriber's friends have adopted iPad 3G. The results shows that if you have one iPad friend the probability to adopt iPad is 14 times higher than with zero friends. If you have 2 friends the probability is 41 times higher, 3 friends 96 times. We observe a steep monotonic growth which should indicate strong social spreading effects. The adoption is highly dependent on the number of friends.

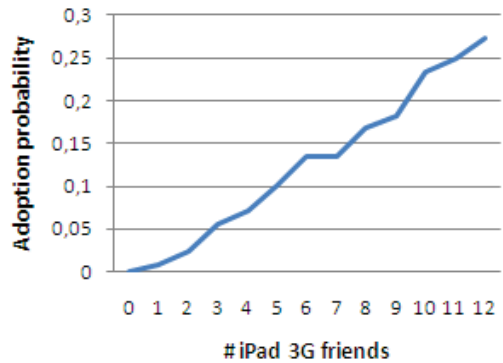

Figure 17.  Adoption probability $p_k$ vs the number $k$ of adopting friends, for iPad 3G. We observe a monotonic growth of $p_k$ with $k$— indicating that some kind of social spreading is occurring.

## XII. SUMMARY AND FUTURE WORK

All of our results support a simple and fairly consistent interpretation:

- The iPhone has very strong social spreading, and has truly taken off

- The Doro handsets have only very weak social spreading. This device will probably never take off in the same sense as the iPhone.

- Video telephony use also has strong social effects, and started spreading very strongly; however its early takeoff was stopped by an external factor— here, a new price model.

- Preliminary results from studying the newly released iPad 3G reveals even stronger network-dependent adoption patterns than iPhone.

Standard whole-network measures, such as total number of users, or total traffic over time, can also give useful information on these same questions. We believe however that our measurement methods give new and useful insight into how and why these services have performed so differently.

We have also constructed two different link-based κ-tests for the VAN, because the VAN has a fundamental constraint—that adoption occurs only in pairs—that our simple κ-test does not capture. We conclude from these κ-tests that the remaining subscribers are well clustered due to the increase of κ over time – both for the test based on the clustering coefficient and κ based on adjacent links. These tests should be useful for any transactional graph with the pair constraint.

Also, we suspect that social spreading effects for the Doro handsets may be more visible at the two-hop level. A typical scenario might be an adult child would buy this type of phone for one or more of their elderly parents when, for example, use of a more traditional mobile phone becomes difficult because of the size of the display or keypad, or the complexity of the device. In such a scenario, it may well be there is little or no direct communication among the adopters, but a strong two-hop connection via the younger generation. We plan to test this idea in the near future.

ACKNOWLEDGMENT

All of our visualizations were produced using the open-source visualization platform cytoscape.org. We would also like to thank Dr. Ellen Altenborg at Telenor ASA, and professor Øystein D. Fjeldstad of the Norwegian School of Management, for fruitful discussions and inspiration related to network economy and network analysis.

# Paper 7

# Big Data-Driven Marketing: How machine learning outperforms marketers' gut-feeling

Pål Sundsøy[1], Johannes Bjelland[1], Asif M Iqbal[1],
Alex Sandy Pentland[2], Yves-Alexandre de Montjoye[2]

[1]Telenor Research, Norway
{pal-roe.sundsoy,johannes.bjelland,asif.iqbal}@telenor.com
[2]Massachusetts Institute of Technology, The Media Laboratory, USA
{pentland,yva}@mit.edu

**Abstract.** This paper shows how big data can be experimentally used at large scale for marketing purposes at a mobile network operator. We present results from a large-scale experiment in a MNO in Asia where we use machine learning to segment customers for text-based marketing. This leads to conversion rates far superior to the current best marketing practices within MNOs.

Using metadata and social network analysis, we created new metrics to identify customers that are the most likely to convert into mobile internet users. These metrics falls into three categories: discretionary income, timing, and social learning. Using historical data, a machine learning prediction model is then trained, validated, and used to select a treatment group. Experimental results with 250 000 customers show a 13 times better conversion-rate compared to the control group. The control group is selected using the current best practice marketing. The model also shows very good properties in the longer term, as 98% of the converted customers in the treatment group renew their mobile internet packages after the campaign, compared to 37% in the control group. These results show that data-driven marketing can significantly improve conversion rates over current best-practice marketing strategies.

**Keywords:** Marketing, Big Data, Machine learning, social network analysis, Metadata, Asia, Mobile Network Operator, Carrier

## 1   Introduction

For many people in Asia, mobile phones are the only gateway to the Internet. While many people have internet capable phones, they are often not aware of their capabilities. The overall penetration of internet in these countries is very small which causes a large digital discrepancy [1]. In the market of this study, internet penetration is less than 10%.

Mobile Network Operators (MNOs) commonly use texts as a way to raise customer awareness of new products and services - and to communicate with their customers. In Asian markets, MNOs typically run thousands of text campaigns a year, resulting in customers receiving several promotional texts per month. Making sure they are not seen as spammers by customers but rather as providing useful information is a major concern for MNOs. For this particular operator, the policy is to not send more than one text per customer every 14 days. This limit is currently maxed and is preventing new campaigns from being launched.

Targeting, deciding which offer to send to which customer, often relies on the marketing team's "gut-feeling" of what the right audience is for this campaign. In a recent IBM study, 80% of marketers report making such decisions based on their "gut-feeling" [2]. A data-driven approach might lead to an increased efficiency of text-based campaigns by being one step closer to delivering "the right offer to the right customer". For example, previous research showed that data-driven approaches can reliably predict mobile phone and Facebook user's personality [3-5,18], sexual orientation [6], or romantic relationships [7].

Our data-driven approach will be evaluated against the MNO's current best-practice in a large-scale "internet data" experiment [8,9]. This experiment will compare the conversion rates of the treatment and control groups after one promotional text.

We show that this data-driven approach using machine learning and social network analysis leads to higher conversation rates than best-practice marketing approach. We also show that historical natural adoption data can be used to train models when campaign response data is unavailable

## 2   Best practice

The current best practice in MNOs relies on the marketing team's experience to decide which customers should receive a text for a specific campaign. The marketing team typically selects customers using a few simple metrics directly computed from metadata such as call sent, call received, average top-up, etc. For this particular "internet data" campaign, the marketing team recommended to use the number of text sent and received per month, a high average revenue per user (ARPU) [10], and to focus on prepaid customers. Table 1 shows the variables used to select the control group, the customers that, according to the marketing team, are the most likely to convert.

The control group is composed of 50 000 customers selected at random amongst the selected group.

**Table 1:** Variables used to select the control group

| |
|---|
| Sending at least four text per month |
| Receiving at least four text per month |
| Using a data-enabled handset |
| 'Accidental data usage' (less than 50kb per month) |
| Customer in medium to high ARPU segment (spending at least 3.5 USD per month) |

## 3 Data-driven approach

### 3.1 Features

For each subscriber in the experiment, we derive over 350 features from metadata, subscription data, and the use of value added services. Earlier work show the existence of influence between peers [11,12] and that social influence plays an important role in product adoption [13,14]. We thus inferred a social graph between customers to compute new features. We only considered ties where customers interacted more than 3 times every month. The strength of ties in this social graph is a weighted sum of calls and texts over a 2-month period. Using this social graph, we computed around 40 social features. These include the percentage of neighbors that have already adopted mobile internet, the number of mobile data users among your closest neighbors (ranked by tie strength), or the total and average volume of data used by neighbors.

### 3.2 Model

We develop and train the model using 6 months of metadata. As the outcomes of previous mobile internet campaigns were not stored, we train our model using natural adopters. We then compare these natural adopters, people who just started using mobile internet, to people who over the same period of time did not use mobile internet. Our goal is to identify the behavior of customers who 1) might be interested in using internet and who 2) would keep using mobile internet afterwards. We then select 50 000 natural adopters and 100 000 non-internet users at random out of these groups. Note that natural converters are only a way for us to extract characteristics of customers who are likely to convert. The conversion rates are likely to have been better if we had access to data about previous campaigns and previously persuaded adopters.

**Table 2.** Training set

| Sample size | Classifier | Definition |
| --- | --- | --- |
| 50k | Natural adopters | Less than 50KB of data per month from December to March (accidental data usage). More than 1MB of data per month in April and May |
| 100k | Reference users not using internet | No internet usage |

We tested several modeling algorithms such as support vector machine and neural networks to classify natural converters. The final model is a bootstrap aggregated (bagging) decision tree [15] where performance is measured by accuracy and stability. The final model is a trade-off between accuracy and stability where, based on earlier experience, we put more weight on stability. The accuracy of the final model is slightly lower than other considered models. The bagging decision tree however turned out to be more stable when tested across different samples.
The final cross-validated model only relies on a few key variables. Only 20 features out of the initial 350 where selected for the final model. Table 3 shows the top 10 most useful features to classify natural converters as ranked by the IBM SPSS

modeler data mining software. The features tagged as binned are handled by the
software optimal binning feature.

**Table 3.** Top 10 most useful features to classify natural converters. Ranked by importance in the model.

| Rank | Type | Description |
|---|---|---|
| 1 | Social learning | Total spending on data among close social graph neighbors |
| 2 | Discretionary income | Average monthly spending on text (binned) |
| 3 | Discretionary income | Average monthly number of text sent (binned) |
| 4 | Discretionary income | Average monthly spending on value added services over text (binned) |
| 5 | Social Learning | Average monthly spending on data among social graph neighbors |
| 6 |  | Data enabled handset according to IMEI (Yes/No) |
| 7 | Social Learning | Data volume among social graph neighbors |
| 8 | Social Learning | Data volume among close social graph neighbors |
| 9 | Timing | Most used handset has changed since last month |
| 10 |  | Amount of 'accidental' data usage |

### 3.3 Out-of sample validation

Before running the experiment, we validated our model on natural adopters in a new, previously unseen, sample using other customers and another time period. The performance on historical data is measured using lift curves. Fig. 1 shows a lift of around 3 among the 20% highest scored customers. This means that if we were to select the 20% highest scored customers, the model would predict 3 times better than selecting at random from the sample.

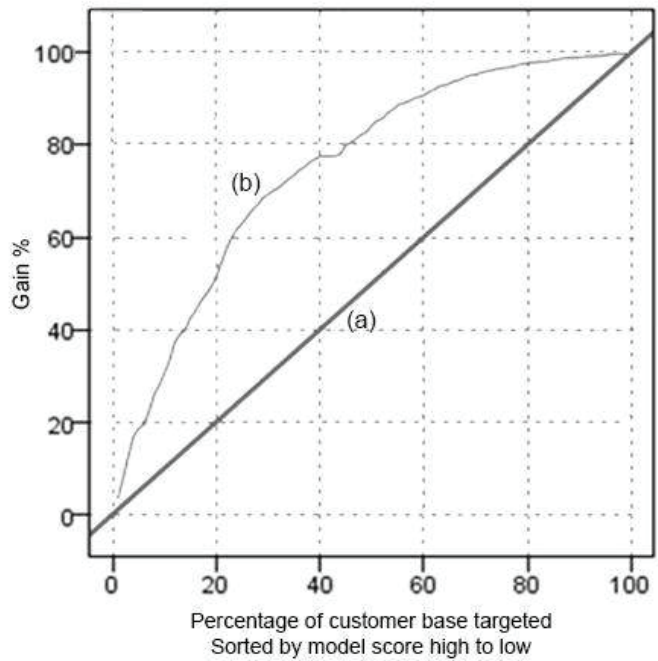

**Fig 1.** Out-of-sample validation of the model using lift curves. **(a)** is the gain if customer were selected at random while **(b)** shows selection by using the model.

We then select the treatment group using our model. We let the marketing team pick their best possible (control) group first and then specifically excluded them when selecting our treatment group. The treatment group is composed of the top 200 000 customers with the highest score. This represents approximately well under 1% of the total customer base.

## 4   Experiment

A large-scale experiment is then run to compare our data driven approach to the current best-practice in MNOs. The approaches will be compared using the conversion rates of the control and treatment group.

In this experiment, the selected customers receive a text saying that they can activate a 15MB bundle of data usage for half of the usual price. The 15MB have a limited validity and are only valid for 15 days. The customer can activate this offer by sending a text with a code to a short-number. This is a common type of campaign and is often used in this market. The text contains information about the offer and instructions on how to activate it.

The conversion rates between treatment and control group were striking. The conversion rate in the treatment group selected by our model is 6,42% while the conversion rate of the control group selected using the best-practice approach is only 0.5%, as shown in Fig. 2. The difference is highly significant (p-value < 2.2e-16). Our data-driven approach leads to a conversion rate 13 times larger than the best-practice approach.

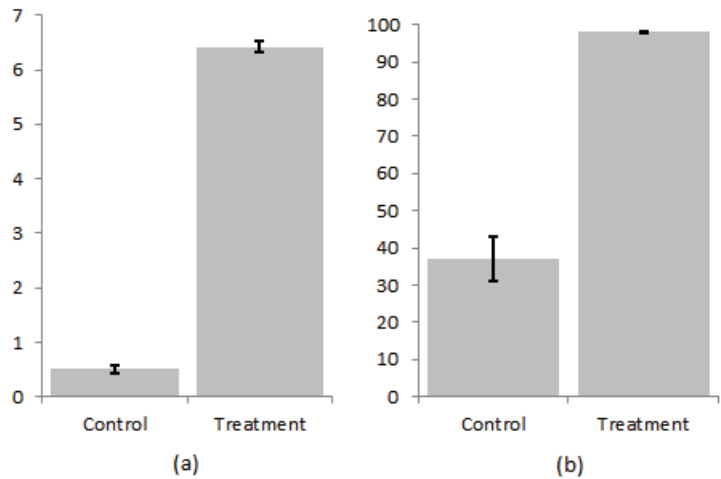

**Fig 2. (a)** Conversion rate in the control (best practice) and treatment (data-driven approach) groups. **(b)** the percentage of converted people who renewed their data plan after using the volume included in the campaign offer. Error bars are the 95% confidence interval on the mean using the blaker method

The goal of this campaign is not only to have customers take this offer but also to have them renew it after the trial period. We thus compare the renewal rate, customers buying another data plan after using the half-priced package, between the converted people in the two groups. Here too, the results are striking and highly significant (p-value < 2.2e-16). We find that while 98% of converted people in the treatment group buy a second, full price package, only 37% of the converted people in the control group renew their plan. This means that 6.29 % of the treatment group is converted in month two, compared to 0.19% of the control group.

## 5    Discussion

Although it was not a goal for our data-driven approach to be interpretable, a posteriori categorization of the features selected by our model leads to some interesting qualitative insights. Indeed, most of the features fall under three categories: discretionary income, timing, and social learning, see Table 3.

Discretionary income was expected by the marketing team to be important overall. They hypothetised that customers with a high total ARPU would be more likely to convert. The model does however not select total spending as an important variable to help predict conversion. In fact, looking at the ARPU of those who received an SMS and then adopt, we see that the low ARPU segment is slightly overrepresented. Our text and data focused discretionary spending variables are however selected as important by the model. Text and data focused spending variables seem to contain relevant information to help predict adoption more than overall spending.
Timing measured through using a new phone is our 9th most useful feature.

Finally, the social learning features we computed for this study turn out to be very helpful to help classify the natural converters. The total spending on data among the closest social graph neighbors is our most predictive feature.

Using social features in selecting customers for the offer might have improved the retention rates. We speculate that the value customers derive from mobile data increases when their neighbors are also mobile data users. In other words that we expect that a network externality effect exists in mobile internet data. This means that selecting customers whose closest neighbors are already using mobile data might have locally used this network effect to create the lasting effect we observe with very high retention rate in the second month.

The success of this pilot study triggered new technical developments and campaign data are now being recorded and stored. Future models might be refined using this experimental data and new insights might be uncovered using statistical models or interpretable machine learning algorithms. We also plan to refine the different attribute categories; social learning, discretionary income, and timing and to use them a priori in further model building.

The marketing team was impressed by the power of this method and is now looking into how this can be implemented in the MNO's operations and used more systematically for future campaigns. They see such data-driven approach to be particularly useful to help launch new products where little prior experience exists. This was the case with this campaign as the overall mobile internet penetration rate in the country is very low. The marketing team usually learns what the right segment for a new product through extensive trial-and-error.

When performing research on sensitive data, privacy is a major concern. [16,17] showed that large scale simply anonymized mobile phone dataset could be uniquely characterized using as little as four pieces of outside information. All sensitive information in this experiment was hashed and only the local marketing team had the contacts of the control and treatment groups.

We believe our findings open up exciting avenues of research within data-driven marketing and customer understanding. Using behavioral patterns, we increased the conversion rate of an internet data campaign by 13 times compared to current best-practice. We expect such an approach will greatly reduce spamming by providing the customer with more relevant offers.

**Acknowledgments.** This research was partially sponsored by the Army Research Laboratory under Cooperative Agreement Number W911NF-09-2-0053, and by the Media Laboratory Consortium. The conclusions in this document are those of the authors and should not be interpreted as representing the social policies, either expressed or implied, of the sponsors